\documentclass[twocolumn]{aa}
\usepackage{epsfig}
\def\simlt{\ \raise -2.truept\hbox{\rlap{\hbox{$\sim$}}\raise5.truept   %
\hbox{$<$}\ }}
\def\simgt{\ \raise -2.truept\hbox{\rlap{\hbox{$\sim$}}\raise5.truept   %
\hbox{$>$}\ }}                                                          %

\def\be{\begin{equation}}
\def\ee{\end{equation}}
\def\newline{\hfil\break}

\def\ergcm2s{{erg~cm$^{-2}$s$^{-1}$~}}

\def\la{\mathrel{\hbox{\rlap{\hbox{\lower4pt\hbox{$\sim$}}}\hbox{$<$}}}}
\def\ga{\mathrel{\hbox{\rlap{\hbox{\lower4pt\hbox{$\sim$}}}\hbox{$>$}}}}

\def\mug{$\mu$G ~}
\def\be{\begin{equation}}
\def\ee{\end{equation}}
\def\cm3{cm$^{-3}$}
\newcommand{\ds}{{\sffamily DarkSUSY}}

\def\msun{M_{\odot}{\ }}
\newcommand{\code}[1]{{\tt #1}}
\def\farcm{\hbox{$.\mkern-4mu^\prime$}}
\begin{document}

\title{Multi-frequency analysis of neutralino dark matter annihilations in the Coma cluster}

   \author{S. Colafrancesco \inst{1}, S. Profumo \inst{2} and P. Ullio \inst{3}}

   \offprints{S. Colafrancesco}

\institute{   INAF - Osservatorio Astronomico di Roma,
              via Frascati 33, I-00040 Monteporzio, Italy and\\
              Istituto Nazionale di Fisica Nucleare,
              Sezione di Roma 2, I-00133 Roma, Italy\\
              Email: cola@mporzio.astro.it
 \and
              Department of Physics, Florida State University,
              505 Keen Bldg., Tallahassee, FL 32306, U.S.A.\\
              Email: profumo@hep.fsu.edu
 \and
              Scuola Internazionale Superiore di Studi Avanzati,
              Via Beirut 2-4, I-34014 Trieste, Italy and\\
              Istituto Nazionale di Fisica Nucleare,
              Sezione di Trieste, I-34014 Trieste, Italy\\
              Email: ullio@sissa.it
             }
\date{Received: / Accepted: }

\authorrunning {S. Colafrancesco et al.}

\titlerunning {DM annihilations in Coma}

\abstract{We study the astrophysical implications of neutralino dark matter annihilations
in galaxy clusters, with a specific application to the Coma cluster. We first address the
determination of the dark halo models for Coma, starting from structure formation models
and observational data, and we discuss in detail the role of sub-halos. We then perform a
thorough analysis of the transport and diffusion properties of neutralino annihilation
products, and investigate the resulting multi-frequency signals, from radio to gamma-ray
frequencies. We also study other relevant astrophysical effects of neutralino
annihilations, like the DM-induced Sunyaev-Zel'dovich effect and the intracluster gas
heating. As for the particle physics setup, we adopt a two-fold approach, resorting both
to model-independent bottom-up scenarios and to benchmark, GUT-motivated frameworks. We
show that the Coma radio-halo data (the spectrum and the surface brightness) can be
nicely fitted by the neutralino-induced signal for peculiar particle physics models and
for magnetic field values, which we outline in detail. Fitting the radio data and moving
to higher frequencies, we find that the multi-frequency spectral energy distributions are
typically dim at EUV and X-ray frequencies (with respect to the data), but show a
non-negligible gamma-ray emission, depending on the amplitude of the Coma magnetic field.
A simultaneous fit to the radio, EUV and HXR data is not possible without violating the
gamma-ray EGRET upper limit. The best-fit particle physics models yields substantial
heating of the intracluster gas, but not sufficient energy injection as to explain the
quenching of cooling flows in the innermost region of clusters. Due to the specific
multi-frequency features of the DM-induced spectral energy distribution in Coma, we find
that supersymmetric models can be significantly and optimally constrained either in the
gamma-rays or at radio and microwave frequencies.

}

 \maketitle


\section{Introduction}

Most of the matter content of the universe is in form of dark matter, whose presence is
indicated by several astrophysical evidences (e.g., gravitational lensing, galaxy
rotation curves, galaxy clusters masses) but whose nature is still elusive.
On the cosmological side, the most recent results of observational
cosmology, {\em i.e.} WMAP vs. distant SN Ia, indicate that the matter content
of the universe is $\Omega_m h^2 = 0.135^{+0.008}_{-0.009}$ with a baryon
density of $\Omega_b h^2 = 0.0224 \pm 0.0009$ (\cite{Spergeletal2003}).
The combination of the available data on large scale structures (Ly-$\alpha$ forest
analysis of the SDSS, the SDSS galaxy clustering) with the latest SNe and with the 1-st
year WMAP CMB anisotropies can improve the determination of the cosmological parameters
(\cite{seljak}) and hence allow us to set a {\em concordance cosmological model}.\\
We refer, in this paper, to a flat $\Lambda$CDM cosmology with parameters chosen
according to the global best fitting results derived in \cite{seljak} (see their Table~1,
third column): we assume, in fact, that the present matter energy density is $\Omega_m =
0.281$, that the Hubble constant in units of 100 km s$^{-1}$ Mpc$^{-1}$ is $h=0.71$, that
the present mean energy density in baryons is $\Omega_b  = 0.0233/h^2$, with the only
other significant extra matter term in cold dark matter $\Omega_{CDM} = \Omega_m
-\Omega_b$, that our Universe has a flat geometry and a cosmological constant $\Lambda$,
{\em i.e.} $\Omega_\Lambda = 1-\Omega_m$, and, finally, that the primordial power
spectrum is scale invariant and is normalized to the value $\sigma_8 = 0.897$.
This choice sets our framework, but  it is not actually crucial for any of the results
presented in the paper, which can be easily rescaled in case of a re-assessment of
best-fit values of the cosmological parameters, and in particular of the value of
$\Omega_{CDM} $ (the present concordance cosmological model, while widely used, has been
also criticized and questioned in the light of still unexplored systematics, see, e.g.,
\cite{Myersetal2003,Sadatetal2005,LieuMittaz2005,Copietal2003})

The Coma cluster has been the first astrophysical laboratory for dark matter (DM) since
the analysis of F. Zwicky (\cite{Zwicky1933}). In this respect, we can consider the Coma
cluster as an astrophysical {\em benchmark case-study} for DM.
Modern observations have led to an increasingly sophisticated exploration of the DM
distribution in the universe, now confirmed to be a dominant component (relative to the
baryonic material) over scales ranging from those of galaxy halos to that of the particle
horizon.

The nature of DM is not yet known and several detection techniques have been used so far.
Obviously, direct detection is the cleanest and most decisive discriminant (see {\em
e.g.} \cite{Munoz:2003gx} for a review). However, it would be interesting if astronomical
techniques were to reveal some of the fundamental properties of DM particles.
In fact, if DM is supposed to consist of fundamental particles which are weakly
interacting, then their own interaction will lead to a number of astrophysical signatures
(e.g., high-energy gamma-rays, electrons and positrons, protons and neutrinos and hence
by their emission/interaction properties) indicative of their nature and composition.

These facts provide the basic motivations for our study, which is aimed to:
{\it i)} describe the multi-wavelength signals of the presence of DM through the emission features of the
secondary products of neutralino annihilation. These signals are of non-thermal nature and cover the
entire electro-magnetic spectrum, from radio to gamma-ray frequencies;
{\it ii)} indicate the best frequency windows where it will be possible to catch physical indications for the
nature of DM;
{\it iii)} apply this analysis to the largest bound containers of DM in
the universe, {\em i.e.} galaxy clusters. We shall focus here on the
case of the Coma cluster, a particularly rich and suitable
laboratory for which an extended observational database is at
hand.

\subsection{The fundamental physics framework}

Several candidates have been proposed as fundamental DM constituents, ranging from axions
to light, MeV DM, from KK particles, branons, primordial BHs, mirror matter to
supersymmetric WIMPs (see, {\em e.g.}, \cite{Baltz:2004tj}, \cite{Bertone:2004pz}, and
\cite{Bergstrom:2000pn} for recent reviews). In this paper we will assume that the main
DM constituent is the lightest neutralino of the minimal supersymmetric extension of the
Standard Model (MSSM). Although no experimental evidence in favor of supersymmetry has
shown up to date, several theoretical motivations indicate that the MSSM is one of the
best bets for new physics beyond the Standard Model. Intriguingly enough, and contrary to
the majority of other particle physics candidates for DM, supersymmetry can unambiguously
manifest itself in future accelerator experiments. Furthermore, provided neutralinos are
stable and are the lightest supersymmetric particles, next generation direct detection
experiments feature good chances to explore most of the neutralino-proton scattering
cross section range predicted by supersymmetry.

A long standing issue in phenomenological studies of low-energy supersymmetry is traced to the
parameterization of the supersymmetry breaking terms (see \cite{Chung} for a recent review). In this
respect, two somehow complementary attitudes have been pursued. On the one hand, one can appeal to a
(set of) underlying high energy principles to constrain the form of supersymmetry breaking term,
possibly at some high energy (often at a grand unification) scale (see {\em e.g.} \cite{Baer:2000gf}).
The low energy setup is then derived through the renormalization group evolution of the supersymmetry
breaking parameters down to the electroweak scale. Alternatively, one can directly face the most
general possible low energy realization of the MSSM, and try to figure out whether general properties
of supersymmetry phenomenology can be derived (see {\em e.g.} \cite{Profumo:2004at}).

In this paper we will resort to both approaches. We will show that the final state
products of neutralino pair annihilations show relatively few spectral patterns, and that
any supersymmetric configuration can be thought as an interpolation among the extreme
cases we shall consider here. The huge number of free parameters of the general MSSM are
therefore effectively decoupled, and the only relevant physical properties are the final
state products of neutralino pair annihilations, and the mass of the neutralino itself.
We will indicate this first strategy as a {\em bottom-up approach} (see
Sect.\ref{sec:radio} for details).

Since most phenomenological studies have been so far based on GUT-motivated models, and a
wealth of results on accelerator physics, direct and indirect detection has accumulated
within these frameworks, we decided to work out here, as well, the astrophysical
consequences, for the system under consideration, of a few {\em benchmark models}. The
latter have been chosen among the minimal supergravity (mSUGRA) models indicated in
\cite{Battaglia:2003ab} with the criterion of exemplifying the widest range of
possibilities within that particular theoretical setup (see Sect.\ref{sec:dm} for
details).

\subsection{The astrophysical framework}

To make our study quantitative, we will compare the predictions of the above mentioned
neutralino models with the observational set-up of the Coma cluster, which represents the
largest available observational database for a galaxy cluster. The total mass of Coma
found within $10 h^{-1}$ Mpc from its center is $M_{< 10 h^{-1} Mpc} \approx 1.65 \times
10^{15} M_{\odot}$ (\cite{GDK}).
The assumption of hydrostatic equilibrium of the thermal intra-cluster gas in Coma
provides a complementary estimate of its total mass enclosed in the radius $r$. A value
$M \approx 1.85 \times 10^{15} M_{\odot}$ within $5 h^{-1}_{50}$ Mpc from the cluster
center has been obtained from X-ray data (\cite{hughes}).\\
X-ray observations of Coma also yield detailed information about the thermal electrons
population. We know that the hot thermal electrons are at a temperature $k_B T_e=8.2 \pm
0.4$ keV (\cite{Arnaudetal2001}) and have a central density
$n_{0}=(3.42\pm0.047)\,h_{70}^{1/2}\times10^{-3}$ \cm3, with a spatial distribution
fitted by a $\beta$-model, $n(r)=n_{0}(1+r^2/r_c^2)^{-3\beta/2}$, with parameters
$r_c=10.5'\pm0.6'$ and $\beta\approx 0.75$ (\cite{BHB}).
Assuming spherical symmetry and the previous parameter values, the optical depth of the
thermal gas in Coma is $\tau_{th} \simeq 5.54 \times 10^{-3}$ and the pressure due to the
thermal electron population is $P_{th}\simeq 2.80 \cdot 10^{-2}$ keV \cm3. The hot
intra-cluster gas produces also a thermal SZ effect (\cite{sz1,sz2}; see
\cite{Birkinshaw1999} for a general review) which has been observed over a wide frequency
range, from $32$ to $245$ GHz (see \cite{DePetris2003} and references therein).\\
Beyond the presence of DM and thermal gas, Coma also shows hints for the presence of
relativistic particles in its atmosphere. The main evidence for the presence of a
non-thermal population of relativistic electrons comes from the observation of the
diffuse radio halo at frequencies $\nu_r \sim 30 MHz - 5 GHz$ (\cite{Deiss1997};
\cite{Thierbach2003}). The radio halo spectrum can be fitted by a power-law spectrum
$J_{\nu} \sim \nu^{-1.35}$ in the range $30$ MHz-$1.4$ GHz with a further steepening of
the spectrum at higher radio frequencies.
The radio halo of Coma has an extension of $R_h \approx 0.9 h_{70}^{-1}$ Mpc, and its
surface brightness is quite flat in the inner 20 arcmin with a radial decline at larger
angular distances (e.g., \cite{CMP2005}).\\
Other diffuse non-thermal emissions have been reported for Coma (as well as for a few
other clusters) in the extreme UV (EUV) and in the hard X-ray (HXR) energy bands.
The Coma flux observed in the $65 - 245$ eV band (\cite{Lieuetal1996}) is $\sim 36 \%$
above the expected flux from the thermal bremsstrahlung emission of the $k_BT \approx
8.2$ keV IC gas (\cite{Ensslin1998}) and it can be modeled with a power-law spectrum with
an approximately constant slope $\approx 1.75$, in different spatial regions
(\cite{Lieu1999}, \cite{Ensslin1999}, \cite{Bowyeretal2004}).
The EUV excess in Coma has been unambiguously detected and it does not depend much on the
data analysis procedure. The integrated flux in the energy band $0.13 - 0.18$ keV is
$F_{EUV} \approx (4.1 \pm 0.4) \cdot 10^{-12}$ \ergcm2s (\cite{Bowyeretal2004}).
According to the most recent analysis of the EUVE data (\cite{Bowyeretal2004}), the EUV
excess seems to be spatially concentrated in the inner region ($\theta \simlt 15-20$
arcmin) of Coma (see also \cite{Bonamenteetal2003}). The nature of this excess is not
definitely determined since both thermal and non-thermal models are able to reproduce the
observed EUV flux. However, the analysis of \cite{Bowyeretal2004} seems to favour a
non-thermal origin of the EUV excess in Coma generated by an additional population of
secondary electrons.
A soft X-ray (SXR) excess (in the energy range $\approx 0.1 - 0.245$ keV) has been also
detected in the outer region ($20' < \theta < 90'$) of Coma
(\cite{Bonamenteetal2003,Finoguenovetal2003}). The spectral features of this SXR excess
seem to be more consistent with a thermal nature of the SXR emission, while a non-thermal
model is not able to reproduce accurately the SXR data (e.g., \cite{Bonamenteetal2003}).
The SXR emission from the outskirts of Coma has been fitted in a scenario in which the
thermal gas at $k_B T_e \sim 0.2$ keV with $\sim 0.1$ solar abundance (see, e.g.
\cite{Finoguenovetal2003} who identify the warm gas with a WHIM component) resides in the
low-density filaments predicted to form around clusters as a result of the evolution of
the large-scale web-like structure of the universe (see \cite{Bonamenteetal2003}). It has
been noticed, however, that the WHIM component cannot reproduce, by itself, the Coma SXR
excess because it would produce a SXR emission by far lower (see \cite{Mittazetal2004}),
and thus one is forced to assume a large amount of warm gas in the outskirts of Coma.
Thus, it seems that the available EUV and SXR data indicate (at least) two different
electron populations: a non-thermal one, likely yielding the centrally concentrated EUV
excess and a thermal (likely warm) one, providing the peripherically located SXR excess.
In this paper, we will compare our models with the EUV excess only, which is intimately
related to Coma being spatially concentrated towards the inner region of the cluster.\\
There is also evidence of a hard X-ray (HXR) emission observed towards the direction of
Coma  with the BeppoSAX-PDS (\cite{Fusco1999}, \cite{Fuscoetal2004}) and with the
ROSSI-XTE experiments (\cite{RGB1999}).  Both these measurements indicate an excess over
the thermal emission which amounts to $F_{(20-80)keV}=(1.5\pm0.5)\cdot10^{-11}$ erg
s$^{-1}$ cm$^{-2}$ (\cite{Fuscoetal2004}).
It must be mentioned, for the sake of completeness, that the HXR excess of Coma is still
controversial (see \cite{RM2004}), but, for the aim of our discussion, it could at worst
be regarded as an upper limit.  The nature of the HXR emission of Coma is not yet fully
understood.\\
Finally, a gamma-ray upper limit of $F(>100 MeV) \approx 3.2 \times 10^{-8}$ pho
cm$^{-2}$ has been derived for Coma from EGRET observations (\cite{Sreekumar1996},
\cite{Reimer2003}).

The current evidence for the radio-halo emission features of Coma has been interpreted as
synchrotron emission from a population of primary relativistic electrons which are
subject to a continuous re-acceleration process supposedly triggered by merging shock
and/or intracluster turbulence (e.g., \cite{Brunetti2003}). The EUV and HXR emission
excesses are currently interpreted as Inverse Compton scattering (ICS) emission from
either primary or secondary electrons. Alternative modeling has been proposed in terms of
suprathermal electron bremsstrahlung emission for the HXR emission of Coma (see
\cite{Ensslin1999,KempnerSarazin2000}, see also \cite{Petrosian2001} for a critical
discussion) and in this case the EUV emission should be produced either by a different
relativistic electron population or by a warm thermal population both concentrated
towards the cluster center. Lastly, models in which the EUV and HXR emission can be
reproduced by synchrotron emission from the interaction of Ultra High Energy cosmic rays
and/or photons (\cite{Timokhin,Inoue}) have also been presented.
The situation is far from being completely clear and several problems still stand on both
the observational and theoretical sides of the issue.

Since DM is abundantly present in Coma and relativistic particles are among the main
annihilation products, we explore here the effect of DM annihilation on the
multi-frequency spectral energy distribution (SED) of Coma.
The plan of the paper is the following. We discuss in Sect.~\ref{sec:DMmodels} the DM
halo models for the Coma cluster, the set of best fitting parameters for the DM
distribution, and the role of sub-halos. The annihilation features of neutralinos and the
main annihilation products are discussed in Sect.~\ref{sec:ann}.  The multi-frequency
signals of DM annihilation are presented and discussed in details in
Sect.~\ref{sec:multiwave} and \ref{sec:discussion}, while the details of the transport
and diffusion properties of the secondary particles are described in the
Appendix~\ref{sec:secondary}, together with the derivation of the equilibrium spectrum of
relativistic particles in Coma.  The conclusions of our analysis and the outline for
forthcoming astrophysical searches for DM signals in galaxy clusters are presented in the
final Sect.\ref{sec:conclusions}.

\section{A $\Lambda$CDM model for the Coma cluster}\label{sec:DMmodels}

To describe the DM halo profile of the Coma cluster we refer, as a general setup, to the
$\Lambda$CDM model for structure formation, implementing results of galaxy cluster
formation obtained from N-body simulations. Free parameters are fitted against the
available dynamical information and are compared to the predictions of this scheme.
Substructures will play a major role when we will discuss the predictions for signals of
DM annihilations. In this respect, the picture derived from simulations is less clean
and, hence, we will describe in details our set of assumptions.

\subsection{The dark matter halo profile}

To describe the DM halo profile of the Coma cluster we consider the limit in which the
mean DM distribution in Coma can be regarded as spherically symmetric and represented by
the parametric radial density profile:
\begin{equation}
  \rho (r)=\rho^{\prime} g(r/a)\,.
\label{eq:profi}
\end{equation}
Two schemes are adopted to choose the function $g(x)$ introduced here. In the first one,
we assume that $g(x)$ can be directly inferred as the function setting the universal
shape of DM halos found in numerical N-body simulations of hierarchical clustering. We
are assuming, hence, that the DM profile is essentially unaltered from the stage
preceding the baryon collapse, which is -- strictly speaking -- the picture provided by
the simulations for the present-day cluster morphology. A few forms for the universal DM
profile have been proposed in the literature: we implement here the non-singular form
(which we label as N04 profile) extrapolated by \cite{n04}:
\begin{equation}
  g_{N04}(x) = \exp[-2/\alpha (x^\alpha-1)]
  \;\;\;\;{\rm with} \;\;\;\; \alpha \simeq 0.17\;,
\label{eq:n04}
\end{equation}
and the shape with a mild singularity towards its center proposed by \cite{d05} (labeled
here as D05 profile):
\begin{equation}
  g_{D05}(x) = \frac{1}{x^{\gamma} (1+x)^{3-\gamma}}
  \;\;\;\;{\rm with} \;\;\;\; \gamma \simeq 1.2\;.
\label{eq:d05}
\end{equation}
The other extreme scheme is a picture in which the baryon infall induces a large transfer
of angular momentum between the luminous and the dark components of the cosmic structure,
with significant modification of the shape of the DM profile in its inner region.
According to a recent model (\cite{elzant}), baryons might sink in the central part of DM
halos after getting clumped into dense gas clouds, with the halo density profile in the
final configuration found to be described by a profile (labeled here as B profile) with a
large core radius  (see, e.g., \cite{burkert}):
\begin{equation}
  g_{B}(x) = \frac{1}{(1+x)\,(1+x^2)}\;.
\label{eq:burk}
\end{equation}

Once the shape of the DM profile is chosen, the radial density profile in
Eq.~(\ref{eq:profi}) is fully specified by two parameters: the length-scale $a$ and the
normalization parameter $\rho^{\prime}$. It is, however, useful to describe the density
profile model by other two parameters, {\em i.e.}, its virial mass $M_{vir}$ and
concentration parameter $c_{vir}$. For the latter parameter, we adopt here the definition
by \cite{Bullock}. We introduce the virial radius $R_{vir}$ of a halo of mass $M_{vir}$
as the radius within which the mean density of the halo is equal to the virial
overdensity $\Delta_{vir}$ times the mean background density $\bar{\rho}=\Omega_m
\rho_c$:
\begin{equation}
  M_{vir} \equiv  {4\pi\over 3} \Delta_{vir} \bar{\rho}\, R_{vir}^3.
\end{equation}
We assume here that the virial overdensity can be approximated by the expression (see ~\cite{BN}),
appropriate for a flat cosmology,
\begin{equation}
  \Delta_{vir} \simeq {\frac{(18\pi^2 + 82x- 39 x^2)}{1-x}}\;,
\end{equation}
with $x \equiv \Omega_m(z) -1$. In our cosmological setup we find at $z=0$, $\Delta_{vir} \simeq 343$
(we refer to \cite{Colafrancesco1994}, \cite{Colafrancesco1997} for a general derivation of the virial
overdensity in different cosmological models).
The concentration parameter is then defined as
\begin{equation}
  c_{vir} = \frac{R_{vir}}{r_{-2}} \equiv  \frac{R_{vir}}{x_{-2}\, a}\;,
  \label{eq:cvir}
\end{equation}
with $r_{-2}$ the radius at which the effective logarithmic slope of the profile is $-2$.
We find that $x_{-2} =1$ for the N04 profile (see Eq.~\ref{eq:n04}), $x_{-2} =2-\gamma$
for D05 profile (see Eq.~\ref{eq:d05}), and $x_{-2} \simeq 1.52$ for the Burkert profile
(see Eq.~\ref{eq:burk}).

Since the first numerical results with large statistics became available (\cite{NFW}), it
has been realized that, at any given redshift, there is a strong correlation between
$c_{vir}$ and $M_{vir}$, with larger concentrations found in lighter halos. This trend
may be intuitively explained by the fact that mean overdensities in halos should be
correlated with the mean background densities at the time of collapse, and in the
hierarchical structure formation model small objects form first, when the Universe was
indeed denser.
The correlation between $c_{vir}$ and $M_{vir}$ is relevant in our context at two levels:
{\em i}) when discussing the mean density profile of Coma and, {\em ii}) when including
substructures. Hence, we will review this relevant issue here and we will apply it to the
present case of Coma. \cite{Bullock} proposed a model to describe this correlation,
improving on the toy model originally outlined in~\cite{NFW}. A collapse redshift $z_c$
is assigned, on average, to each halo of mass $M$ at the epoch $z$ through the relation
$M_{\star}(z_c) \equiv F M$. Here it is postulated that a fixed fraction $F$ of $M$
(following \cite{Wechsleretal} we choose $F=0.015$) corresponds to the typical collapsing
mass $M_{\star}$, as defined implicitly by $\sigma\left(M_{\star}(z)\right)=
\delta_{sc}(z)$, with $\delta_{sc}$ being the critical overdensity required for collapse
in the spherical model and $\sigma(M)$ being the present-day rms density fluctuation in
spheres containing a mean mass $M$ (see, {\em e.g.}, \cite{Peebles1980}). An expression
for $\delta_{sc}$ is given, {\em e.g.}, in \cite{ECF}. The rms fluctuation $\sigma(M)$ is
related to the fluctuation power spectrum $P(k)$ (see {\em e.g.} \cite{Peebles}) by
\begin{equation}
  \sigma^2(M) \equiv \int d^3k \;
  \tilde{W}^2(k\,R) \, P(k) \, ,
\end{equation}
where $\tilde{W}$ is the top-hat window function on the scale $R^3=3M/4\pi \bar{\rho}$
with $\bar{\rho}$ the mean (proper) matter density, {\em i.e.} $\bar{\rho} = \Omega_m
\rho_c$ with $\rho_c$ the critical density. The power spectrum $P(k)$ is parametrized as
$P(k) \propto k^n T^2(k)$ in terms of the primordial power-spectrum shape $\propto k^n$
and of the transfer function $T^2(k)$ associated to the specific DM scenario. We fix the
primordial spectral index $n=1$ and we take the transfer function $T^2(k)$ given by
\cite{Bardeenetal} for an adiabatic CDM model, with the shape parameter modified to
include baryonic matter according to the prescription in, {\em e.g.}~\cite{Peacock} (see
their eqs.15.84 and 15.85) and introducing a multiplicative exponential cutoff at large
$k$ corresponding to the free-streaming scale for WIMPs (\cite{greenetal}).
The spectrum $P(k)$ is normalized to the value $\sigma_8=0.897$ as was quoted above.

\begin{figure}[!t]
\begin{center}
\includegraphics[scale=0.58]{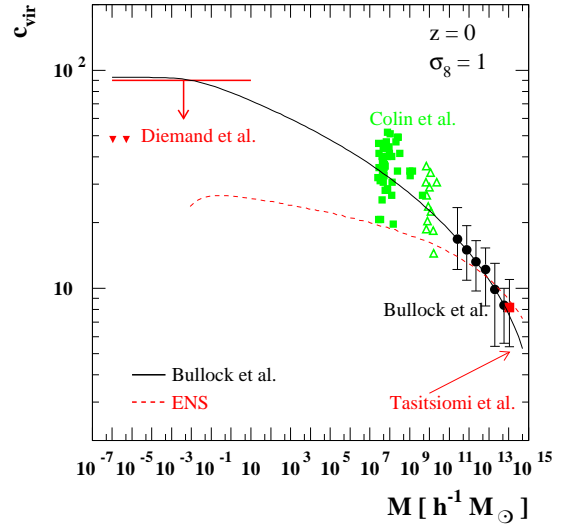}
\end{center}
 \caption{The dependence of $c_{vir}$ on the halo mass $M$, at $z=0$, as in the Bullock et al. toy model (solid line)
and in the ENS toy model (dashed line); predictions are compared to a few sets of simulation results
in different mass ranges. A flat, vacuum-dominated cosmology with $\Omega_M = 0.3$, $\Omega_{\Lambda}
= 0.7$, $h = 0.7$ and $\sigma_8 = 1$ is assumed here.}
 \label{fig:cvir}
\end{figure}

The toy model of Bullock et al. (2001) prescribes a one to one correspondence between the
density of the Universe at the collapse redshift $z_c$ of the DM halo and a
characteristic density of the halo at the redshift $z$; it follows that, on average, the
concentration parameter is given by
\begin{equation}
  c_{vir}(M,z) = K \frac{1+z_c}{1+z}  =
  \frac{c_{vir}(M,z=0)}{(1+z)} \;,
\label{cvir}
\end{equation}
with $K$ being a constant ({\em i.e.}\ independent of $M$ and cosmology) to be fitted to the results of the
N-body simulations.
We plot in Fig.~\ref{fig:cvir} the dependence of $c_{vir}$ on the halo mass $M$, at
$z=0$, according to the toy model of Bullock et al. (2001) as extrapolated down to the
free-streaming mass scale for DM halos made of WIMPs, {\em i.e.} around $10^{-6}\,\msun$
(see \cite{greenetal}). The predictions are compared to the results of a few sets of
N-body simulations: we use ``data'' points and relative error bars from~\cite{Bullock}
(representing a binning in mass of results for a large sample of simulated halos; in each
mass bin, the marker and the error bars correspond, respectively, to the peak and the
68\% width in the $c_{vir}$ distribution) to determine the parameter $K$. The same value
will be used to infer the mean $c_{vir}$ predicted in our cosmological setup. Other
``datasets'' refer actually to different values of $\sigma_8$ and different redshifts $z$
($z=26$ for the two minihalos fitted in Fig.~2 of~\cite{moorenature} and for the upper
bound in the range up to $10\,\msun$ quoted in the same paper; $z=3$ for the sample from
\cite{colinetal.}) and have been extrapolated, consistently with our prescriptions, to
$z=0$ and $\sigma_8 = 1$. Since small objects tend to collapse all at the same redshift,
the dependence on mass of the concentration parameters flattens at small masses; the mean
asymptotic value we find is slightly larger than the typical values found in
\cite{moorenature}, but it is still consistent with that analysis.

An alternative toy-model to describe the relation between $c_{vir}$ and $M$ has been discussed by Eke,
Navarro and Steinmetz (\cite{ENS}, hereafter ENS model). The relation they propose has a similar
scaling in $z$, but with a different definition of the collapse redshift $z_c$ and a milder dependence
of $c_{vir}$ on $M$. In our notation, they define $z_c$ through the equation
\begin{equation}
  D(z_c) \sigma_{\rm eff}(M_p)=  {1 \over C_{\sigma}}
 \label{zcens}
\end{equation}
where $D(z)$ represents the linear theory growth factor, and $\sigma_{\rm eff}$ is an `effective'
amplitude of the power spectrum on scale $M$:
\begin{equation}
  \sigma_{\rm eff}(M)=\sigma(M) \,
  \left(-\frac{d\ln(\sigma)}{d \ln(M)}(M)\right)
  = - \frac{d\sigma}{dM} M
 \label{sigeff}
\end{equation}
which modulates $\sigma(M)$ and makes $z_c$ dependent on both the amplitude and on the
shape of the power spectrum, rather than just on the amplitude, as in the toy model of
Bullock et al. (2001). Finally, in Eq.~(\ref{zcens}), $M_p$ is assumed to be the mass of
the halo contained within the radius at which the circular velocity reaches its maximum,
while $C_{\sigma}$ is a free parameter (independent of $M$ and cosmology) which we will
fit again to the ``data'' set in~\cite{Bullock}. With such a definition of $z_c$ it
follows that, on average, $c_{vir}$ can be expressed as:
\begin{equation}
 c_{vir}(M,z) =
  \left(\frac{\Delta_{vir}(z_c)\,\Omega_M(z)}
  {\Delta_{vir}(z)\,\Omega_M(z_c)}\right)^{1/3}
  \frac{1+z_c}{1+z}\;.
\label{cvirens}
\end{equation}
As shown in Fig.~\ref{fig:cvir}, the dependence of $c_{vir}$ on $M$ given by
Eq.(\ref{cvirens}) above is weaker than that obtained in the Bullock et al. (2001)
toy-model, with a significant mismatch in the extrapolation already with respect to the
sample from \cite{colinetal.} and an even larger mismatch in the low mass end. Moreover,
the extrapolation breaks down when the logarithmic derivative of the $\sigma(M)$ becomes
very small, in the regime when $P(k)$ scales as $k^{-3}$. Note also that predictions in
this model are rather sensitive to the specific spectrum $P(k)$ assumed (in particular
the form in the public release of the ENS numerical code gives slightly larger values of
$c_{vir}$ in its low mass end, around a value $c_{vir}\approx 40$ (we checked that
implementing our fitting function for the power spectrum, we recover our trend).

\subsection{Fitting the halo parameters of Coma}

For a given shape of the halo profile we make a fit of the parameters $M_{vir}$ and $c_{vir}$ against
the available dynamical constraints for Coma. We consider two bounds on the total mass of the cluster
at large radii, as inferred with techniques largely insensitive to the details of the mass profile in
its inner region. In \cite{GDK}, a total mass
\begin{equation}
  M(r<10\;{\rm h}^{-1}\, {\rm Mpc}) =
  (1.65 \pm 0.41)\, 10^{15} \;{\rm h}^{-1} \,M_{\odot} \;
  \label{eq:mass1}
\end{equation}
is derived mapping the caustics in redshift space of galaxies infalling in Coma on nearly radial
orbits. Several authors derived mass budgets for Coma using optical data and applying the virial
theorem, or using X-ray data and assuming hydrostatic equilibrium. We consider the bound derived by
\cite{hughes}, cross-correlating such techniques:
\begin{equation}
  M(r<5\;{\rm h}_{50}^{-1}\, {\rm Mpc}) = (1.85 \pm 0.25)\, 10^{15} \;{\rm h}_{50}^{-1} \,M_{\odot} \;,
  \label{eq:mass2}
\end{equation}
where $h_{50}$ is the Hubble constant in units of 50 km s$^{-1}$ Mpc$^{-1}$.

In our discussion some information on the inner shape of the mass profile in Coma is also important:
we implement here the constraint that can derived by studying the velocity moments of a given tracer
population in the cluster. As the most reliable observable quantity one can consider the projection
along the line of sight of the radial velocity dispersion of the population; under the assumption of
spherical symmetry and without bulk rotation, this is related to the total mass profile $M(r)$ by the
expression (\cite{BM,LM}):
\begin{eqnarray}
    \sigma_{\rm los}^2 (R) & = & \frac{2 G}{I(R)} \int_{R}^{\infty}
    \, {\rm d} r^\prime \ \nu (r^\prime) M(r^\prime) (r^\prime)^{2 \beta - 2}    \nonumber \\
    & &  \times
    \int_{R}^{r^\prime}  \,{\rm d} r \left( 1-\beta \frac{R^2}{r^2} \right)
    \frac{r^{-2 \beta +1}}{\sqrt{r^2 - R^2}} \ ,
    \label{eq:los}
\end{eqnarray}
where $\nu(r)$ is the density profile of the tracer population and $I(R)$ represents its
surface density at the projected radius $R$. In the derivation of Eq.~(\ref{eq:los}), a
constant-over-radius anisotropy parameter $\beta$ defined as
\begin{equation}
    \beta \equiv 1-\frac{\sigma_\theta^2(r)}{\sigma_r^2(r)}\;,
\end{equation}
has been assumed with $\sigma_r^2$ and $\sigma_\theta^2$ being, respectively, the radial
and tangential velocity dispersion ($\beta=1$ denotes the case of purely radial orbits,
$\beta=0$ that of system with isotropic velocity dispersion, while $\beta \rightarrow -
\infty$ labels circular orbits). Following \cite{LM}, we take as tracer population that
of the E-S0 galaxies, whose line of sight velocity dispersion has been mapped, according
to Gaussian distribution, in nine radial bins from $4^\prime$ out to $190^\prime$ (see
Fig. 3 in~\cite{LM}), and whose density profile can be described by the fitting function:
\begin{equation}
    \nu(r) \propto \frac{1}{(r/r_{\rm S}) (1+r/r_{\rm S})^2} \ ,
\end{equation}
with $r_{\rm S}=7\farcm05$.
Constraints to the DM profile are obtained through its contribution to $M(r)$, in which
we include the terms due to spiral and E-S0 galaxies (each one with the appropriate
density profile normalized  to the observed luminosity through an appropriate
mass-to-light ratio), and the gas component (as inferred from the X-ray surface
brightness distribution) whose number density profile can be described by the fitting
function:
\begin{equation}
    n(r) = n_0  \left[ 1+ \left(\frac{r}{r_{\rm c}} \right)^2 \right]^{- 1.5 \; b}\, ,
    \label{eq:gas}
\end{equation}
with $n_0 = 3.42 \times 10^{-3}$~cm$^{-3}$, $r_{\rm c}=10\farcm5$ and $b=0.75$ (\cite{BHB}).

To compare a model with such datasets, we build a reduced $\chi^2$-like variable of the form:
\begin{eqnarray}
  \chi^2_r  & = & \frac{1}{2}
  \bigg[ \frac{1}{2} \sum_{i =1}^2 \frac{\left(M(r<r_i)-M_i\right)^2}{(\Delta M_i)^2}
  \nonumber \\
  & &  +  \frac{1}{9} \sum_{j =1}^9 \frac{\left( \sigma_{\rm los}(R_j)-\sigma_{\rm los}^j
  \right)^2}{(\Delta \sigma_{\rm los}^j)^2} \bigg]
\label{chi2}
\end{eqnarray}
where the index $i$ in the first sum runs over the constraints given in Eqs.~(\ref{eq:mass1}) and
(\ref{eq:mass2}), while, in the second sum, we include the nine radial bins over which the line of sight
velocity dispersion of E-S0 galaxies and its standard deviation has been estimated.
Weight factors have been added to give the same statistical weight to each of the two classes
of constraints, see, e.g., \cite{DB} where an analogous procedure has been adopted.

\begin{figure}[!t]
\begin{center}
\includegraphics[scale=0.55]{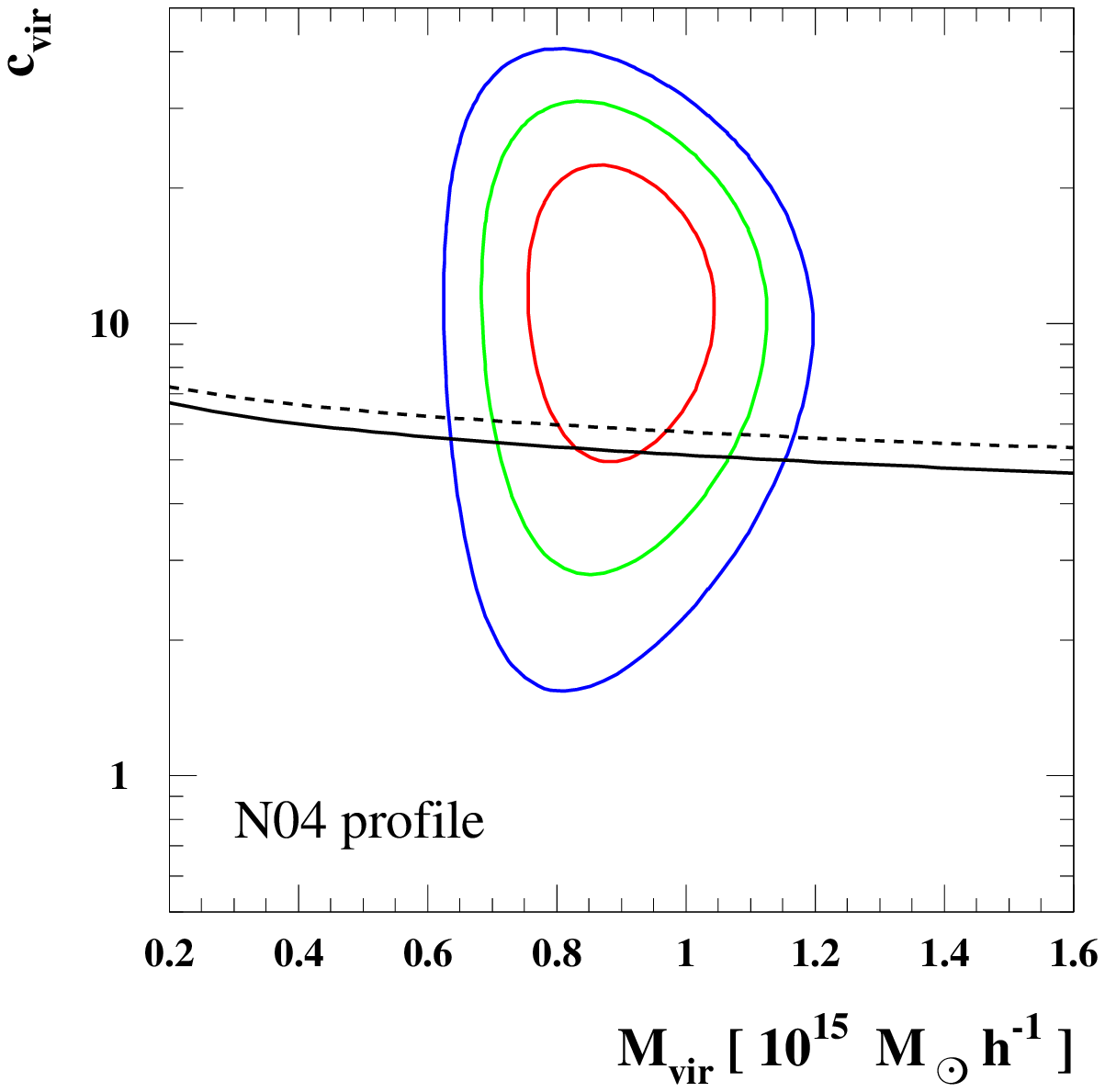}
\quad\includegraphics[scale=0.55]{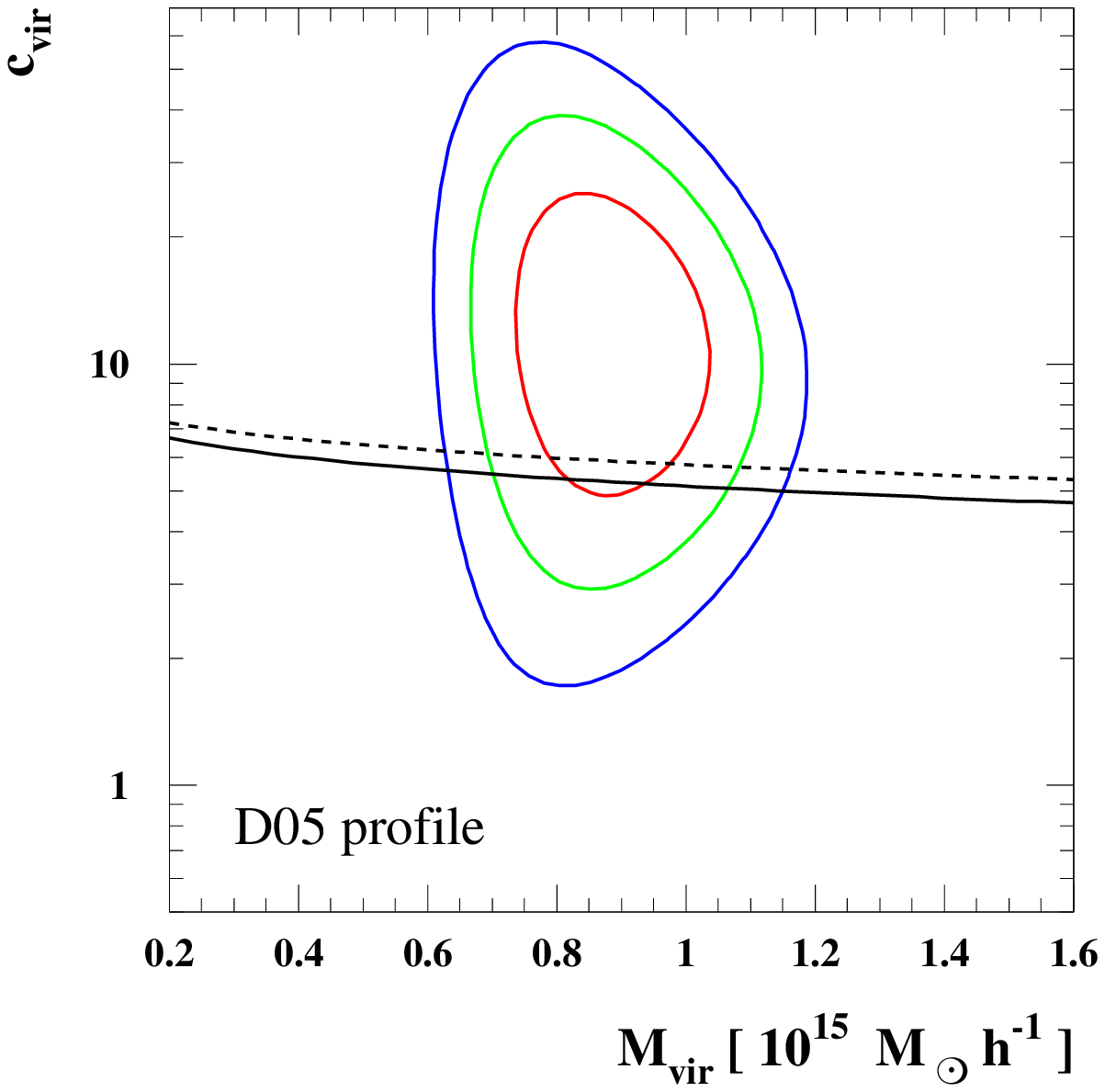}\\
\end{center}
\caption{We show the 1~$\sigma$,  2~$\sigma$ and 3~$\sigma$ contours as derived from the
$\chi^2_r$ variable in eq.(19), for  the Navarro et al. halo profile,
Eq.~(\protect{\ref{eq:n04}}), (upper panel)  and for the Diemand et al. halo profile,
Eq.~(\protect{\ref{eq:d05}}) (lower panel). Also shown are mean values for the
correlation between $M_{vir}$ and $c_{vir}$ as in the toy models of
\protect{\cite{Bullock}} (solid line) and that of \protect{\cite{ENS}} (dashed line).}
\label{fig:halo}
\end{figure}

\begin{figure}[t]
\begin{center}
\includegraphics[scale=0.55]{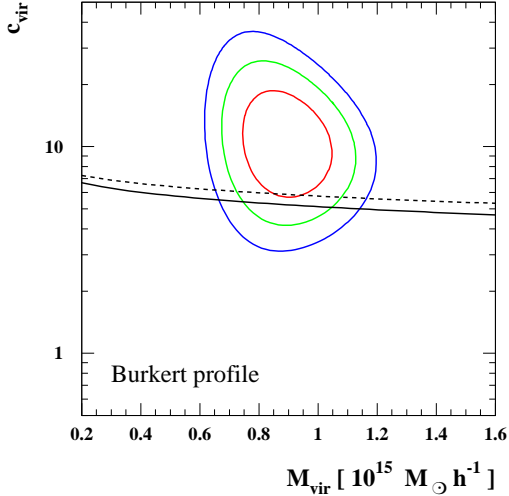}
\end{center}
\caption{We show here the 1~$\sigma$,  2~$\sigma$ and 3~$\sigma$ contours as derived from the
$\chi^2_r$ variable introduced in the text, for the Burkert profile, Eq.~(\protect{\ref{eq:burk}}).}
\label{fig:halo2}
\end{figure}

Nonetheless, we have derived in Fig.~\ref{fig:halo} the 1~$\sigma$,  2~$\sigma$ and
3~$\sigma$ contours in the $(M_{vir},\;c_{vir})$ plane for the Navarro et al. halo
profile (Eq.~(\ref{eq:n04}) and for the Diemand et al. halo profile (Eq.~\ref{eq:d05}).
In Fig.~\ref{fig:halo2} we show the analogous contours for the Burkert profile
(Eq.~(\ref{eq:burk})). In all these cases we have performed the fit of the line-of-sight
radial velocity dispersion of E-S0 galaxies assuming that this system has an isotropic
velocity dispersion, {\em i.e.} we have taken $\beta=0$. Best fitting values are found at
$M_{vir} \simeq 0.9 \cdot 10^{15} \msun h^{-1}$ and $c_{vir} \simeq 10$ (that we
consider, hence, as reference values in the following analysis), not too far from the
mean value expected from models sketching the correlation between these two parameters in
the $\Lambda$CDM picture. We show in Figs.~\ref{fig:halo} and \ref{fig:halo2} the
predictions of such correlation in the models of Bullock et al. (solid line) and of
\cite{ENS} (dashed line).

\subsection{Substructures in the Coma cluster}

Since the astrophysical signals produced by WIMP pair annihilation scale with the square
of the WIMP density, any local overdensity does play a role (see {\em e.g.}
\cite{Bergstrom:1998jj} and references therein). To discuss substructures in the Coma
cluster, analogously to the general picture introduced above for DM halos, we label a
subhalo through its virial mass $M_s$ and its concentration parameter $c_s$ (or
equivalently a typical density and length scale, $\rho^{\prime}_s$ and $a_s$). The
subhalo profile shape is considered here to be spherical and of the same form as for the
parent halo. Finally, as for the mean DM density profile, the distribution of subhalos in
Coma is taken to be spherically symmetric. The subhalo number density probability
distribution can then be fully specified through $M_s$, $c_s$ and the radial coordinate
for the subhalo position $r$. To our purposes, it is sufficient to consider the
simplified case when the dependence on these three parameters  can be factorized, {\em
i.e.}:
\begin{equation}
    \frac{dn_s}{dr^3\,dM_s\,dc_s} =
    p_s(r) \, \frac{dn_s}{dM_s}(M_s) \, {\cal{P}}_s(c_s)\;.
\label{eq:probsub}
\end{equation}
Here we have introduced a subhalo mass function, independent of radius, which is assumed
to be of the form:
\begin{equation}
    \frac{dn_s}{dM_s} = \frac{A(M_{vir})}{M_s^{1.9}}
    \exp{\left[-\left(\frac{M_s}{M_{cut}}\right)^{-2/3}\right]}\;,
\end{equation}
\cite{moorenature} where $M_{cut}$ is the free streaming cutoff mass (\cite{greenetal}),
while the normalization $A(M_{vir})$ is derived imposing that the total mass in subhalos
is a fraction $f_s$ of the total virial mass $M_{vir}$ of the parent halo, {\em i.e.}
\begin{equation}
    \int_{M_{cut}}^{M_{vir}} dM_s \frac{dn_s}{dM_s} M_s = f_s M_{vir} \;.
\end{equation}
According to \cite{moorenature}, $f_s$ is about 50\% for a Milky Way size halo, and we
will assume that the same holds for Coma. The quantity ${\cal{P}}_s(c_s)$ is a log-normal
distribution in concentration parameters around a mean value set by the substructure
mass; the trend linking the mean $c_s$ to $M_s$ is expected to be analogous to that
sketched above for parent halos with the Bullock et al. or ENS toy models, except that,
on average, substructures collapsed in higher density environments and suffered tidal
stripping. Both of these effects go in the direction of driving larger concentrations, as
observed in the numerical simulation of \cite{Bullock}, where it is shown that, on
average and for $M \sim 5\cdot 10^{11} \msun$ objects, the concentration parameter in
subhalos is found to be a factor of $\approx 1.5$ larger than for halos. We make here the
simplified ansazt:
\begin{equation}
    \langle c_s(M_s) \rangle = F_s \langle c_{vir}(M_{vir}) \rangle
  \;\;\;\;{\rm with} \;\;\;\; M_s = M_{vir}\;,
\end{equation}
where, for simplicity, we assume that the enhancement factor $F_s$ does not depend on
$M_s$. Following again \cite{Bullock}, the $1 \sigma$ deviation $\Delta(\log_{10} c_s)$
around the mean in the log-normal distribution ${\cal{P}}_s(c_s)$, is assumed to be
independent of $M_s$ and of cosmology, and to be, numerically, $\Delta(\log_{10} c_s)=
0.14$.\\
Finally, we have to specify the spatial distribution of substructures within the cluster.
Numerical simulations, tracing tidal stripping, find radial distributions which are
significantly less concentrated than that of the smooth DM component. This radial bias is
introduced here assuming that:
\begin{equation}
    p_s(r)  \propto g(r/a^{\prime})\;,
\end{equation}
with $g$ being the same functional form introduced above for the parent halo, but with
$a^{\prime}$ much larger than the length scale $a$ found for Coma. Following \cite{nk},
we fix $a^{\prime}/a \simeq 7$.
Since the fraction $f_s$ of DM in subhalos refers to structures within the virial radius,
the normalization of $p_s(r)$ follows from the requirement
\begin{equation}
    4 \pi \int_0^{R_{vir}} r^2 p_s(r) = 1\;.
\label{eq:normps}
\end{equation}

\section{Neutralino annihilations in Coma}\label{sec:ann}

\subsection{Statistical properties}

Having set the reference particle physics framework and specified the distribution of DM
particles, we can now introduce the source function from neutralino pair annihilations. For any stable
particle species $i$, generated promptly in the annihilation or produced in the decay and
fragmentation processes of the annihilation yields, the source function $Q_i(r,E)$ gives the number of
particles per unit time, energy and volume element produced locally in space:
\begin{equation}
Q_i(r,E) = \langle\sigma v\rangle_0 \sum_f  \frac{dN_{i}^f}{dE}(E) B_f\; {\cal N}_{\rm pairs}(r)\;,
\end{equation}
where $\langle\sigma v\rangle_0$ is the neutralino annihilation rate at zero temperature.
The sum is over all kinematically allowed annihilation final states $f$, each with a
branching ratio $B_f$ and a spectral distribution $dN_i^f/dE$, and  ${\cal N}_{\rm
pairs}(r)$ is the number density of neutralino pairs at a given radius $r$ ({\em i.e.},
the number of DM particles pairs per volume element squared). The particle physics
framework sets the quantity $\langle\sigma v\rangle_0$ and the list of $B_f$. Since the
neutralino is a Majorana fermion light fermion final states are suppressed, while --
depending on mass and composition -- the dominant channels are either those with heavy
fermions or those with gauge and Higgs bosons. The spectral functions $dN_i^f/dE$ are
inferred from the results of MonteCarlo codes, namely  the \code{Pythia} (\cite{pythia})
6.154, as included in the \ds\ package (\cite{ds}). Finally, ${\cal N}_{\rm pairs}(r)$ is
obtained by summing the contribution from the smooth DM component, which we write here as
the difference between the cumulative profile and the term that at a given radius is
bound in subhalos, and the contributions from each subhalo, in the limit of unresolved
substructures and in view of the fact that we consider only spherically averaged
observables:
\begin{eqnarray}
  {\cal N}_{\rm pairs}(r) & = &   \bigg[
  \frac{\left(\rho^{\prime} g(r/a)
  - f_s \, M_{vir} \,p_s(r)\right)^2}{2\,M_{\chi}^2} \nonumber \\
  &&  +  p_s(r) \,\int dM_s \frac{dn_s}{dM_s} \int dc_s ^{\,\prime}\;
  {\cal{P}}_s\left(c_s ^{\,\prime}(M_s)\right) \nonumber \\
  & & \times  \int d^3r_s\;
  \frac{\left(\rho^{\prime}_s \,g(r_s/a_s)\right)^2}
  {2\,M_{\chi}^2} \bigg] \;.
\end{eqnarray}
This quantity can be rewritten in the more compact form:
\begin{eqnarray}
   {\cal N}_{\rm pairs}(r) & = & \frac{\bar{\rho}^2}{2\,M_{\chi}^2}  \bigg[
  \frac{\left(\rho^{\prime} g(r/a)
  - f_s \, \tilde{\rho}_s \,g(r/{a^{\prime}})\right)^2}{\bar{\rho}^2} \nonumber \\
  & &  + f_s \Delta^2
  \,\frac{\tilde{\rho}_s \,g(r/{a^{\prime}})}{\bar{\rho}} \bigg]\;,
 \label{eq:pairs}
\end{eqnarray}
where we have normalized densities to the present-day mean matter density in the Universe
$\bar{\rho}$, and we have defined the quantity:
\begin{eqnarray}
  f_s \Delta^2 & \equiv &
  \frac{\int dM_s \frac{dn_s}{dM_s} M_s  \Delta_{M_s}^2(M_s)}{M_{vir}} \\
  & = & f_s \frac{\int dM_s \frac{dn_s}{dM_s} M_s  \Delta_{M_s}^2(M_s)}
  {\int dM_s \frac{dn_s}{dM_s} M_s}\;,
\end{eqnarray}
with
\begin{equation}
  \Delta_{M_s}^2(M_s) \equiv
  \frac{\Delta_{vir}(z)}{3}\,\int dc_s ^{\,\prime}\;
  {\cal{P}}_s\left(c_s ^{\,\prime}\right)
  \frac{I_2(c_s ^{\,\prime}\,x_{-2})}
  {\left[I_1(c_s ^{\,\prime}\,x_{-2})\right]^2}
  (c_s ^{\,\prime}\,x_{-2})^3 \;
\end{equation}
and
\begin{equation}
  I_n(x) = \int_{0}^{x} dy\, y^2 \left[g(y)\right]^n\;.
\end{equation}
Such definitions are useful since $\Delta_{M_s}^2$ gives the average enhancement in the source due to
a subhalo of mass $M_s$, while $\Delta^2$ is the sum over all such contributions weighted over the
subhalo mass function times mass. Finally, in Eq.~(\ref{eq:pairs}) we have also introduced the quantity:
\begin{equation}
  \tilde{\rho}_s \equiv \frac{M_{vir}}{4\,\pi\,(a^{\prime})^3
               I_1\left(R_{vir}/a^{\prime}\right)}\;.
\end{equation}
In the limit in which the radial distribution of substructures traces the DM profile, {\em i.e.}
$a^{\prime}=a$, $\tilde{\rho}_s$ becomes equal to the halo normalization parameter $\rho^{\prime}$.

\begin{figure}[t]
\begin{center}
\includegraphics[scale=0.55]{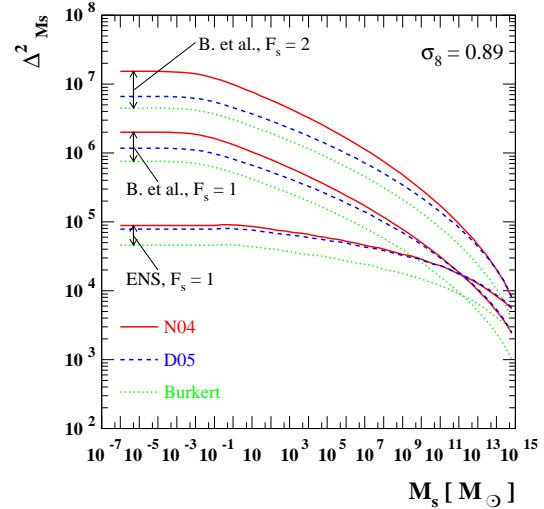}
\end{center}
\caption{Scaling of the average enhancement in source functions due to a subhalo of mass
$M_s$. We show result implementing the three halo profiles introduced, {\em i.e.} the N04, D05 and
Burkert profile, the two toy models for the scaling of concentration parameter with mass,
{\em i.e.} the Bullock et al. and the ENS, and two sample values of the ratio between concentration
parameter in subhalos to concentration parameter in halos at equal mass $F_s$.}
\label{fig:delta2ms}
\end{figure}

\begin{figure*}[t]
\begin{center}
\includegraphics[scale=0.55]{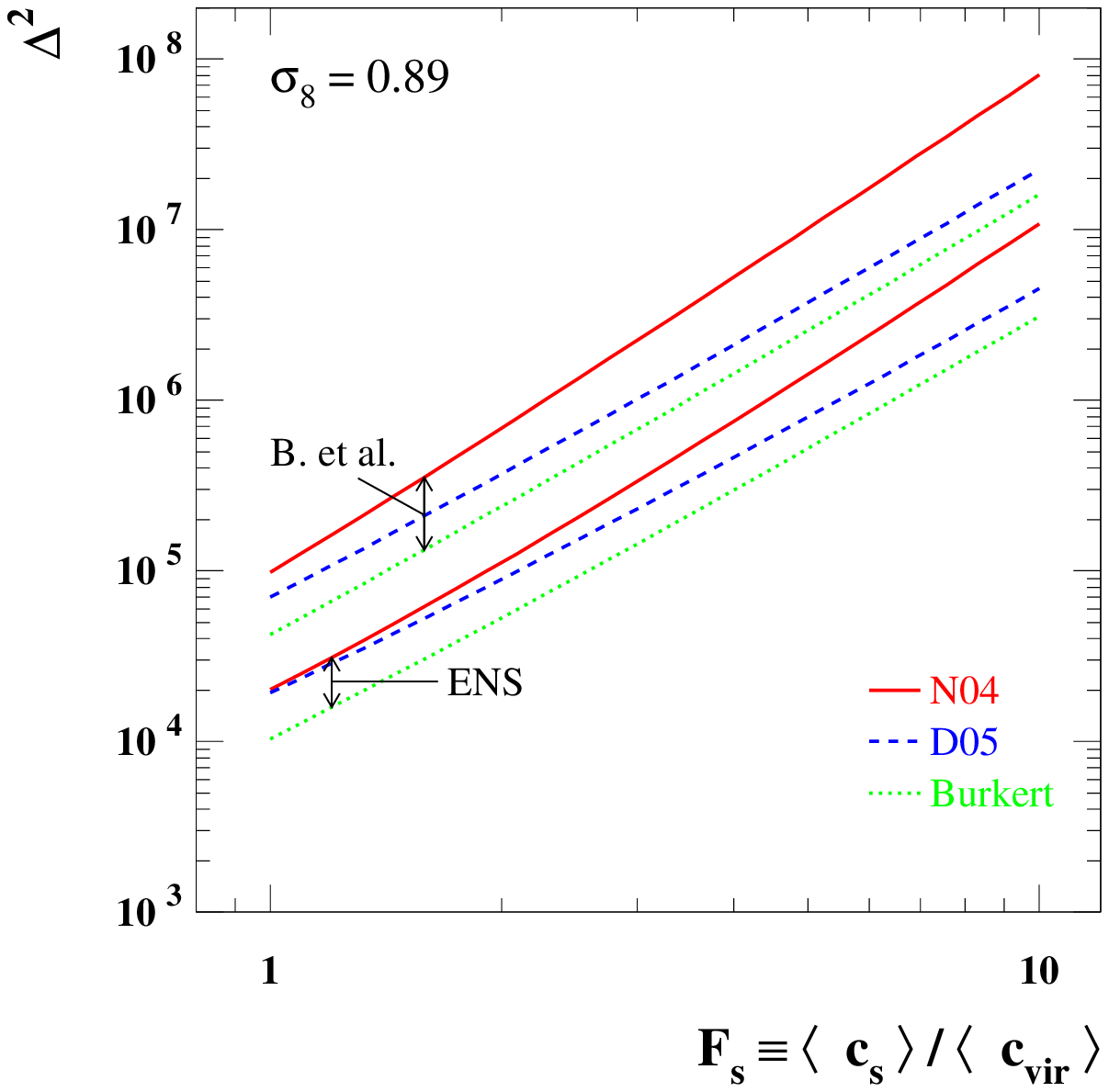}
\quad\includegraphics[scale=0.55]{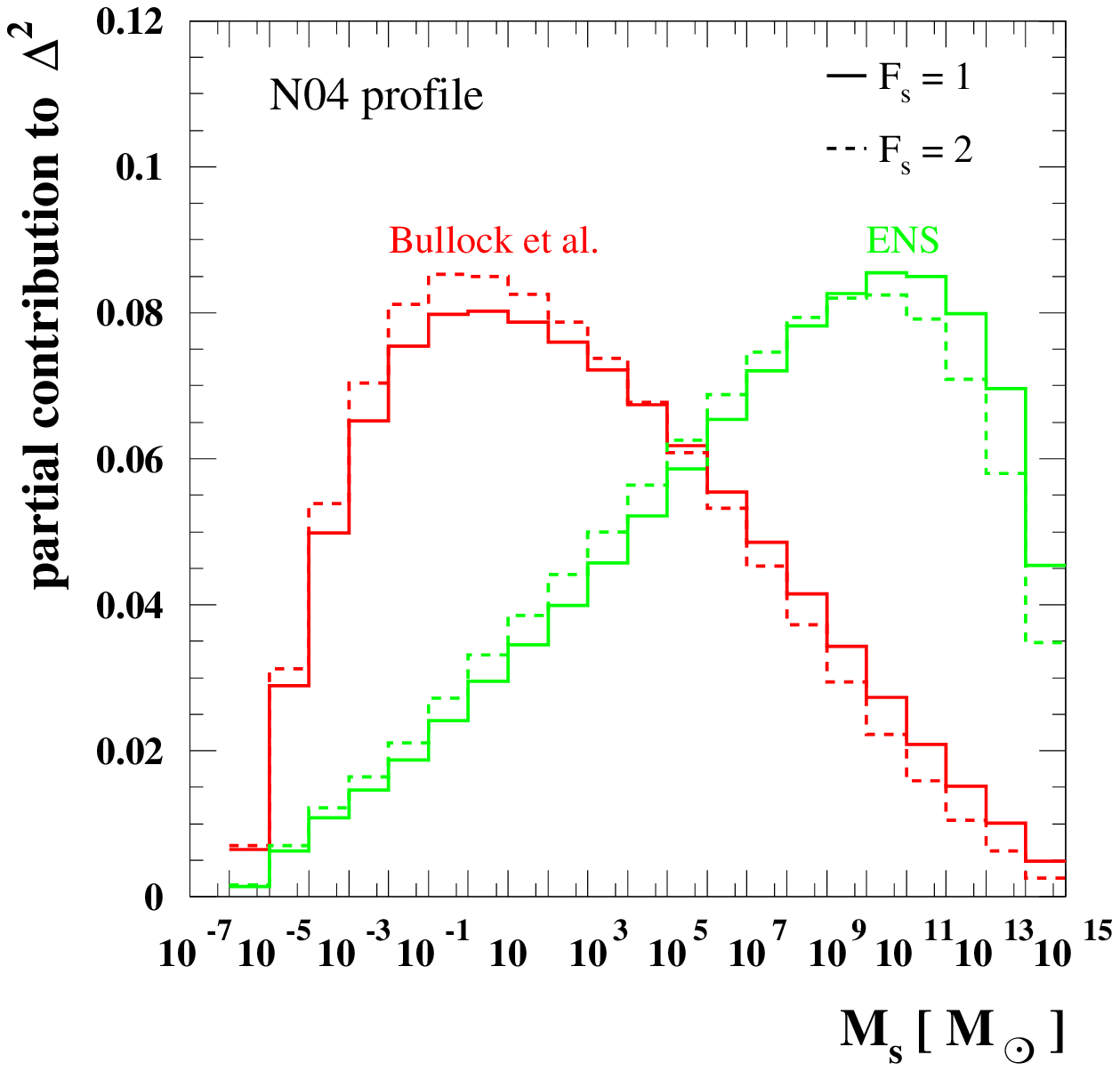}\\
\end{center}
\caption{Left: scaling of the weighted enhancement in source functions due to subhalos versus the
ratio between concentration parameter in subhalos to concentration parameter in halos at equal mass
$F_s$. Results are shown the three halo profiles introduced, {\em i.e.} the N04, D05 and Burkert profile,
the two toy models for the scaling of concentration parameter with mass, {\em i.e.} the Bullock et al. and
the ENS. Right: fractional contribution per logarithmic interval in subhalo mass $M_s$ to $\Delta^2$
in four sample cases. A normalization of the fluctuation spectrum $\sigma_8 = 0.89$ is adopted.}
\label{fig:delta2}
\end{figure*}

We show in Fig.~\ref{fig:delta2ms} the scaling of the average enhancement
$\Delta_{M_s}^2$ in the source function versus the subhalo mass $M_s$. We have considered
the three halo models introduced in the previous Section, {\em i.e.} the N04, D05 and
Burkert profiles, for the two toy models describing the scaling of concentration
parameter with mass, {\em i.e.} the Bullock et al. and the ENS schemes, as well as two
sample values for the ratio $F_s$ between the average concentration parameter in subhalos
and that in halos of equal mass. In each setup, going to smaller and smaller values of
$M_s$, the average enhancement $\Delta_{M_s}^2$ increases and then flattens out at the
mass scale below which all structures tend to collapse at the same epoch, and hence have
equal concentration parameter.

In Fig.~\ref{fig:delta2} we show the scaling of the weighted enhancement $\Delta^2$ in
the source function due to subhalos versus the ratio between concentration parameter in
subhalos to concentration parameter in halos at equal mass $F_s$; we give results for the
usual set of halo profiles considered in our approach. Analogously to the enhancement for
a fixed mass shown in the previous plot, $\Delta^2$  is very sensitive to the scaling of
the concentration parameter and  hence we find a sharp dependence of $\Delta^2$ on $F_s$.
The fractional contribution per logarithmic interval in subhalo mass $M_s$ to $\Delta^2$
is also shown in Fig.~\ref{fig:delta2} for four sample cases. Note that, although the
factorization in the probability distribution for clumps in the radial coordinate and
mass (plus the assumption that $F_s$ does not depend on mass) are a crude approximation,
what we actually need in our discussion is $F_s$ and the radial distribution for subhalos
at the peak of the distribution shown in Fig.~\ref{fig:delta2}: unfortunately we cannot
read out this from numerical simulations.

\begin{figure}[t]
\begin{center}
\includegraphics[scale=0.55]{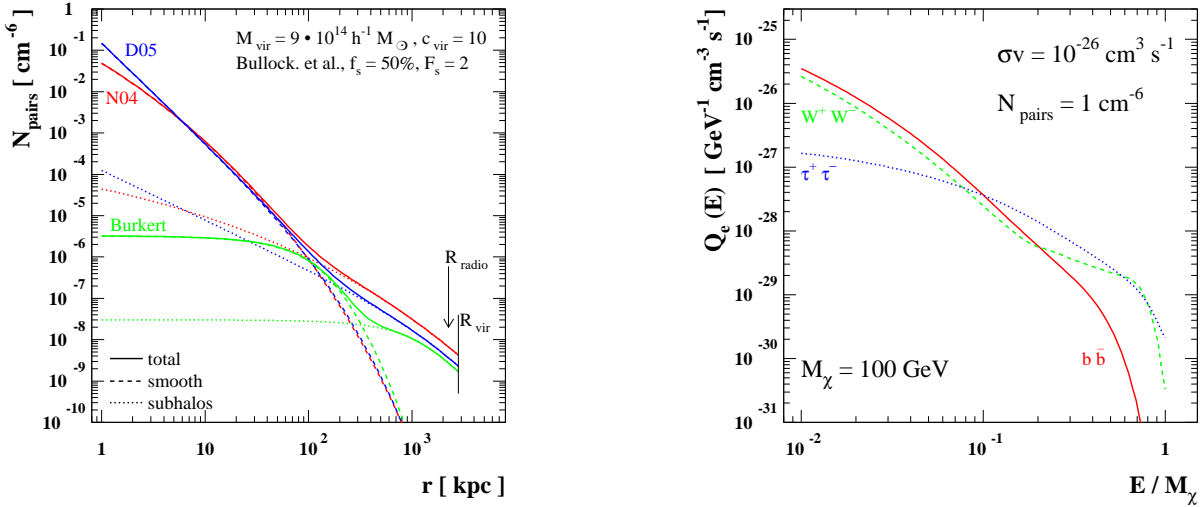}
\end{center}
\caption{The number density of neutralino pairs (neutralino mass set to $M_{\chi}=100$
GeV) as a function distance for the center of Coma for the three halo profiles
introduced, {\em i.e.} the N04, D05 and Burkert profile in their best fit model, and a
sample configuration for the subhalo parameters. } \label{fig:npairs}
\end{figure}

Fig.~\ref{fig:npairs} shows the number density of neutralino pairs (we set here the neutralino mass to
$M_{\chi}=100$ GeV) as a function of the distance from the center of Coma for the three representative
halo profiles introduced here, {\em i.e.} the N04, D05 and Burkert profile in their best fit model,
and a sample configuration for the subhalo parameters. For the D05 and N04 profiles, the central
enhancement increases the integrated source function by a factor $\approx 6$ with respect to the
Burkert profile, but this takes place on such a small angular scale that from the observational point
of view it is like adding a point source at the center of the cluster. The enhancement of the
annihilation signals from subhalos comes instead from large radii. This means that the enhancement
from subhalos largely influences the results when the neutralino source is extended. This is the case
of galaxy clusters, and more specifically of the Coma cluster which is our target in this paper.

\begin{figure}[t]
\begin{center}
\includegraphics[scale=0.55]{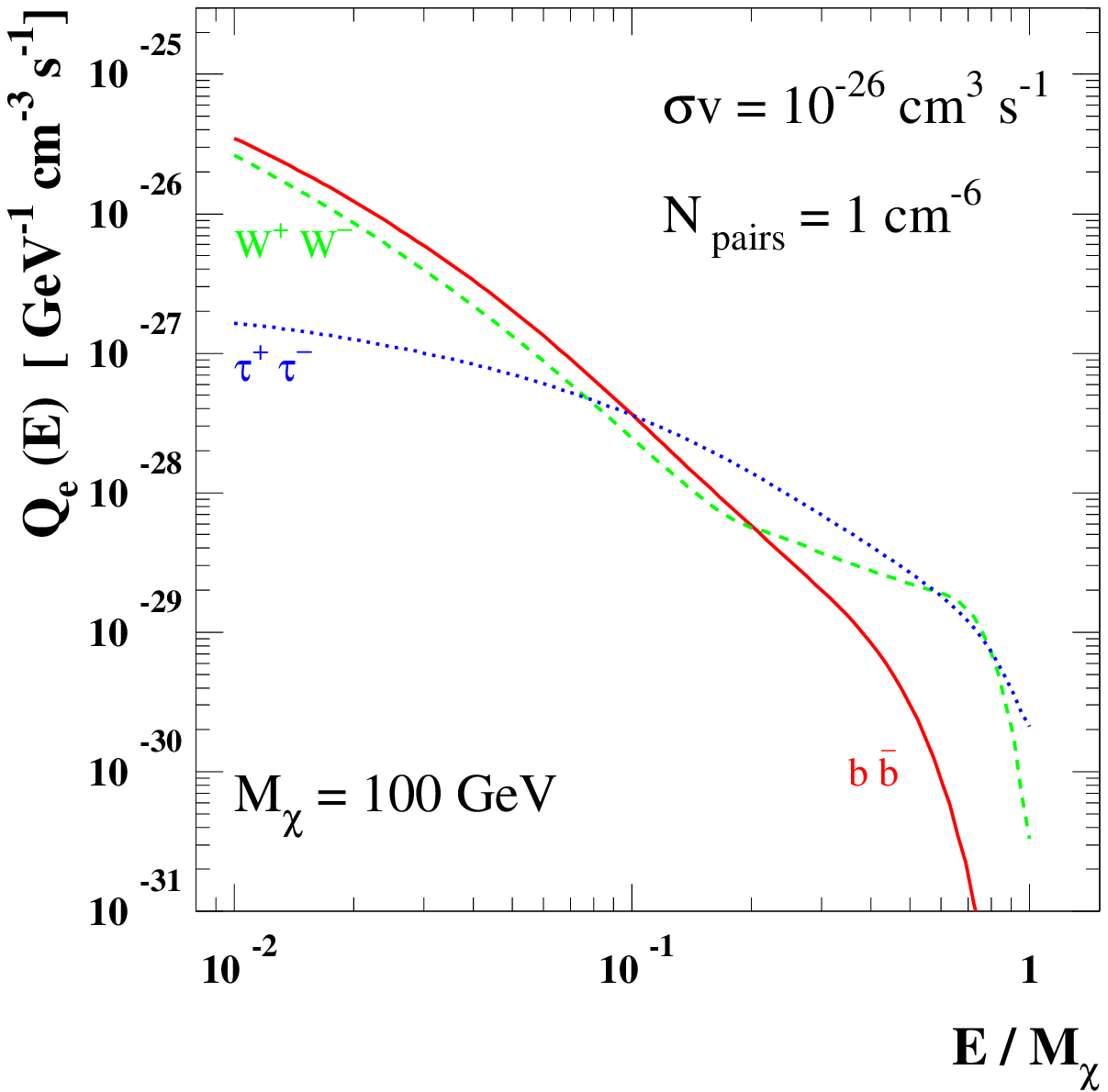}
\end{center}
\caption{The spectral shape of the electron source function in case of three sample final
states (see text for details.}
 \label{fig:esource}
\end{figure}

\subsection{Source functions spectral properties: generalities and supersymmetric benchmarks}\label{sec:dm}

The spectral properties of secondary products of DM annihilations depend only, prior to
diffusion and energy losses, on the DM particle mass $M_{\chi}$ and on the branching
ratio $BR(\chi\chi\rightarrow f)$ for the final state $f$ in the DM pair-annihilation.
The DM particle physics model further sets the magnitude of the thermally averaged pair
annihilation cross section times the relative DM particles velocity, $\langle \sigma
v\rangle_0$ at $T=0$.

The range of neutralino masses and pair annihilation cross sections in the most general
supersymmetric DM setup is extremely wide. Neutralinos as light as few GeV (see
\cite{Bottino:2003iu}) and as heavy as hundreds of TeV (see \cite{Profumo:2005xd}) can
account for the observed CDM density through thermal production mechanisms, and
essentially no constraints apply in the case of non-thermally produced neutralinos.

Turning to the viable range of neutralino pair annihilation cross sections,
coannihilation processes do not allow us to set any {\em lower} bound, while on purely
theoretical grounds a general upper limit on $\langle\sigma v\rangle_0 \simlt
10^{-22}\left(\frac{M_{\chi}}{{\rm TeV}}\right)^{-2}{\rm cm}^3/{\rm s}$ has been recently
set (\cite{Profumo:2005xd}). The only general argument which ties the relic abundance of
a WIMP with its pair annihilation cross section is given by the naive relation
\begin{equation}
\Omega_\chi h^2 \simeq {3\times 10^{-27}{\rm cm}^3/{\rm s} \over  \langle \sigma v
\rangle_0}
\end{equation}
(see \cite{Jungman:1995df}, Eq.3.4), which points at a fiducial value for $\langle\sigma
v\rangle_0 \simeq 3\times 10^{-26}{\rm cm}^3/{\rm s}$ for our choice of cosmological
parameters. The above mentioned relation can be, however, badly violated in the general
MSSM, or even within minimal setups, such as the minimal supergravity scenario (see
\cite{Profumo:2005xd}).

Since third generation leptons and quarks Yukawa couplings are always much larger than
those of the first two generations, and being the neutralino a Majorana fermion, the
largest $BR(\chi\chi\rightarrow f)$ for annihilations into a fermion-antifermion pair are
in most cases\footnote{Models with non-universal Higgs masses at the GUT scale can give
instances of exceptions to this generic spectral pattern, featuring light first and
second generation sfermions (see {\em e.g.} \cite{nuhm}).} into the third generation
final states $b\overline b$, $t\overline t$ and $\tau^+\tau^-$. In the context of
supersymmetry, if the supersymmetric partners of the above mentioned fermions are not
significantly different in mass, the $\tau^+\tau^-$ branching ratio will be suppressed,
with respect to the $b\overline b$ branching ratio by a color factor equal to 1/3, plus a
possible further Yukawa coupling suppression, since the two final states share the same
$SU(2)$ quantum number assignment. Further, the fragmentation functions of third
generation quarks are very similar, and give rise to what we will dub in the following as
a ``{\em soft spectrum}''. A second possibility, when kinematically allowed, is the pair
annihilation into massive gauge bosons\footnote{The direct annihilation into photons is
loop suppressed in supersymmetric models (see {\em e.g.} \cite{Bergstrom:1988fp} and
\cite{Bergstrom:1997fh}).}, $W^+W^-$ and $Z_0Z_0$. Again, the fragmentation functions for
these two final states are mostly indistinguishable, and will be indicated as giving a
``{\em hard spectrum}''. The occurrence of a non-negligible branching fraction into
$\tau^+\tau^-$ or into light quarks will generally give raise to intermediate spectra
between the "hard" and "soft" case.\\
Fig.\ref{fig:esource} shows the spectral shape of the electron source function in the case of the
three sample final states $b\overline b$, $\tau^+\tau^-$ and $W^+W^-$ for $M_{\chi}=100$ GeV, and
clarifies the previous discussion. In what follows we will therefore employ sample DM configurations
making use of either soft ($b\overline b$) or hard ($W^+W^-$) spectra, keeping in mind that other
possibilities would likely fall in between these two extrema.

In order to make a more stringent contact with supersymmetry phenomenology, we will
however also resort to {\em realistic benchmark SUSY models}: by this we mean thoroughly
defined SUSY setups which are fully consistent with accelerator and other
phenomenological constraints, and which give a neutralino thermal relic abundance exactly
matching the central cosmologically observed value. To this extent, we refer to the
so-called minimal supergravity model (\cite{msugra}), perhaps one of the better studied
paradigms of low-energy supersymmetry, which enables, moreover, a cross-comparison with
numerous dedicated studies, ranging from colliders (\cite{msugracolliders}) to DM
searches (\cite{msugradmdet}).

The assumptions of universality in the gaugino and in the scalar (masses and trilinear
couplings) sectors remarkably reduce, in this model, the number of free parameters of the
general soft SUSY breaking Lagrangian (\cite{Chung}) down to four continuous parameters
($m_0,\ M_{1/2}, A_0, \tan\beta$) plus one sign (${\rm sign}(\mu)$). The mSUGRA parameter
space producing a sufficiently low thermal neutralino relic abundance $\Omega_\chi h^2$
has been shown to be constrained to a handful of ``regions'' featuring effective
$\Omega_\chi h^2$ suppression mechanisms (\cite{Ellis:2003cw}). The latter are
coannihilations of the neutralino with the next-to-lightest SUSY particle
(``Coannihilation'' region), rapid annihilations through $s$ channel Higgs exchanges
(``Funnel'' region), the occurrence of light enough neutralino and sfermions masses
(``Bulk'' region) and the presence of a non-negligible bino-higgsino mixing (``Focus
Point'' region).

\begin{figure*}[!t]
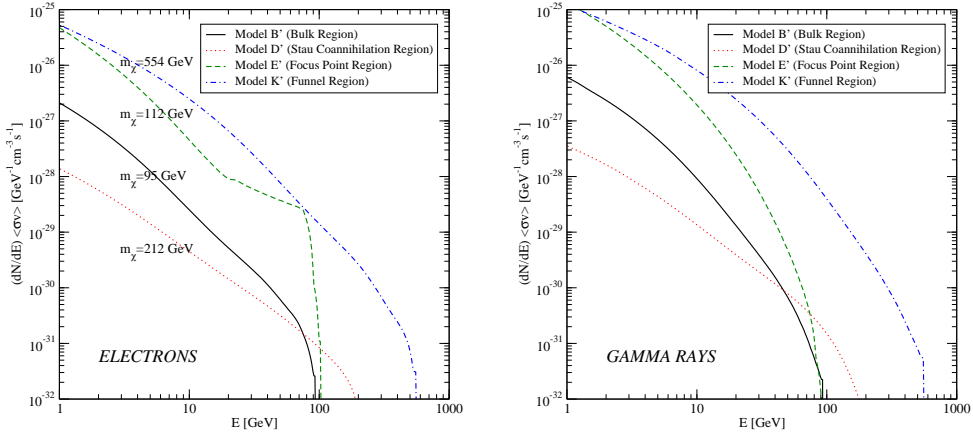

\begin{center}
\hspace*{-0.5cm}\includegraphics[scale=0.3]{E_ele.eps}\quad\quad
\includegraphics[scale=0.3]{E_gam.eps}
\end{center}
\caption{(Left): The electrons flux $(dN_e/dE)\langle \sigma v\rangle_0$ as a
function of the electron energy. (Right): The gamma-rays flux $(dN_\gamma/dE)\langle
\sigma v\rangle_0$ as a function of the photon energy.}
\label{fig:benchmark}
\end{figure*}

\begin{figure*}[!t]
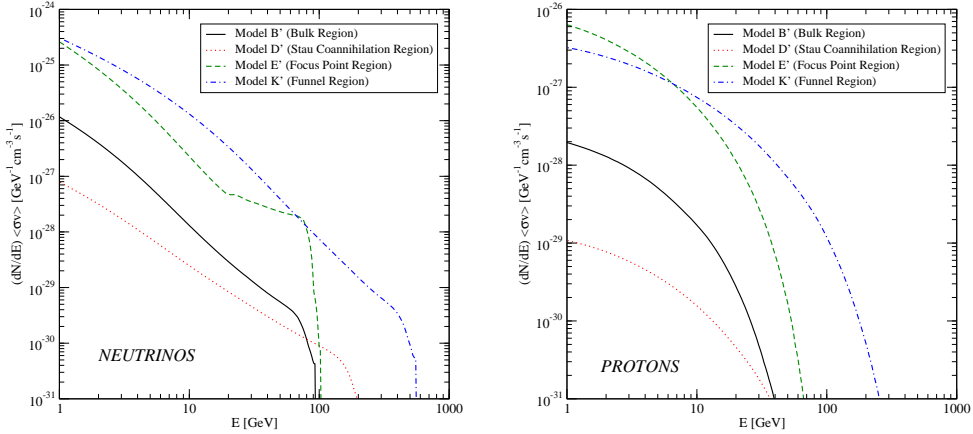

\begin{center}
\hspace*{-0.5cm}\includegraphics[scale=0.3]{E_neu.eps}\quad\quad\includegraphics[scale=0.3]{E_pro.eps}
\end{center}
\caption{(Left): The neutrinos flux $(dN_\nu/dE)\langle \sigma v\rangle_0$ as a
function of the neutrino energy. (Right): The protons flux $(dN_p/dE)\langle
\sigma v\rangle_0$ as a function of the proton energy.}
\label{fig:benchmark2}
\end{figure*}

With the idea of allowing a direct comparison with the existing research work in a wealth
of complementary fields, we restrict ourselves to the ``{\em updated post-WMAP benchmarks
for supersymmetry}'' proposed and studied by \cite{Battaglia:2003ab}. All of those setups
are tuned so as to feature a neutralino thermal relic density giving exactly the central
WMAP-estimated CDM density\footnote{We adjusted here the values of $m_0$ given in
\cite{Battaglia:2003ab} in order to fulfill this requirement making use of the latest
Isajet v.7.72 release and of the \ds package (\cite{Edsjo:2003us}, see
Table~\ref{tab:models}).}.
As a preliminary step, we computed the electrons, neutrinos, gamma-rays and protons
source spectra for all the 13 ${\bf A}^\prime$-${\bf M}^\prime$ models. Remarkably
enough, although the SUSY particle spectrum is rather homogeneous throughout the mSUGRA
parameter space, the resulting spectra exhibit at least three qualitatively different
shapes, according to the dominant final state in neutralino pair annihilation processes.
In particular, in the Bulk and Funnel regions the dominant final state is into $b\bar b$,
and, with a sub-dominant variable contribution, $\tau^+\tau^-$. The latter channel is
instead dominant, for kinematic reasons, in the stau Coannihilation region. Finally, a
third, and last, possibility is a dominant gauge bosons final state, which is the case
along the Focus Point region. In this respect, in the effort to reproduce all of the
mentioned spectral modes, and to reflect every cosmologically viable mSUGRA region, we
focused on the four models indicated in Table~\ref{tab:models}, a subset of the
benchmarks of \cite{Battaglia:2003ab} (to which we refer the reader for further details).
\footnotesize{
\begin{table}[!b]
\begin{center}
\begin{tabular}{l|c|c|c|c|c|}
{\bf Model}&$M_{1/2}$&$m_0$&$\tan\beta$&${\rm sign}(\mu)$&$m_t$\\ \hline ${\bf B}^\prime$
({\em Bulk})& 250 & 57 & 10  & $>0$ & 175\\ ${\bf D}^\prime$ ({\em Coann.})& 525 & 101 &
10  & $>0$ & 175\\ ${\bf E}^\prime$ ({\em Focus P.})& 300 & 1653 & 10  & $>0$ & 171\\
${\bf K}^\prime$ ({\em Funnel})& 1300 & 1070 & 46  & $<0$ & 175\\ \hline
\end{tabular}
\end{center}
\caption{The input parameters of the four mSUGRA benchmark models we consider here. The
units for the mass parameters are GeV, and the universal trilinear coupling $A_0$ is set
to 0 for all models (see \cite{Battaglia:2003ab} for details). }\label{tab:models}
\end{table}
}
\footnotesize{
\begin{table}[!b]
\begin{center}
\begin{tabular}{l|c|c|c|c|c|}
{\bf Model}& BR($b\bar b$) & BR($\tau^+\tau^-$) &  BR($W^+W^-$) & BR($Z^0Z^0$) & $\langle
\sigma v\rangle_0$ \\ \hline ${\bf B}^\prime$ ({\em Bulk})& 74\% & 19\% & 4\% & 0\% &
$7.8\times10^{-28}$\\ ${\bf D}^\prime$ ({\em Coann.})& 21\% & 61\% & 0\% & 0\% &
$8.9\times10^{-29}$\\ ${\bf E}^\prime$ ({\em Focus P.})& 1\% & 0\% & 90\% & 8\% &
$1.7\times10^{-26}$\\ ${\bf K}^\prime$ ({\em Funnel})& 88\% & 11\% & 0\% & 0\%
&$1.1\times10^{-26}$\\ \hline
\end{tabular}
\end{center}
\caption{The branching ratios into various final states for the four mSUGRA benchmark
models of tab.~\ref{tab:models}; in the last column we also indicate $\langle \sigma
v\rangle_0$ in units of ${\rm cm}^3{\rm s}^{-1}$. }\label{tab:bf}
\end{table}
}
%
We collect in Table~\ref{tab:bf} the branching ratios for the final states of neutralino
pair annihilations. In the last column of this table we also provide the
thermally-averaged pair annihilation cross section times the relative velocity, at $T=0$,
$\langle \sigma v\rangle_0$. Table~\ref{tab:bf} is an accurate guideline to interpret the
resulting source spectra for the four benchmarks under consideration here, which are
shown in Figs.~\ref{fig:benchmark} and \ref{fig:benchmark2}. Fig.\ref{fig:benchmark}
shows in particular the differential electron (left) and photon (right) yields per
neutralino annihilation multiplied by $\langle \sigma v\rangle_0$, {\em {\em i.e.}} the
source function $Q(r,E)$ divided by the number density of neutralino pairs ${\cal N}_{\rm
pairs}(r)$ as a function of the particles' kinetic energy. As mentioned above, the Bulk
and Funnel cases are very similar between each other, though in the latter case one has a
heavier spectrum and a larger value of $\langle \sigma v\rangle_0$.
Fig.\ref{fig:benchmark2} shows the same quantity for neutrinos and protons.

The products of the neutralino annihilation which are more relevant to our discussion are
secondary electrons and pions.
The secondary particles produced by neutralino annihilation are subject to various
physical mechanisms: {\it i)} decay (which is especially fast for pions and muons); {\it
ii)} energy losses which can be suffered by stable particles, like electrons and
positrons; {\it iii)} spatial diffusion of these relativistic particles in the atmosphere
of the cluster.
Gamma-rays produced by neutral pion decay, $\pi^0 \to \gamma \gamma$, generate most of the continuum
spectrum at energies $E \simgt 1$ GeV and this emission is directly radiated since the $\pi^0 \to
\gamma \gamma$ e.m. decay is very fast.
This gamma-ray emission is dominant at high energies, $\simgt 0.3 - 0.5$ of the
neutralino mass, but needs to be complemented by other two emission mechanisms which
produce gamma-rays at similar or slightly lower energies: these are the ICS and the
bremsstrahlung emission by secondary electrons. We will discuss the full gamma-ray
emission of Coma induced by DM annihilation in Sect.~\ref{sec:multiwave} below.
Secondary electrons are produced through various prompt generation mechanisms and by the
decay of charged pions (see, {\em e.g.}, \cite{ColafrancescoMele2001}).
In fact, charged pions decay through $\pi^{\pm}\to \mu^{\pm} \nu_{\mu}(\bar{\nu}_{\mu})$, with
$\mu^{\pm}\to e^{\pm} + \bar{\nu}_{\mu}(\nu_{\mu}) + \nu_e (\bar{\nu}_e)$ and produce $e^{\pm}$, muons
and neutrinos.
Electrons and positrons are produced abundantly by neutralino annihilation (see
Fig.~\ref{fig:benchmark}, left) and are subject to spatial diffusion and energy losses.
Both spatial diffusion and energy losses contribute to determine the evolution of the
source spectrum into the equilibrium spectrum of these particles, {\em i.e.} the quantity
which will be used to determine the overall multi-wavelength emission induced by DM
annihilation.
The secondary electrons eventually produce radiation by synchrotron in the magnetized atmosphere of
Coma, Inverse Compton Scattering of CMB (and other background) photons and bremsstrahlung with protons
and ions in the atmosphere of the Coma cluster (see, {\em e.g.}, \cite{ColafrancescoMele2001} and
\cite{Colafrancesco2003,Colafrancesco2005b} for a review). These secondary particles also produce
heating of the intra-cluster gas by Coulomb collisions with the intra-cluster gas particles and SZ
effect (see, {\em e.g.} \cite{Colafrancesco2003}, \cite{Colafrancesco2005b}).
Other fundamental particles which might have astrophysical relevance are also produced in DM
annihilation.
Protons are produced in a smaller quantity with respect to $e^{\pm}$ (see
Fig.~\ref{fig:benchmark2}, right), but do not loose energy appreciably during their
lifetime while they can diffuse and be stored in the cluster atmosphere. These particles
can, in principle, produce heating of the intra-cluster gas and $pp$ collisions
providing, again, a source of secondary particles (pions, neutrinos, $e^{\pm}$, muons,
...) in complete analogy with the secondary particle production by neutralino
annihilation. Neutrinos are also produced in the process of neutralino annihilation (see
Fig.~\ref{fig:benchmark2}, left) and propagate with almost no interaction with the matter
of the cluster. However, the resulting flux from Coma is found to be unobservable by
current experiments.

To summarize, the secondary products of neutralino annihilation which have the most relevant
astrophysical impact onto the multi-frequency spectral energy distribution of DM halos are neutral
pions and secondary electrons.

\section{Neutralino-induced signals}\label{sec:multiwave}

A complete description of the emission features induced by DM must take, consistently,
into account the diffusion and energy-loss properties of these secondary particles.
These mechanisms are taken into account in the following diffusion equation ({\em i.e.}
neglecting convection and re-acceleration effects):
\begin{eqnarray}
\frac{\partial}{\partial t}\frac{dn_e}{dE} & = & \nabla \left[ D(E,\vec{x})
\nabla\frac{dn_e}{dE}\right] + \frac{\partial}{\partial E} \left[ b(E,\vec{x})
\frac{dn_e}{dE}\right] \nonumber \\
 & & + Q_e(E,\vec{x})\;,
 \label{diffeq}
\end{eqnarray}
where $dn_e/dE$ is the equilibrium spectrum, $D(E,\vec{x})$ is the diffusion coefficient,
$b(E,\vec{x})$ is the energy loss term and $Q_e(E,\vec{x})$ is the source function.
The analytical solution of this equation for the case of the DM source function is
derived in the Appendix \ref{sec:secondary}.\\
In the limit in which electrons and positrons lose energy on a timescale much shorter
than the timescale for spatial diffusion, {\em i.e.} the regime which applies to the case
of galaxy clusters, the first term on the r.h.s. of Eq.~(\ref{diffeq}) can be neglected,
and the expression for equilibrium number density becomes:
\begin{equation}
\left({\frac{dn_e}{dE}}\right)_{nsd}\left(r,E \right) =  \frac{1}{b(E)} \int_E^{M_\chi}
dE'  \; Q_e(r,E')\;, \label{eq:nodiff}
\end{equation}
(see the Appendix \ref{sec:secondary} for a general discussion of the role of spatial
diffusion and of the regimes in which it is relevant).

The derivation of the full solution of the diffusion equation (Eq.~\ref{diffeq}) and the
effects of diffusion and energy losses described in the Appendix \ref{sec:secondary}, set
us in the position to discuss the multi-frequency emission produced by the DM
(neutralino) component of the Coma cluster. We will present the overall DM-induced
spectral energy distribution (hereafter SED) from low to high observing frequencies.

We describe here our reference setup for the numerical calculations.
Our reference halo setup is the N04 profile and other parameters/choice of extrapolation schemes as in
Fig.~\ref{fig:npairs}.
We consider the predictions of two particle models, one with a branching ratio equal to 1 in $b
\bar{b}$, {\em i.e.} a channel with a soft production spectrum, and the second one with a branching
ratio equal to 1 into $W^+ W^-$, {\em i.e.} a channel with hard spectrum.
Since we have previously shown that diffusion is not relevant in a Coma-like cluster of galaxy, we
neglect, in our numerical calculations, the spatial diffusion for electrons and
positrons: this is the limit in which the radial dependence and frequency dependence can
be factorized in the expression for the emissivity.

\subsection{Radio emission}\label{sec:radio}

At radio frequencies, the DM-induced  emission is dominated by the synchrotron radiation of the
relativistic secondary electrons and positrons of energy $E= \gamma m_e c^2$, living in a magnetic
field $B(r)$ and a background plasma with thermal electron density $n(r)$, and in the limit of
frequency $\nu$ of the emitted photons much larger than the non-relativistic gyro-frequency $\nu_0= e
B/(2\pi m c) \simeq 2.8 B_{\mu}$~Hz and the plasma frequency $\nu_p = 8980 \left( n(r)/1
cm^{-3}\right)^{1/2}$~Hz. Averaging over the directions of emission, the spontaneously emitted
synchrotron power at the frequency $\nu$ is given by (\cite{book}):
\begin{equation}
P_{\rm synch}\left(\nu,E,r\right) = \int_0^\pi d\theta \frac{\sin\theta}{2} 2 \pi \sqrt{3} r_0 m c
\nu_0 \sin\theta F\left(x/\sin\theta\right) \;,
\end{equation}
where we have introduced the classical electron radius $r_0= e^2/(m c^2) = 2.82 \cdot 10^{-13}$~cm,
and we have defined the quantities $x$ and $F$ as:
\begin{equation}
x \equiv \frac{2\nu}{3\nu_0 \gamma^2}
\left[ 1+ \left( \frac{\gamma \nu_p}{\nu}\right)^2\right]^{3/2}\;,
\end{equation}
and
\begin{equation}
F\left(t\right) \equiv t \int_t^\infty dz K_{5/3}(z) \simeq
1.25 t^{1/3} \exp{(-t)} \left[ 648 + t^2\right]^{1/12}\;.
\end{equation}
Folding the synchrotron power with the spectral distribution of the equilibrium number
density of electrons and positrons, we get the local emissivity at the frequency $\nu$:
\begin{equation}
j_{\rm synch}\left(\nu,r\right) = \int_{m_e}^{M_{\chi}} dE\, \left(\frac{dn_{e^-}}{dE} +
\frac{dn_{e^+}}{dE} \right) P_{\rm synch}\left(\nu,E,r\right)\,.
\end{equation}
This is the basic quantity we need in order to compare our predictions with the available
data. In particular, we will compare our predictions with measurements of the integrated
(over the whole Coma radio halo size) flux density spectrum:
\begin{equation}
S_{\rm synch}(\nu)= \int d^3r \, \frac{ j_{\rm synch}\left(\nu,r\right)}{4 \pi\, D_{\rm Coma}^2}\;,
\end{equation}
where $D_{\rm Coma}$ is the luminosity distance of Coma, and with the azimuthally averaged surface
brightness distribution at a given frequency and within a beam of angular size $\Delta\Omega$ (PSF):
\begin{equation}
I_{\rm synch}(\nu,\Theta,\Delta\Omega)= \int_{\Delta\Omega} d\Omega \int_{l.o.s.} dl \,  \frac{j_{\rm
synch}\left(\nu,l\right)}{4 \pi} \;,
\end{equation}
where the integral is performed along the line of sight (l.o.s.) $l$, within a cone of
size $\Delta\Omega$ centered in a direction forming an angle $\Theta$ with the direction
of the Coma center.

\begin{figure}[!t]
\begin{center}
\hspace*{-0.5cm}\includegraphics[scale=0.55]{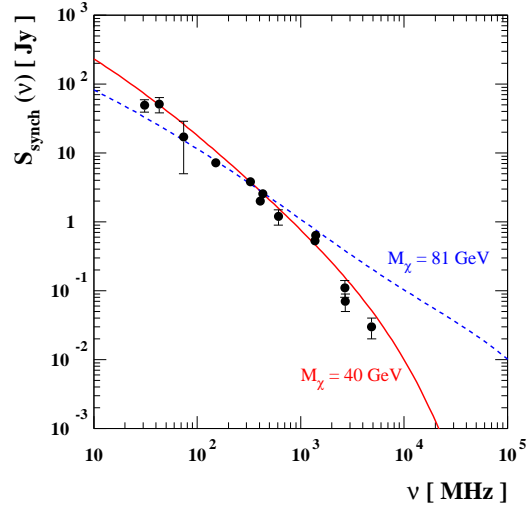}
\\
\end{center}
\caption{
Best fit models for the radio flux density spectrum, in case of a soft spectrum due to a
$b \bar{b}$ annihilation final state (solid line, model with  $M_{\chi} = 40$~GeV) and of
a hard spectrum due to a $W^+ W^-$ channel (dashed line, model with  $M_{\chi} =
81$~GeV); values of all parameters setting the model are given in the text.
The datasets is from \protect{\cite{Thierbach2003}}.
 } \label{fig:bestfit}
\end{figure}

We started from the full dataset on the radio flux density spectrum
(\cite{Thierbach2003}) and minimized the fit with respect to the WIMP mass (with the
bound $M_{\chi} \ge 10$~GeV for the $b \bar{b}$ case, and mass above threshold for the
$W^+ W^-$ case), the strength of the magnetic field (with the bound $B_\mu \ge 1 \mu G$)
and the annihilation rate $\langle \sigma v \rangle_0$. The spectrum predicted by two
models with the lowest values of $\chi^2_r$ are shown in Fig.~\ref{fig:bestfit}.
In both cases the best fit corresponds to the lowest  neutralino mass allowed, since this
is the configuration in which the fall-off of the flux density at the highest observed
frequency tends to be better reproduced. For the same reason, the fit in the case of a
soft spectrum is  favored with respect to the one with a hard spectrum (we have checked
that in case of $\tau^+ \tau^-$ again does not give a bend-over in the spectrum where
needed).
The values of the annihilation rates required by the fit are fairly large: $\langle\sigma
v\rangle_0= 4.7 \cdot 10^{-25}$~cm$^3$~s$^{-1}$ for $b \bar{b}$ case, and about one order
of magnitude smaller, $\langle\sigma v\rangle_0= 8.8 \cdot 10^{-26}$~cm$^3$~s$^{-1}$, for
the $W^+W^-$ case, despite the heavier neutralino mass, since the best fit values
correspond to different values of the magnetic field of about $1.2 \mu G$ and $8 \mu G$,
respectively.

\begin{figure*}[!t]
\begin{center}
\hspace*{-0.5cm}\includegraphics[scale=0.55]{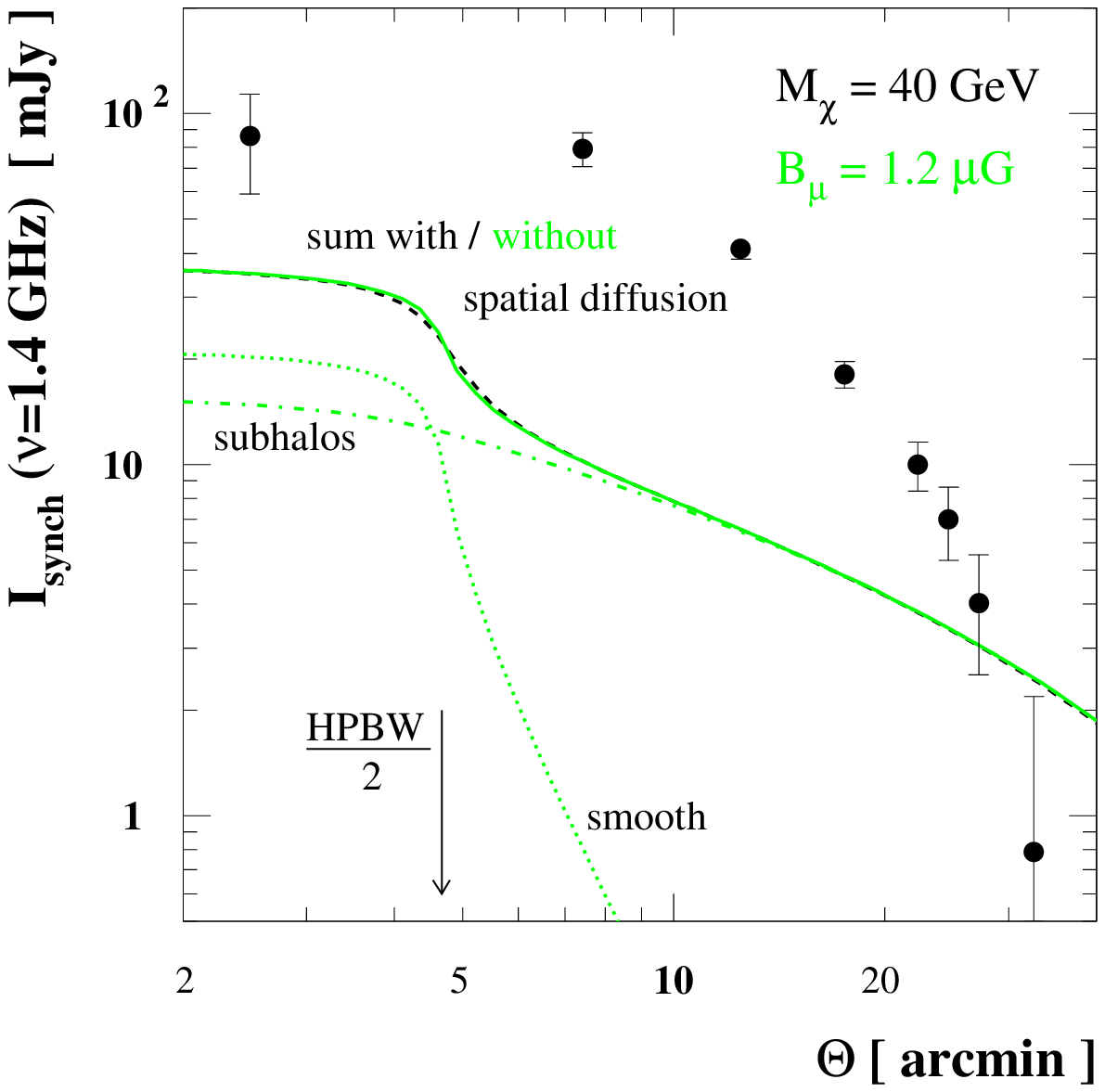}
\quad\includegraphics[scale=0.55]{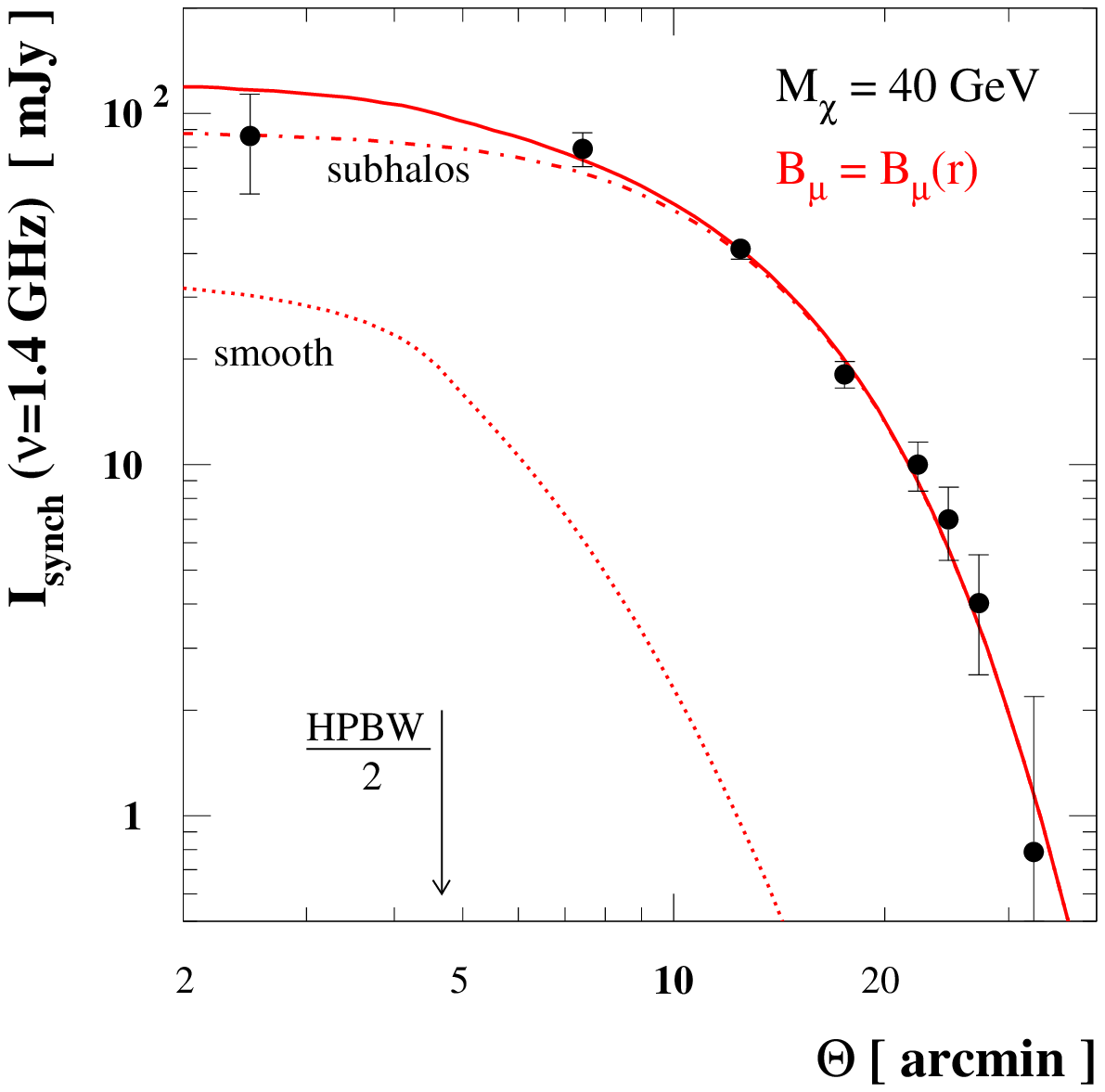}
\\
\end{center}
\caption{Surface brightness distribution at frequency $\nu = 1.4$~GHz, within a beam
equal to $9\farcm35$ (HPBW), for the lightest WIMP model displayed in
Fig.~\protect{\ref{fig:bestfit}}. In the left panel we show the predictions for a model
with magnetic field that does not change with radius, and in the limit in which spatial
diffusion for electrons and positrons has been neglected (solid line) or included (dashed
line). In the right panel we consider a model with magnetic field taking a radial
dependence $ B(r) = B_0 \bigg( 1 + {r \over r_{c1}}\bigg)^2 \cdot \bigg[ 1 + \bigg( {r
\over r_{c2}}\bigg)^2\bigg]^{-\beta}$ with $B_0 = 0.55$~\mug, $\beta = 2.7$,
$r_{c1}=3^\prime$ $r_{c2}=17\farcm5$ in order to reproduce the measured surface
brightness. In both cases, the contributions from the smooth dark matter halo component
only and from substructures only are also displayed. The dataset is from
\protect{\cite{Deiss1997}}.} \label{fig:bestfit2}
\end{figure*}

\begin{figure}[!t]
 \begin{center}
 \hspace*{-0.5cm}\includegraphics[scale=0.35]{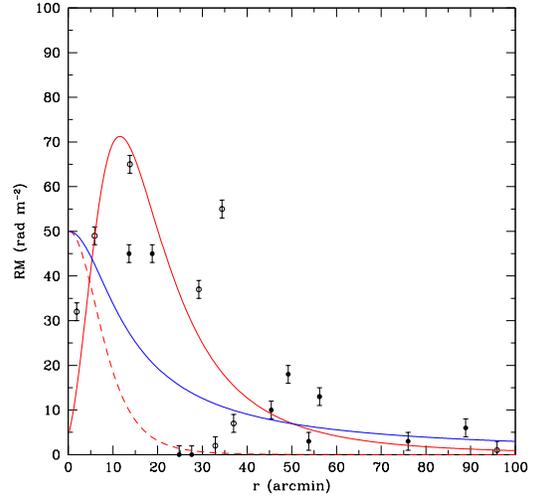}
 \end{center}
 \caption{\footnotesize{The observed absolute RM of background sources in the Coma field
 are shown as a function of projected radius $\theta$ in arcmin.
 The blu curve is the prediction of a model with $B=$ const. The red dashed curve is the
 prediction of a model with $B(r)$ decreasing monotonically towards large radii and the solid
 red curve is the prediction of our model that fits the Coma radio-halo surface brightness.
 Data on positive (filled dots) and negative (empty dots) RMs are from \cite{Kimetal1990}.}}
 \label{fig.rm}
\end{figure}

In Fig.~\ref{fig:bestfit2} we compare the radio-halo brightness data of \cite{Deiss1997}
with the surface brightness distribution $I_{synch}(r)$ predicted at $\nu =1.4$~GHz,
within a beam equal to the detector angular resolution (HPBW of $9\farcm35$), for the
best fit model with $M_{\chi} = 40$~GeV. In the left panel we plot the predicted surface
brightness considering the case of a uniform magnetic field  equal to  $1.2 \mu G$,
showing explicitly in this case that the assumption we made of neglecting spatial
diffusion for electrons and positrons is indeed justified, since the results obtained
including or neglecting spatial diffusion essentially coincide. The radial brightness we
derive in this case does not match the shape of the radio halo indicated by the data.
However, it is easy to derive a phenomenological setup with a magnetic field $B(r)$
varying with radius in which a much better fit can be obtained, while leaving unchanged
the total radio flux density $S_{synch.}(\nu)$. We show in the right panel of
Fig.\ref{fig:bestfit2} the predictions for $I_{synch}(r)$ considering a radial dependence
of the magnetic field of the form:
 $$
 B(r) = B_0 \bigg( 1 + {r \over r_{c1}}\bigg)^2 \cdot \bigg[ 1 + \bigg( {r \over r_{c2}}\bigg)^2\bigg]^{-\beta}\;,
 $$
which is observationally driven by the available information on the Faraday rotation
measures (RM) for Coma (see Fig.16).
Such $B(r)$ profile starts at a slightly smaller value in the center of the Coma, rises
at a first intermediate scale $r_{c1}$ and then drops rather rapidly at the scale
$r_{c2}$.
The basic information we provide here is that a radial dependence of the magnetic
field like the previous one is required in DM annihilation models to reproduce the
radio-halo surface brightness distribution.
The specific case displayed is for best-fit values $B_0 = 0.55$~\mug, $\beta = 2.7$,
$r_{c1}=3^\prime$ $r_{c2}=17\farcm5$, and it provides an excellent fit to the surface
brightness radial profile (see Fig.\ref{fig:bestfit2}). In that figure we also plot
separately the contributions to the surface brightness due to the smooth DM component
(essentially a point-like source in case of this rather poor angular resolution) and the
term due to subhalos (which extends instead to larger radii).
It is interesting to note that the surface brightness profile can only be fitted by
considering the extended sub-halo distribution which renders the DM profile of Coma more
extended than the smooth, centrally peaked component. This means that any peaked and
smooth DM profile is unable to fit this observable for Coma.\\
A decrease of $B(r)$ at large radii is expected by general considerations of the
structure of radio-halos in clusters and, more specifically, for Coma (see
\cite{CMP2005}) and it is also predicted by numerical simulations (see, e.g.,
\cite{Dolagetal2002}): thus it seems quite natural and motivated.
At small radii, the mild central dip of $B(r)$ predicted by the previous formula is what
is phenomenologically required by the specific DM model we worked out in our paper.
Finally, we notice that our specific phenomenological model for the spatial distribution
of $B(r)$ is able to reproduce the spatial distribution of Faraday rotation measures
(RMs) observed in Coma (see \cite{Kimetal1990}), as shown in Fig.\ref{fig.rm}.
It is evident that models in which $B$ is either constant or decreases monotonically
towards large radii seem to be difficult to reconcile with the available RM data. The RM
data at $\theta \simlt 20$ arcmin seem to favour, indeed, a model for $B(r)$ with a
slight rise at intermediate angular scales followed by a decrease at large scales, like
the one we adopt here to fit the radio-halo surface brightness of Coma. In this respect,
it seems that our choice for $B(r)$ is, at least, an observationally driven result.

The synchrotron signal produced by the annihilation of DM depends, given the fundamental physics and
astrophysics framework, on two relevant quantities: the annihilation rate and the magnetic field.
Thus, it is interesting to find the best-fitting region of the $\langle\sigma v\rangle_0-B$ plane which is
consistent with the available dataset for Coma.

\begin{figure*}[!t]
\begin{center}
\hspace*{-0.5cm}\includegraphics[scale=0.5]{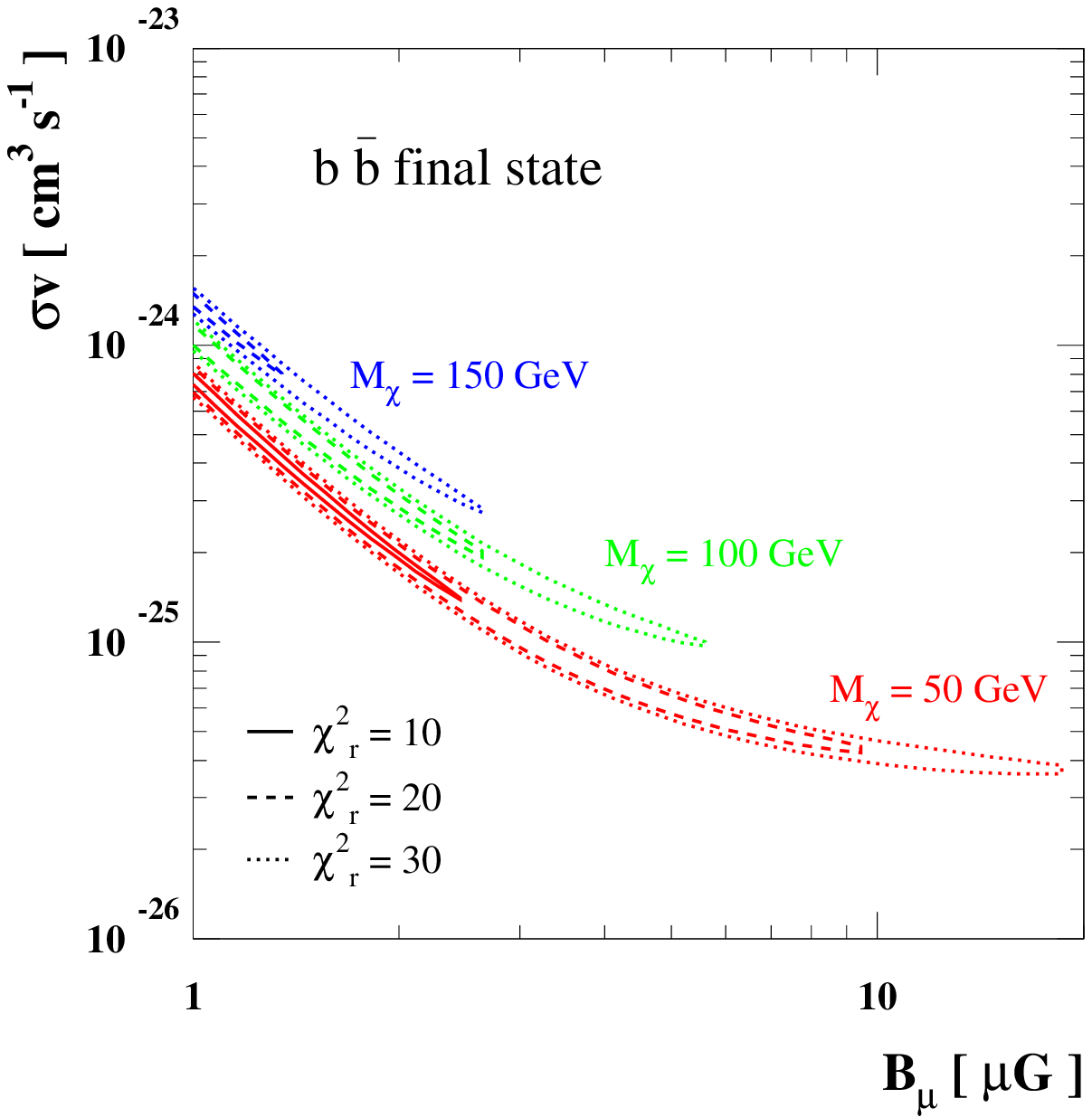}
\quad\includegraphics[scale=0.5]{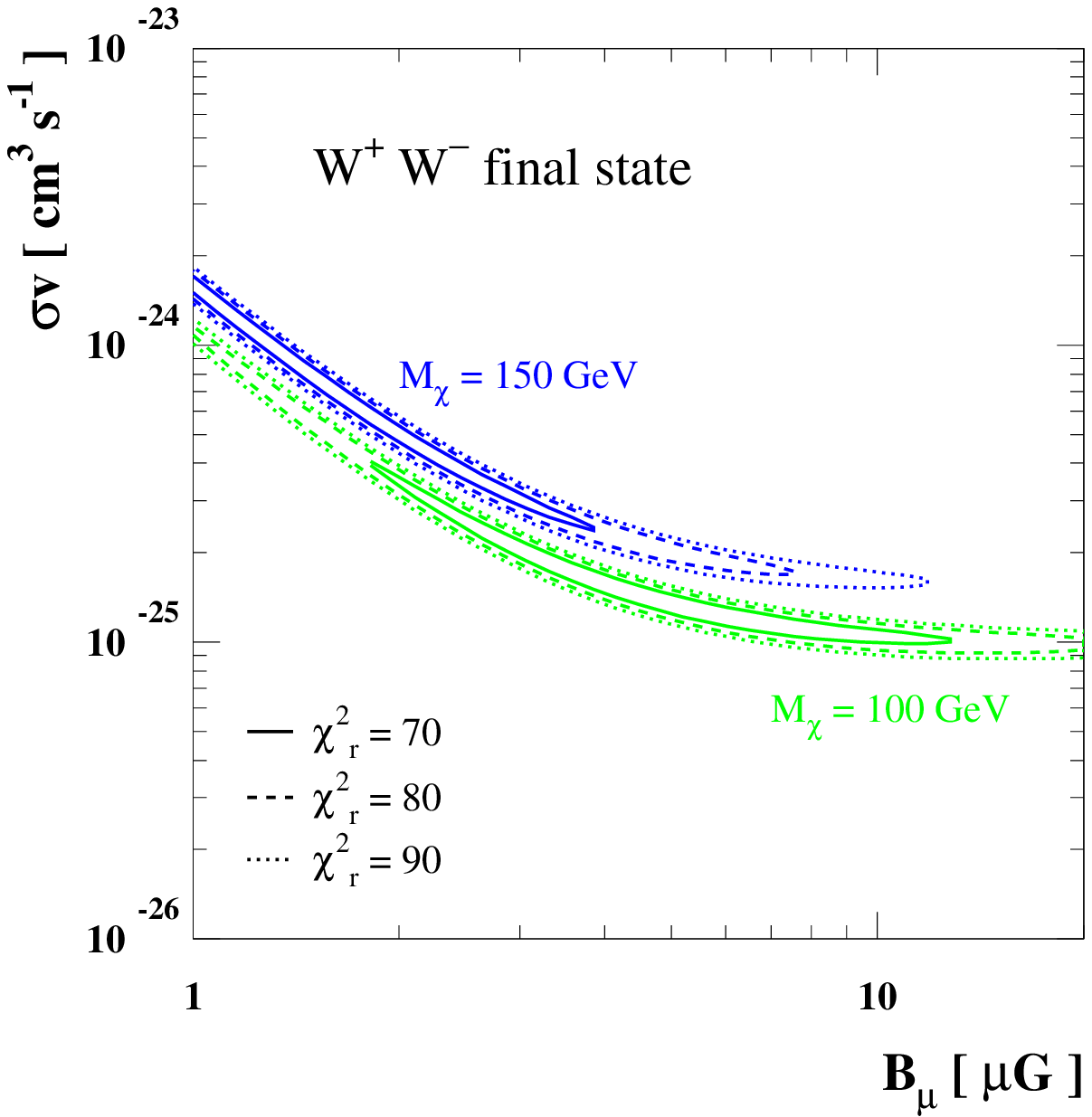}\\
\end{center}
\caption{Isolevel curves for reduced $\chi^2$ from the fit of the Coma radio flux density
spectrum, for a given mass (50 - 100 -150 GeV) and annihilation channel. The halo profile
is the best fit N04 profile: $M_{vir} = 0.9 \, 10^{15} \msun h^{-1}$ and $c_{vir} = 10$,
with subhalo setup as in Fig.~\protect{\ref{fig:npairs}}.} \label{fig:fit1}
\end{figure*}

\begin{figure*}[!t]
\begin{center}
\hspace*{-0.5cm}\includegraphics[scale=0.5]{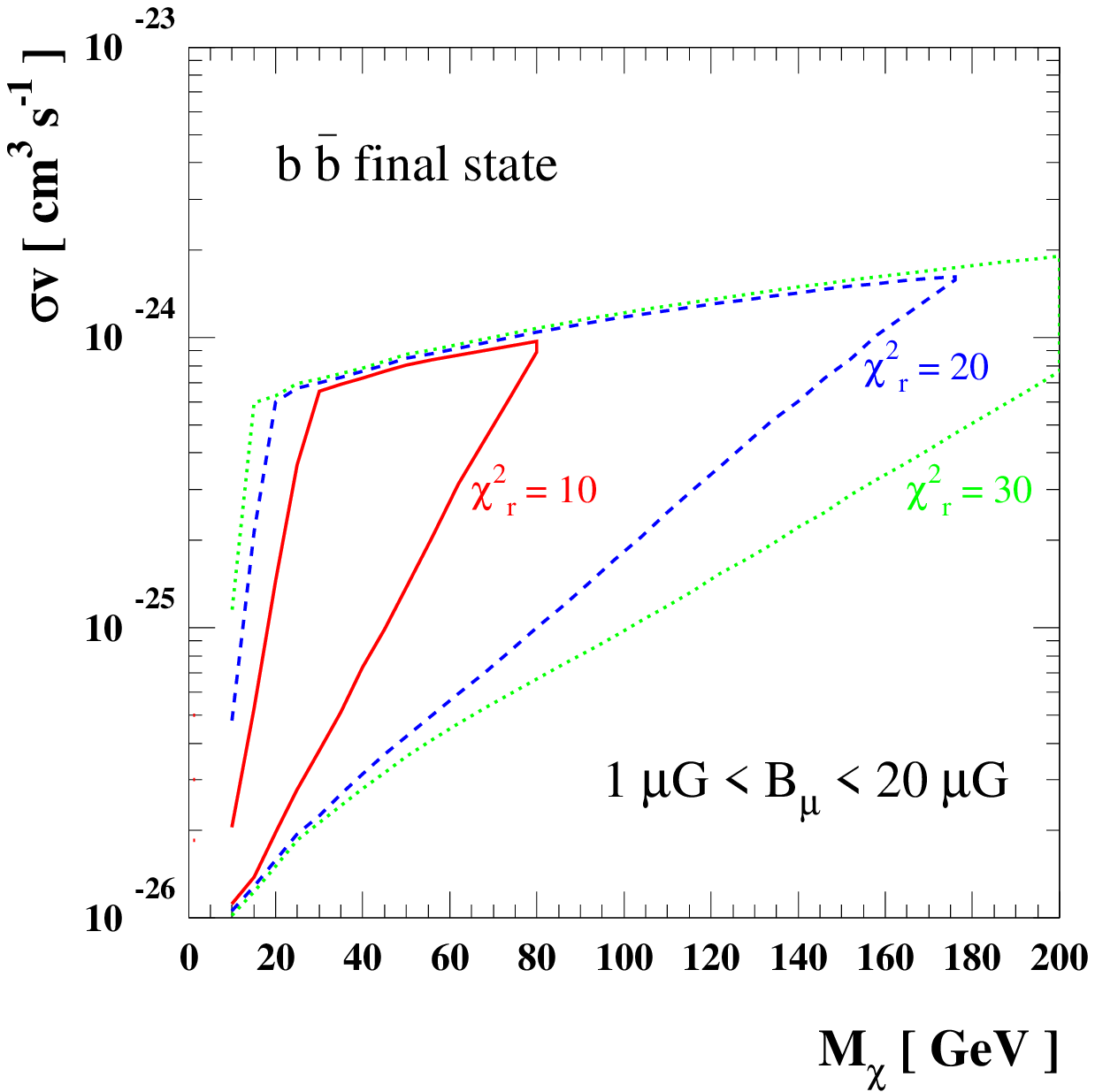}
\quad\includegraphics[scale=0.5]{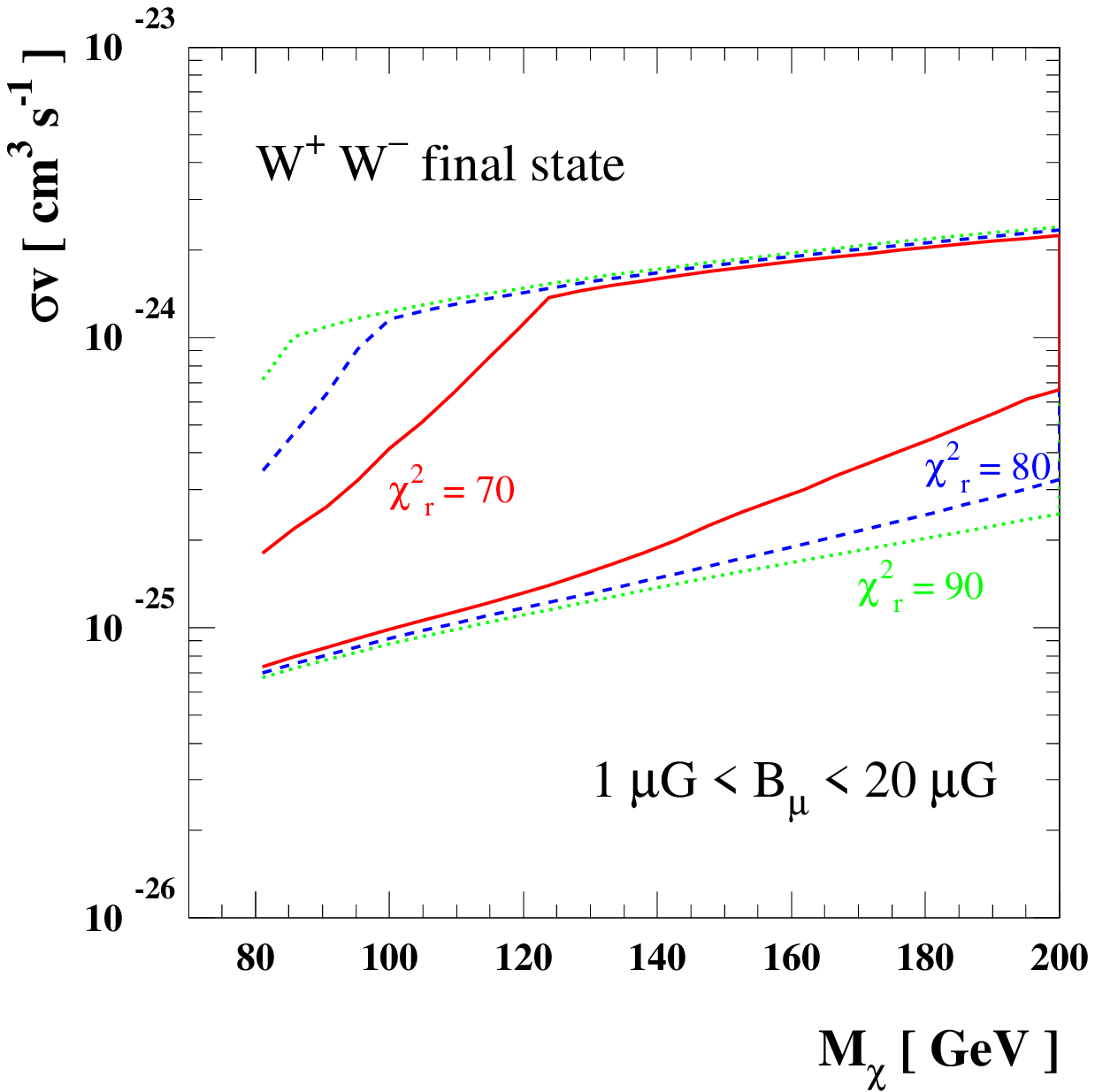}\\
\end{center}
\caption{Isolevel curves for minimum reduced $\chi^2$ from the fit of the Coma radio flux
density spectrum, obtained by varying the magnetic field within $1 \mu G \le B_\mu \le 20
\mu G$, for a given annihilation channel. The halo profile is the best fit N04 profile:
$M_{vir} = 0.9 \, 10^{15} \msun h^{-1}$ and $c_{vir} = 10$, with subhalo setup as in
Fig.~\protect{\ref{fig:npairs}}.} \label{fig:fit2}
\end{figure*}

Since the data on the radio flux density spectrum (\cite{Thierbach2003}) is a compilation
of measurements performed with different instruments (possibly with different
systematics), it is difficult to decide a cut on the $\chi^2_r$ value which defines an
acceptable fit. In Fig.~\ref{fig:fit1} we plot sample isolevel curves for $\chi^2_r$,
spotting the shape of the minima of $\chi^2_r$ , in the plane $B - \langle \sigma v
\rangle_0$, for the two sample annihilation channels and a few sample values of the WIMP
mass (note that values labeling isolevel curves are sensibly different in the two
panels).
In Fig.~\ref{fig:fit2} we show the analogous $\chi^2_r$ isolevels in the WIMP mass --
annihilation rate plane, and taking at each point the minimum $\chi^2_r$ while varying
the magnetic field strength between $1 \mu G$ and $20 \mu G$: the curves converge to a
maximal value enforced by the lower limit of $1 \mu G$, and the upper value does not
enter in defining isolevel curve shapes.

In order to assess whether the outlined radio-data preferred regions are or are not compatible with
supersymmetric DM models, we proceed to a random scan of the SUSY parameter space, in the bottom-up
approach which we outline below.
We relax all universality assumptions, and fully scan the low-energy scale MSSM, imposing
phenomenological as well as cosmological constraints on the randomly generated
models\footnote{We scan all the SUSY parameters linearly over the indicated range.}. We
take values of $\tan\beta$, the ratio between the vacuum expectation values of the two
Higgs doublets, between 1 and 60. The parameters entering the neutralino mass matrix are
generated in the range $1\ {\rm GeV}<m_1,\ m_2,\ \mu<1 \ {\rm TeV}$, and we define
$m_{\rm LSP}\equiv{\rm min}(m_1,\ m_2,\ \mu)$. To avoid flavor changing effects in the
first two lightest quark generations, we assume that the soft-breaking masses in the
first two generations squark sector are degenerate, {\em i.e.} we assume
$m^{(1)}_{\widetilde Q,\ \widetilde U,\ \widetilde D}=m^{(2)}_{\widetilde Q,\ \widetilde
U,\ \widetilde D}$. The scalar masses are scanned over the range
\begin{equation}
m_{\rm LSP}\ < \ m^{(1,3)}_{\widetilde Q}, \ m^{(1,3)}_{\widetilde
U}, \ m^{(1,3)}_{\widetilde U}, \ m^{(1,2,3)}_{\widetilde L}, \
m^{(1,2,3)}_{\widetilde E}, \ m_A \ <\ 2.5\ {\rm TeV}
\end{equation}
The trilinear couplings are sampled in the range
\begin{eqnarray}
-3\cdot{\rm max}(m^{(i)}_{\widetilde Q},\ m^{(i)}_{\widetilde U})\ < & A_U^{(i)} & <\
3\cdot{\rm max}(m^{(i)}_{\widetilde Q},\ m^{(i)}_{\widetilde U})\\ -3\cdot{\rm
max}(m^{(i)}_{\widetilde Q},\ m^{(i)}_{\widetilde D})\ < & A_D^{(i)} &<\ 3\cdot{\rm
max}(m^{(i)}_{\widetilde Q},\ m^{(i)}_{\widetilde D})\\ -3\cdot{\rm
max}(m^{(i)}_{\widetilde L},\ m^{(i)}_{\widetilde E})\ < & A_E^{(i)} &<\ 3\cdot{\rm
max}(m^{(i)}_{\widetilde L},\ m^{(i)}_{\widetilde E})
\end{eqnarray}
Finally, we take the gluino mass in the range $200\ {\rm GeV}<m_{\widetilde g}<3\ {\rm
TeV}$. The mass ranges for squarks and gluino have been chosen following qualitative
criteria (\cite{msugracolliders,Battaglia:2003ab}), so that all viable models generated
should be ``visible'' at the LHC.

\begin{figure*}[!t]
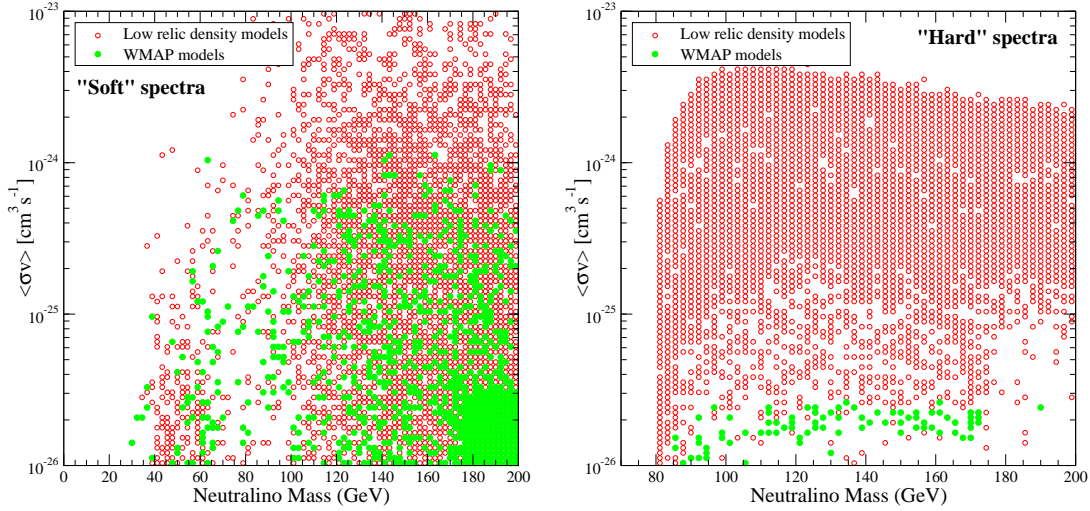

\begin{center}
\hspace*{-0.5cm}\includegraphics[scale=0.35]{grid_soft.eps}
\quad\includegraphics[scale=0.35]{grid_hard.eps}\\
\end{center}
\caption{A scatter plot of SUSY models, consistent with all available phenomenological constraints, giving a relic abundance in the 2-$\sigma$ WMAP range (green filled circles) or below it (low, relic density models, red circles), for soft (left panel) and hard (right panel) annihilation spectra.} \label{fig:scatter}
\end{figure*}

We exclude models giving a relic abundance of neutralinos exceeding $\Omega_\chi
h^2>0.13$. Further, we impose the various colliders mass limits on charginos, gluinos,
squarks and sleptons, as well as on the Higgs masses\footnote{Since we do not impose any
gaugino unification relation, we do not impose any constraint from collider searches on
the neutralino sector.}. Moreover, we also require the BR($b\rightarrow s\gamma$) and all
electroweak precision observables to be consistent with the theoretical and experimental
state-of-the-art (\cite{Eidelman:2004wy}).\\
We classify the models according to the branching ratios of the neutralino pair-annihilations final
states, according to the following criteria: we consider a model having a {\em hard} spectrum if
\begin{equation}
{\rm BR}(W^+W^-)+{\rm BR}(ZZ)>0.8;
\end{equation}
a {\em soft} spectrum is instead attributed to models satisfying the condition
\begin{equation}
\sum_{i=1}^6 \left[{\rm BR}(q_i\bar q_i)+{\rm BR}(q_i\bar q_i
g)\right]+{\rm BR}(gg)>0.8.
\end{equation}
We show, in Fig.\ref{fig:scatter} a scatter plots of the viable SUSY configurations,
indicating with filled green circles those {\em thermally} producing a neutralino relic
abundance within the 2-$\sigma$ WMAP range, and with red circles those producing a relic
abundance {\em below} the WMAP range (whose relic abundance can however be cosmologically
enhanced, in the context of quintessential or Brans-Dicke cosmologies, or which can be
non-thermally produced, as to make up all of the observed CDM (see \cite{nonth}). The low
$\chi^2$ ranges of $\langle \sigma v \rangle_0 - B$ and $\langle \sigma v \rangle_0 -
M_{\chi}$ values indicated in Figs.~\ref{fig:fit1} and \ref{fig:fit2} are therefore shown
to be actually populated by a number of viable SUSY models.

\subsection{From the UV to the  gamma-ray band}

Inverse Compton (IC) scatterings of relativistic electrons and positrons on target  cosmic microwave
background (CMB) photons give rise to a spectrum of photons stretching from below the extreme
ultra-violet up to the soft gamma-ray band, peaking in the soft X-ray energy band. Let $E= \gamma m_e
c^2$ be the energy of electrons and positrons, $\epsilon$ that of the target photons and $E_\gamma$
the energy of the scattered photon. The Inverse Compton power is obtained by folding the differential
number density of target photons with the IC scattering cross section:
\begin{equation}
P_{\rm IC}\left(E_\gamma,E\right) = c \, E_\gamma \int d\epsilon \, n(\epsilon) \,
\sigma(E_\gamma, \epsilon, E)
\label{eq:ICpower}
\end{equation}
where $n(\epsilon)$ is the black body spectrum of the $2.73 K$ CMB photons, while
$\sigma(E_\gamma, \epsilon, E)$ is given by the Klein-Nishina formula:
\begin{equation}
\sigma(E_\gamma, \epsilon, E) = \frac{3 \sigma_T}{4 \epsilon \gamma^2}\,
G\left(q,\Gamma_e\right)
\end{equation}
where $\sigma_T$ is the Thomson cross section and
\begin{equation}
G\left(q,\Gamma_e\right) \equiv \left[  2 q \ln q + (1+2 q)(1-q) + \frac{\left(\Gamma_e q
\right)^2 (1-q)}{2 \left( 1+ \Gamma_e q\right)}\right]
\end{equation}
with
\begin{equation}
 \Gamma_e= 4 \epsilon \gamma / (m_e c^2) \;\;\;\;\; q = E_\gamma /
\left[ \Gamma_e \left( \gamma m_e c^2 - E_\gamma \right)\right]\;.
\end{equation}
Folding the IC power with the spectral distribution of the equilibrium number
density of electrons and positrons, we get the local emissivity of IC photons of energy
$E_\gamma$:
\begin{equation}
j_{\rm IC}\left(E_\gamma, r\right) = \int
dE\, \left(\frac{dn_{e^-}}{dE} + \frac{dn_{e^+}}{dE} \right)
P_{\rm IC}\left(E_\gamma,E\right)\;
\label{eq:ICemiss}
\end{equation}
which we use to estimate the  integrated flux density spectrum:
\begin{equation}
S_{\rm IC}(E_\gamma)= \int d^3r \, \frac{ j_{\rm IC}\left(E_\gamma, r\right)}{4 \pi\, D_{\rm Coma}^2}\;.
\label{eq:ICflux}
\end{equation}
In Eq.~(\ref{eq:ICpower}) and Eq.~(\ref{eq:ICemiss}) the limits of integration over $\epsilon$ and
$E_\gamma$ are set from the kinematics of the IC scattering which restricts $q$ in the range $1/(4
\gamma^2) \le q \le 1$.

The last relevant contribution to the photon emission of Coma due to relativistic
electrons and positrons is the process of non-thermal bremsstrahlung, {\em i.e.} the
emission of gamma-ray photons in the deflection of the charged particles by the
electrostatic potential of intra-cluster gas. Labeling with $E= \gamma m_e c^2$ the
energy of electrons and positrons, and with $E_\gamma$ the energy of the emitted photons,
the local non-thermal bremsstrahlung power is given by:
\begin{equation}
P_{\rm B}\left(E_\gamma,E,r\right) = c \, E_\gamma \sum_j  n_j(r) \,
\sigma_j(E_\gamma, E)\;,
\label{eq:Bpower}
\end{equation}
with the sum including all species $j$ in the intra-cluster medium, each with number density $n_j(r)$
and relative production cross section:
\begin{equation}
\sigma_j(E_\gamma, E) = \frac{3 \alpha \sigma_T}{8 \pi\, E_\gamma} \cdot
\left[\left(1+(1-E_\gamma/E)^2\right)\,\phi_1-\frac{2}{3}(1-E_\gamma/E)\,\phi_2\right]
\end{equation}
where $\alpha$ is the fine structure constant, $\phi_1$ and $\phi_2$ two energy dependent
scattering functions which depend on the species $j$ (see \cite{book} for details). The
emissivity $j_{\rm B}\left(E_\gamma, r\right) $ is obtained by folding the power over the
equilibrium electron/positron number density, {\em i.e.} the analogous of
Eq.~(\ref{eq:ICemiss}), while the integrated flux density $S_{\rm B}(E_\gamma)$ is
obtained by summing over all relevant sources as in Eq.~(\ref{eq:ICflux}). We apply this
scheme to Coma implementing the gas density profile in Eq.~(\ref{eq:gas}) by including
atomic and molecular hydrogen and correcting for the helium component.

As we have already mentioned, a hard gamma-ray component arises also from prompt emission in WIMP pair
annihilations, either in loop suppressed two-body final states giving monochromatic   photons, or
through the production and prompt decay of neutral pions giving gamma-rays with continuous spectrum.
Since photons propagate on straight lines (or actually geodesics), the gamma-ray flux due to prompt
emission is just obtained by summing over sources along the line of sight; we will consider terms
integrated over volume
\begin{equation}
  F_{\gamma}(E_\gamma)= \int d^3r \,
  \frac{ Q_\gamma\left(E_\gamma, r\right)}{4 \pi\, D_{\rm Coma}^2}\;.
\label{eq:gammaflux}
\end{equation}

\subsection{The multi-frequency SED of Coma}

\begin{figure*}[!t]
\begin{center}
\includegraphics[scale=0.5]{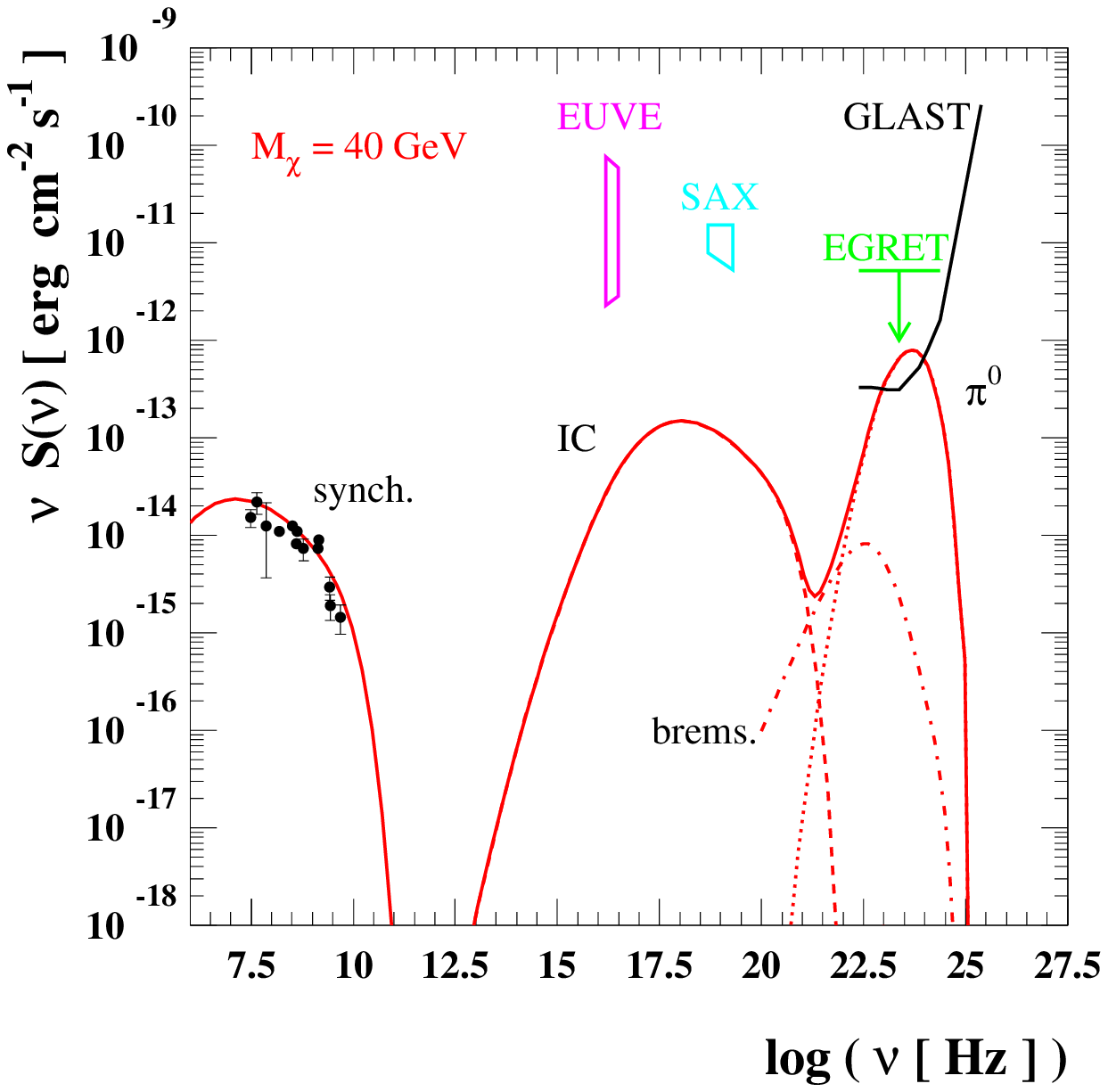}
\quad
\hspace*{-1.cm}
\includegraphics[scale=0.5]{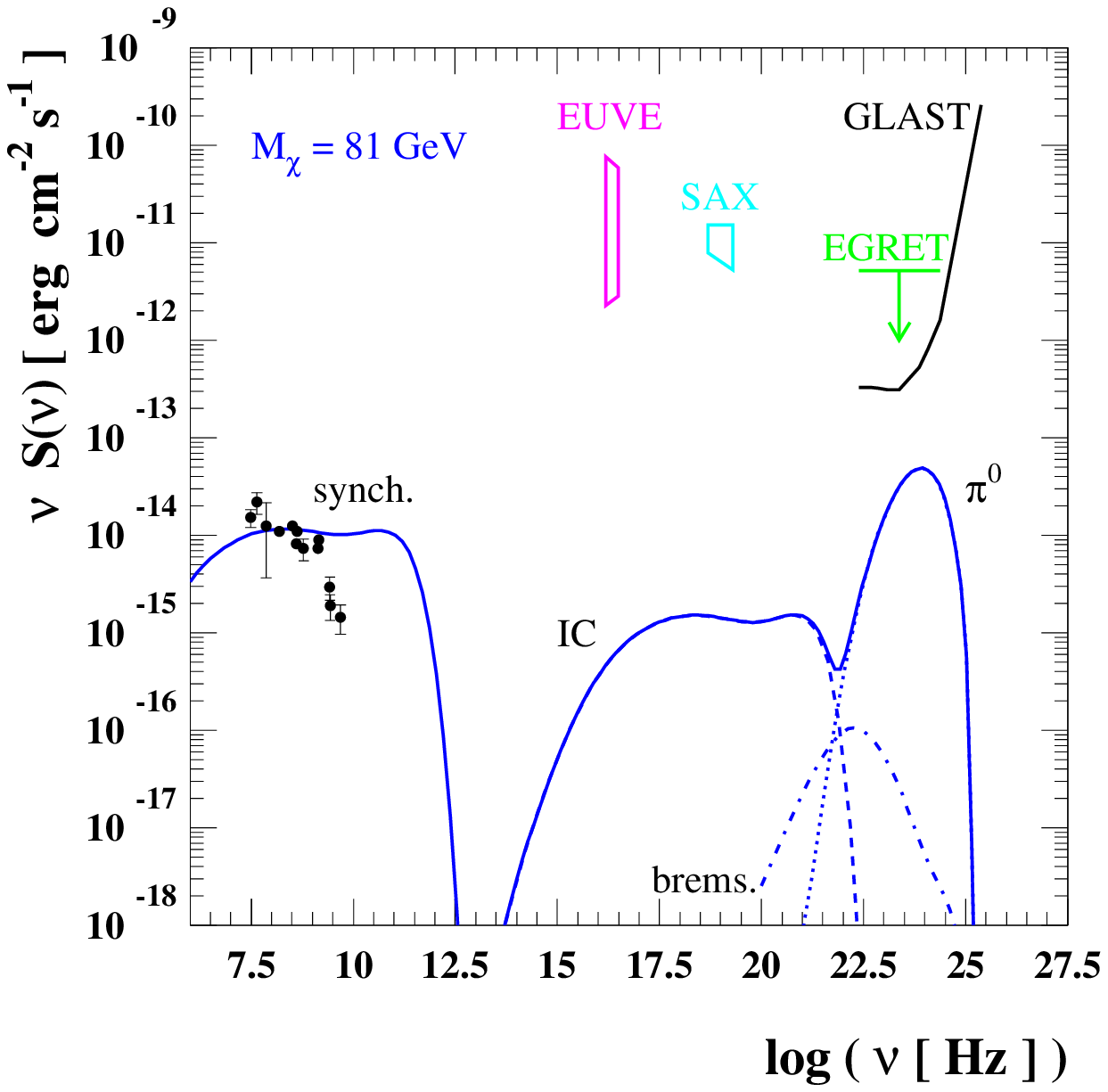}\\
\end{center}
\caption{Multi-wavelength spectrum of the two best fit models for the radio flux shown in
Fig.~\protect{\ref{fig:bestfit}} (see text for details). The halo profile is the best fit
N04 profile: $M_{vir} = 0.9 \, 10^{15} \msun h^{-1}$ and $c_{vir} = 10$, with subhalo
setup as in Fig.~\protect{\ref{fig:npairs}}. }
 \label{fig:multiflux}
\end{figure*}

We show in Fig.~\ref{fig:multiflux} the multi-frequency SED produced by WIMP annihilation
in the two models used to fit the radio halo spectrum of Coma, as shown in
Fig.~\ref{fig:bestfit}.
%

The model with $M_{\chi}=40$ GeV provides the better fit to the radio halo data because
the relative equilibrium electron spectrum is steeper and shows also the high-$\nu$
bending which fits the most recent data (\cite{Thierbach2003}). The IC and bremsstrahlung
branches of the SED are closely related to the synchrotron branch (since they depend on
the same particle population) and their intensity ratio depends basically on the value of
the adopted magnetic field. The relatively high value $B = 1.2$ \mug indicated by the
best fit to the radio data implies a rather low intensity of the IC and bremsstrahlung
emission, well below the EUV and hard X-ray data for Coma. Nonetheless, the gamma-ray
emission due to $\pi^0 \to \gamma \gamma$ decay predicted by this model could be
detectable with the GLAST-LAT detector, even though it is well below the EGRET upper
limit.

The detectability of the multi-frequency SED worsens in the model with $M_{\chi}=81$ GeV,
where the flatness of the equilibrium electron spectrum cannot provide an acceptable fit
to the radio data. Moreover, the adopted value of the magnetic field $B=8$ \mug implies a
very low intensity of the IC, bremsstrahlung and $\pi^0 \to \gamma \gamma$ emission,
which should be not detectable by the next generation HXR and gamma-ray experiments.

The energetic electrons and positrons produced by WIMP annihilation have other
interesting astrophysical effects among which we will discuss specifically in the
following the Sunyaev-Zel'dovich (hereafter SZ) effect produced by DM annihilation and
the heating of the intracluster gas produced by Coulomb collisions.

\subsection{SZ effect}

The energetic electrons and positrons produced by WIMP annihilation interact with the CMB
photons and up-scatter them to higher frequencies producing a peculiar SZ effect (as
originally realized by \cite{Colafrancesco2004}) with specific spectral and spatial
features.

The generalized expression for the SZ  effect  which is valid in the Thomson limit for a
generic electron population in the relativistic limit and includes also the effects of
multiple scatterings and the combination with other electron population in the cluster
atmospheres has been derived by \cite{Colafrancescoetal2003}. This approach is the one
that should be properly used to calculate the specific SZ$_{\rm DM}$ effect induced by
the secondary electrons produced by WIMP annihilation. Here we do not repeat the
description of the analytical technique and we refer to the general analysis described in
\cite{Colafrancescoetal2003}. According to these results, the DM induced spectral
distortion is
 \begin{equation}
\Delta I_{\rm DM}(x)=2\frac{(k_{\rm B} T_0)^3}{(hc)^2}y_{\rm DM} ~\tilde{g}(x) ~,
\end{equation}
where $T_0$ is the CMB temperature and the Comptonization parameter $y_{\rm DM}$ is given by
\begin{equation}
y_{\rm DM}=\frac{\sigma_T}{m_{\rm e} c^2}\int P_{\rm DM} d\ell ~,
\end{equation}
in terms of the pressure $P_{\rm DM}$ contributed by the secondary electrons produced by neutralino
annihilation.
The quantity $y_{\rm DM} \propto \langle \sigma v \rangle_{0}  n^2_{\chi}$ and scales as
$\propto \langle \sigma v \rangle_{o} M^{-2}_{\chi}$, providing an increasing pressure
$P_{\rm DM}$ and optical depth $\tau_{\rm DM} = \sigma_T \int d \ell n_{\rm e}$ for
decreasing values of the neutralino mass $M_{\chi}$.
The function $\tilde{g}(x)$, with $x \equiv h \nu / k_{\rm B} T_0$, can be written as
\begin{equation}
\label{gnontermesatta} \tilde{g}(x)=\frac{m_{\rm e} c^2}{\langle k_{\rm B} T_{\rm e} \rangle} \left\{
\frac{1}{\tau} \left[\int_{-\infty}^{+\infty} i_0(xe^{-s}) P(s) ds- i_0(x)\right] \right\}
\end{equation}
in terms of the photon redistribution function $P(s)$ and of  $i_0(x) = 2 (k_{\rm B} T_0)^3 / (h c)^2
\cdot x^3/(e^x -1)$, where we defined the quantity
\begin{equation}
 \langle k_{\rm B} T_{\rm e} \rangle   \equiv  \frac{\sigma_{\rm T}}{\tau}\int P d\ell
= \frac{\int P d\ell}{\int n_{\rm e} d\ell}
 = \int_0^\infty dp f_{\rm e}(p) \frac{1}{3} p v(p) m_{\rm e} c
 \label{temp.media}
\end{equation}
(see \cite{Colafrancescoetal2003,Colafrancesco2004}), which is the analogous of the
average temperature for a thermal population (for a thermal electron distribution
$\langle k_{\rm B} T_{\rm e} \rangle = k_{\rm B} T_{\rm e}$ obtains, in fact). The photon
redistribution function $P(s)= \int dp f_{\rm e}(p) P_{\rm s}(s;p)$ with $s =
\ln(\nu'/\nu)$, in terms of the CMB photon frequency increase factor $\nu' / \nu = {4
\over 3} \gamma^2 - {1 \over 3}$, depends on the electron momentum ($p$) distribution,
$f_{\rm e}(p)$, produced by WIMP annihilation.
\begin{figure}[!t]
\begin{center}
\includegraphics[scale=0.35]{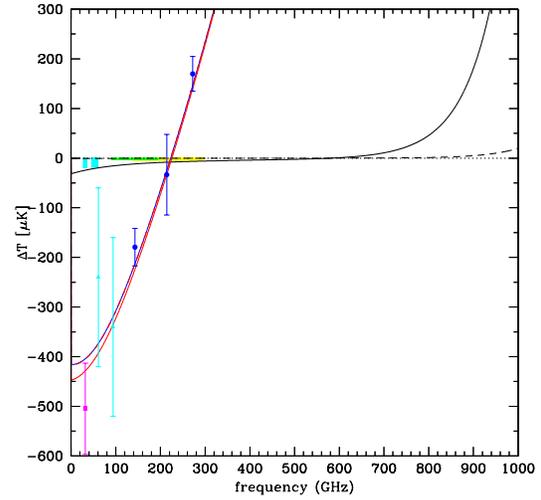}
\end{center}
\caption{The SZ effect produced by the $b \bar{b}$ model with $M_{\chi}=40$ GeV (black
solid curve) and by the $W^+W^-$ model with $M_{\chi}=81$ GeV  (black dashed curve) in
Coma are shown in comparison with the thermal SZ effect of Coma (blue curve). The red
curves represent the overall SZ effect. Notice that the DM-induced SZ effect has a very
different spectral behavior with respect  to the thermal SZ effect. SZ data are from OVRO
(magenta), WMAP (cyan) and MITO (blue). The sensitivity of PLANCK (18 months, 1 $\sigma$)
is shown for the LFI detector at 31.5 and 53 GHz channels (cyan shaded regions) and for
the HFI detector 143 and 217 GHZ channels (green and yellow shaded areas, respectively).
}
 \label{fig:sz_coma_spec}
\end{figure}
\begin{figure}[!t]
\begin{center}
\includegraphics[scale=0.35]{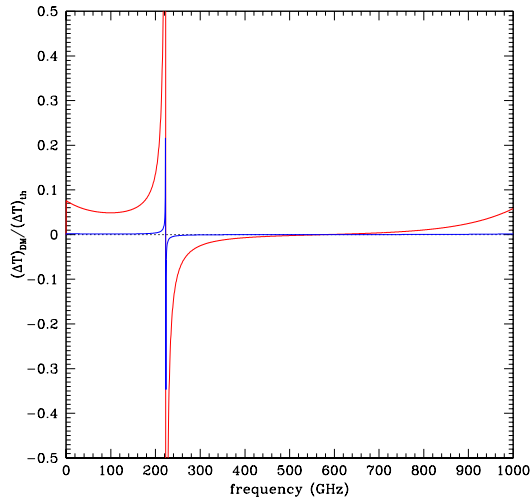}
\end{center}
\caption{The ratio between the DM-induced and the thermal SZ effect in Coma is shown for the model
with $M_{\chi}=40$ GeV (red curve) and for the model with $M_{\chi}=81$ GeV (blue curve). The model
with $M_{\chi}=40$ GeV produces an SZ effect which could be detected with the next coming microwave
experiments.}
 \label{fig:sz_coma_ratio}
\end{figure}

We show in Fig.\ref{fig:sz_coma_spec} the frequency dependence of the CMB temperature
change,
 \be
{\Delta T \over T_{0}} = {(e^x-1)^2 \over x^4 e^x} {\Delta I \over I_0} \,,
 \ee
as produced by the DM-induced SZ effect in the two best fit WIMP models here considered, compared to
the temperature change due to the thermal SZ effect produced by the intracluster gas. The most recent
analysis of the thermal SZ effect in Coma (\cite{DePetris2003}) provides an estimate of the optical
depth of the thermal intracluster gas $\tau_{th} = 4.9 \cdot 10^{-3}$ which best fits the data.
The model with $M_{\chi}=40$ GeV provides a detectable SZ$_{DM}$ effect which has a quite
different spectral shape with respect to the thermal SZ effect: it yields a temperature
decrement at all the microwave frequencies, $\simlt 600$ GHz, where the thermal SZ effect
is observed and produces a temperature increase only at very high frequencies $> 600$
GHz. This behavior is produced by the large frequency shift of CMB photons induced by the
relativistic secondary electrons generated by the WIMP annihilation. As a consequence,
the zero of the SZ$_{DM}$ effect is effectively removed from the microwave range and
shifted to a quite high frequency $\sim 600$ GHz with respect to the zero of the thermal
SZ effect, a result which allows one, in principle, to estimate directly the pressure of
the electron populations and hence to derive constraints on the WIMP model (see
\cite{Colafrancesco2004}).

The presence of a substantial SZ$_{\rm DM}$ effect is likely to dominate the overall SZ signal at
frequencies $x\simgt 3.8-4.5$ providing a negative total SZ effect. It is, however, necessary to
stress that in such frequency range there are other possible contributions to the SZ effect, like the
kinematic effect and the non-thermal effect which could provide additional biases (see, {\em e.g.},
\cite{Colafrancescoetal2003}).
Nonetheless, the peculiar spectral shape of the $SZ_{\rm DM}$ effect is quite different
from that of the kinematic SZ effect and of the thermal SZ effect and this result allows
us to disentangle it from the overall SZ signal.
An appropriate multi-frequency analysis of the overall SZ effect based on observations
performed on a wide spectral range (from the radio to the sub-mm region) is required to
separate the various SZ contributions and to provide an estimate of the DM induced SZ
effect.
In fact, simultaneous SZ observations at low frequencies $\sim 30$ GHz (where there is
the largest temperature decrement due to SZ$_{DM}$), at $\sim 150$ GHz (where the
SZ$_{\rm DM}$ deepens the minimum in $\Delta I/I$ with respect to the dominant thermal SZ
effect), at $\sim 220$ GHz (where the SZ$_{\rm DM}$ dominates the overall SZ effect and
produces a negative signal instead of the expected $\approx$ null signal) and at $\simgt
250$ GHz (where the still negative SZ$_{\rm DM}$ decreases the overall SZ effect with
respect to the dominant thermal SZ effect) coupled with X-ray observations which
determine the gas distribution within the cluster (and hence the associated dominant
thermal SZ effect) can separate the SZ$_{\rm DM}$ from the overall SZ signal, and
consequently, set constraints on the WIMP model.

The WIMP model with $M_{\chi}=40$ GeV produces a temperature decrement which is of the
order of $\sim$ 40 to 15 $\mu$K for SZ observations in the frequency range $\sim$ 30 to
150 GHz (see Fig.\ref{fig:sz_coma_spec}). These signals are still within the actual
uncertainties of the available SZ data for Coma and are below the current SZ sensitivity
of WMAP (see, e.g., \cite{Bennettetal2003} and the results of the analysis of the WMAP SZ
signals from a sample of nearby clusters performed by \cite{Lieuetal2005}).
Nonetheless, such SZ signals could be detectable with higher sensitivity experiments.
The high sensitivity planned for the future SZ experiments can provide much stringent
limits to the additional SZ effect induced by DM annihilation. In this context, the next
coming sensitive bolometer arrays (e.g., APEX), interferometric arrays (e.g., ALMA) and
the PLANCK-HFI experiment, or the planned OLIMPO balloon-borne experiment, have enough
sensitivity to probe the contributions of various SZ effects in the frequency range $\nu
\approx 30 - 250$ GHz, provided that accurate cross-calibration at different frequencies
can be obtained.
The illustrative comparison (see Fig.\ref{fig:sz_coma_spec}) between the model
predictions and the sensitivity of the PLANCK LFI and HFI detectors at the optimal
observing frequencies ($ \nu =31.5$ and $53$ GHz for the LFI detector and $\nu =143$ and
$217$ GHz for the HFI detector) show that the study of the SZ effect produced by DM
annihilation is actually feasible with the next generation SZ experiments.
We show in Fig.\ref{fig:sz_coma_ratio} the expected ratio between the DM-induced SZ
effect and the thermal SZ effect for the two WIMP models here considered. It is evident
that while the model with $M_{\chi}=40$ GeV provides a detectable signal which is a
sensitive fraction of the thermal SZ effect at  $\nu < 250$ GHz, the SZ signal provided
by the model with $M_{\chi}=81$ GeV is by far too small to be detectable at any
frequency.\\
The spectral properties shown by the SZ$_{DM}$ for neutralinos depends on the specific
neutralino model as we have shown in Fig.\ref{fig:sz_coma_spec}: in fact, the SZ effect
is visible for a neutralino with $M_{\chi} = 40$ GeV and not visible for a neutralino
with $M_{\chi} = 81$ GeV. Thus the detailed features of the SZ effect from DM
annihilation depends strongly on the mass and composition of the DM particle, and - in
turn -  on the equilibrium spectrum of the secondary electrons. Each specific DM model
predicts its own spectrum of secondary electrons and this influences the relative SZ
effect. Models of DM which provide similar electron spectra will provide similar SZ
effects.

\subsection{Heating of the intracluster gas}

Low energy secondary electrons produced by WIMP annihilation might heat the intracluster gas by
Coulomb collisions since the Coulomb loss term dominates the energy losses at $E \simlt 200$ MeV (see
Fig.~\ref{fig:losses}).
The specific heating rate is given by
 \be
 {dE \over dt dV} = \int dE {dn_e \over dE} \cdot \bigg({dE \over dt}\bigg)_{Coul}
 \ee
where ${dn_e \over dE}$ is the equilibrium electron spectrum derived in
Sect.~\ref{sec:secondary} and the Coulomb loss rate is $(dE/dt)_{Coul}= b_{Coul}^0 n
\left(1+\log(\gamma/n)/75 \right)$ where $n$ is the mean number density of thermal
electrons in $\rm{cm}^{-3}$ (see Eq.~\ref{eq:gas}, the average over space gives about $n
\simeq 1.3 \; 10^{-3}$), $\gamma \equiv E/m_e$ and $b_{Coul}^0 \simeq 6.13 \cdot
10^{-16}\; \rm{GeV}\, \rm{s}^{-1}$.
\begin{figure}[!t]
\begin{center}
\includegraphics[scale=0.4]{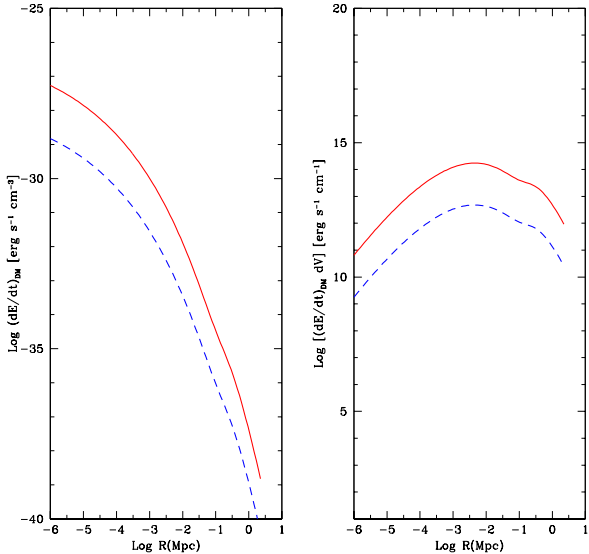}
\end{center}
\caption{Left: the specific heating rate is plotted against the radial distance from the center of
Coma. Right: the specific heating rate multiplied by the volume element is plotted against the radial
distance from the center of Coma. }
 \label{fig:heating_coma}
\end{figure}
Fig.\ref{fig:heating_coma} shows the specific heating rate of Coma as produced in the two
WIMP models explored here.
The non-singular N04 halo model adopted in our analysis does not provide a high specific heating rate
at the cluster center, and thus one might expect an overall heating rate for Coma which is of order of
$\sim 10^{38}$ erg/s ($\sim 10^{36}$ erg/s) for the  WIMP model with $M_{\chi}=40$ GeV ($M_{\chi}=81$
GeV).
We also notice that the region that mostly contributes to the overall heating of Coma is
not located at the center of the cluster. This is again a consequence of the non-singular
N04 DM profile which has been adopted.
The diffusion of electrons in the innermost regions of Coma acts in the same direction and moves the
maximum of the curves shown in the right panel of Fig.\ref{fig:heating_coma} towards the outskirts of
Coma, even in the case of a halo density profile which is steeper than the adopted one.\\
This implies, in conclusion, that WIMP annihilation cannot provide most of the heating of
Coma, even in its innermost regions.
Such a conclusion seems quite general and implies that non-singular DM halo models are
not able to provide large quantities of heating at the center of galaxy clusters so to
quench efficiently the cooling of the intracluster gas (with powers of $\sim 10^{43-45}$
erg/s).
Only very steep halo profiles (even steeper than the Moore profile) and with the possible adiabatic
growth of a central matter concentration (e.g., a central BH) could provide sufficient power to quench
locally ({\em i.e.} in the innermost regions) the intracluster gas cooling (see, {\em e.g.},
\cite{Totani2005}).
However, we stress that the spatial diffusion of the secondary electrons in the innermost regions of
galaxy clusters should flatten the specific heating rate in the vicinity of the DM spike and thus
decrease substantially the heating efficiency by Coulomb collisions. In conclusion, we believe that
the possibility to solve the cooling flow problem of galaxy clusters by WIMP annihilation is still an
open problem.

\section{Discussion}
 \label{sec:discussion}

WIMP annihilation in galaxy cluster is an efficient mechanism to produce relativistic
electrons and high-energy particles which are able, in turn, to produce a wide SED
extended over more than 18 orders of magnitude in frequency, from radio to gamma-rays. We
discuss here the predictions of two specific models which embrace a vast range of
possibilities.

The $b {\bar b}$ model with $M_{\chi}=40$ GeV  and annihilation cross section $\langle
\sigma v \rangle_0 = 4.7 \cdot 10^{-25} cm^3 s^{-1}$ provides a reasonable fit to the
radio data (both the total spectrum and the surface brightness radial distribution) with
a magnetic field whose mean value is $B \approx 1.2 \mu$G.
We remind here that the quite high value of $\langle \sigma v \rangle_0$ is well inside
the range of neutralino masses and annihilation cross-sections provided by the most
general supersymmetric DM setup (see our discussion in Sect.3.2).
Table \ref{tab.multif} provides an illustrative scheme of the radiation mechanisms, of
the particle energies and of the fluxes predicted by this best-fit WIMP model for a wide
range of the physical conditions in the cluster atmosphere.
\begin{table*}[!t]
\begin{center}
\begin{tabular}{|c|c|c|c|c|}
  \hline
                  & $\nu$     & $\nu F(\nu)$           & $E_{particle}$          &  Mechanism    \\
                  & [GHz]     & [$erg s^{-1} cm^{-2}$] & [GeV]                   &               \\
  \hline
  Radio           & $1.4$                   &  $6.4 \cdot 10^{-15}$     & $17.3\cdot B^{-1/2}_{\mu}$& $e^{\pm}$  Synchrotron   \\
  Optical         & $(7.25-14.5)\cdot 10^5$ &    $1.7 \cdot 10^{-15}$      & $(1.9-2.7)\cdot 10^{-2}$& $e^{\pm}$  ICS           \\
  EUV             & $(0.31-0.43)\cdot 10^8$ &    $5.2 \cdot 10^{-14}$      & $0.13-0.15$             & $e^{\pm}$  ICS           \\
  HXR             & $(4.83-19.3)\cdot 10^9$ &    $1.0 \cdot 10^{-13}$      & $1.56-3.13$             &  $e^{\pm}$ ICS           \\
                  &                         &  $1.2 \cdot 10^{-17}$        & $(4-16)\cdot10^{-5}$    &  $e^{\pm}$ Bremsstrahlung\\
  $\gamma$-ray    & $2.42 \cdot 10^{14}$    &   $6.6 \cdot 10^{-13}$       & $2$                     &   $\pi^0 \to \gamma \gamma$ decay\\
                  &                         &   $2.2 \cdot 10^{-15}$       & $2$                     &  $e^{\pm}$  Bremsstrahlung\\
  \hline
\end{tabular}
\end{center}
 \caption{Predicted flux in various frequency ranges for a neutralino $b {\bar b}$ model
 with $M_{\chi}=40$ GeV and $\langle \sigma v \rangle = 10^{-26}$ cm$^3$/s. The value of the magnetic field is $B_{\mu}=1.2 \, \mu$G.  $E_{particle}$  refers to the approximate energy for $e^\pm$ or $\gamma$ sourcing the corresponding flux in the monochromatic limit.}
 \label{tab.multif}
\end{table*}

For the best-fit values of $M_{\chi} = 40$ GeV and $\langle \sigma v \rangle_0 = 4.7
\cdot 10^{-25} cm^3 s^{-1}$ this model yields EUV and HXR fluxes which are more than one
order of magnitude fainter than the Coma data. The gamma-ray flux produced by this model
is dominated by the continuum $\pi^0 \to \gamma \gamma$ component and it is a factor
$\sim 5$ lower than the EGRET upper limit of Coma at its peak frequency (see
Fig.~\ref{fig:multiflux}, left panel). Such gamma-ray flux could be, nonetheless,
detectable by the GLAST--LAT detector (we will discuss more specifically the
detectability of the gamma-ray WIMP annihilation signals from galaxy clusters in a
dedicated, forthcoming paper (Colafrancesco, Profumo \& Ullio 2006b).
The rather low neutralino mass $M_{\chi}= 40$ GeV of this model makes it rather difficult
to be testable by Cherenkov gamma-ray detectors operating at higher threshold energies.

Increasing the neutralino mass does not provide a good fit of the radio-halo spectrum
(see Fig.~\ref{fig:multiflux}, right panel) and yields, in addition, extremely faint EUV,
HXR and gamma-ray fluxes, which turn out to be undetectable even by GLAST and/or by the
next coming high-energy experiments.

\begin{figure*}[!t]
\begin{center}
\includegraphics[scale=0.5]{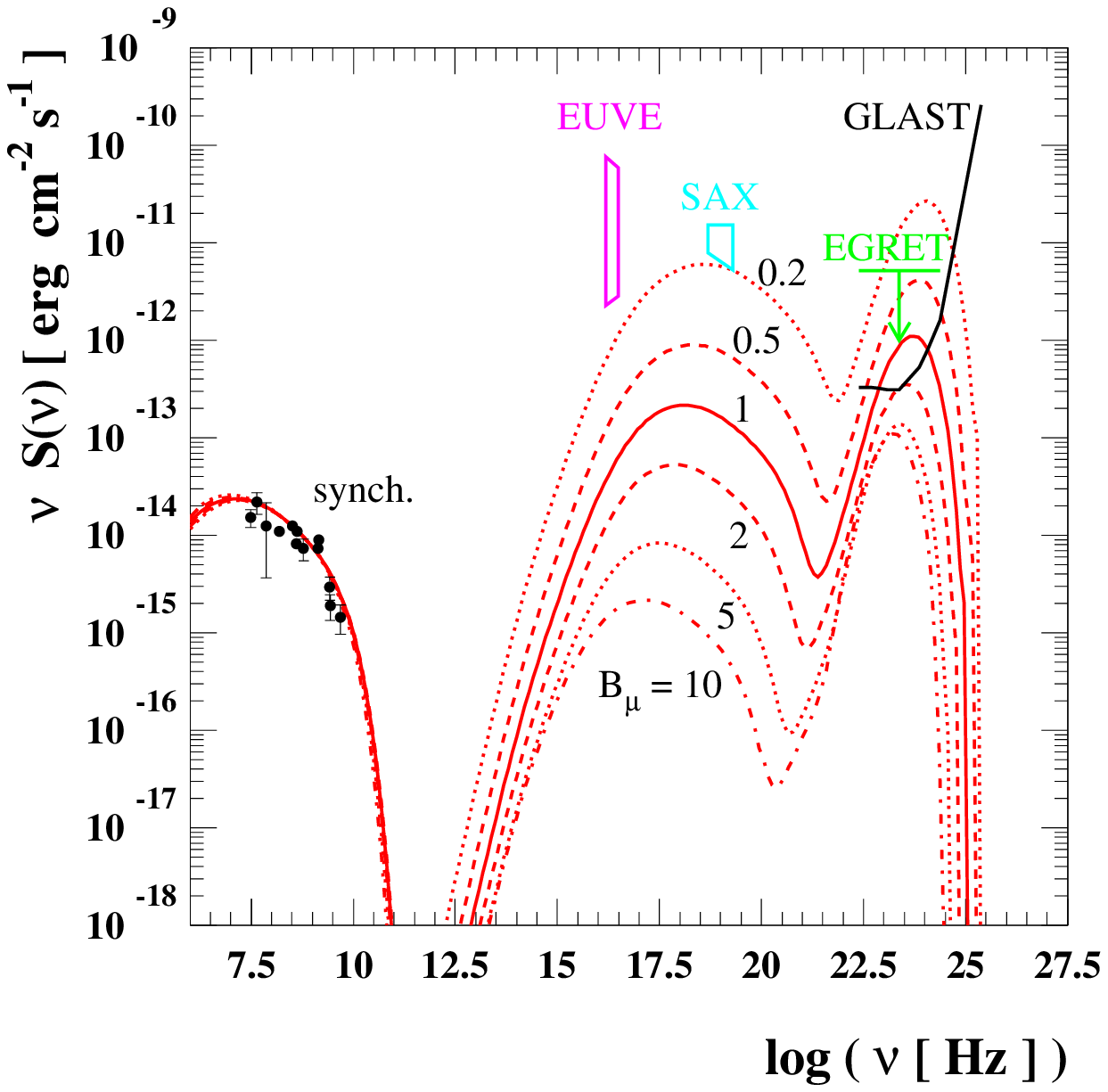}
\quad \hspace*{-1.cm}
\includegraphics[scale=0.5]{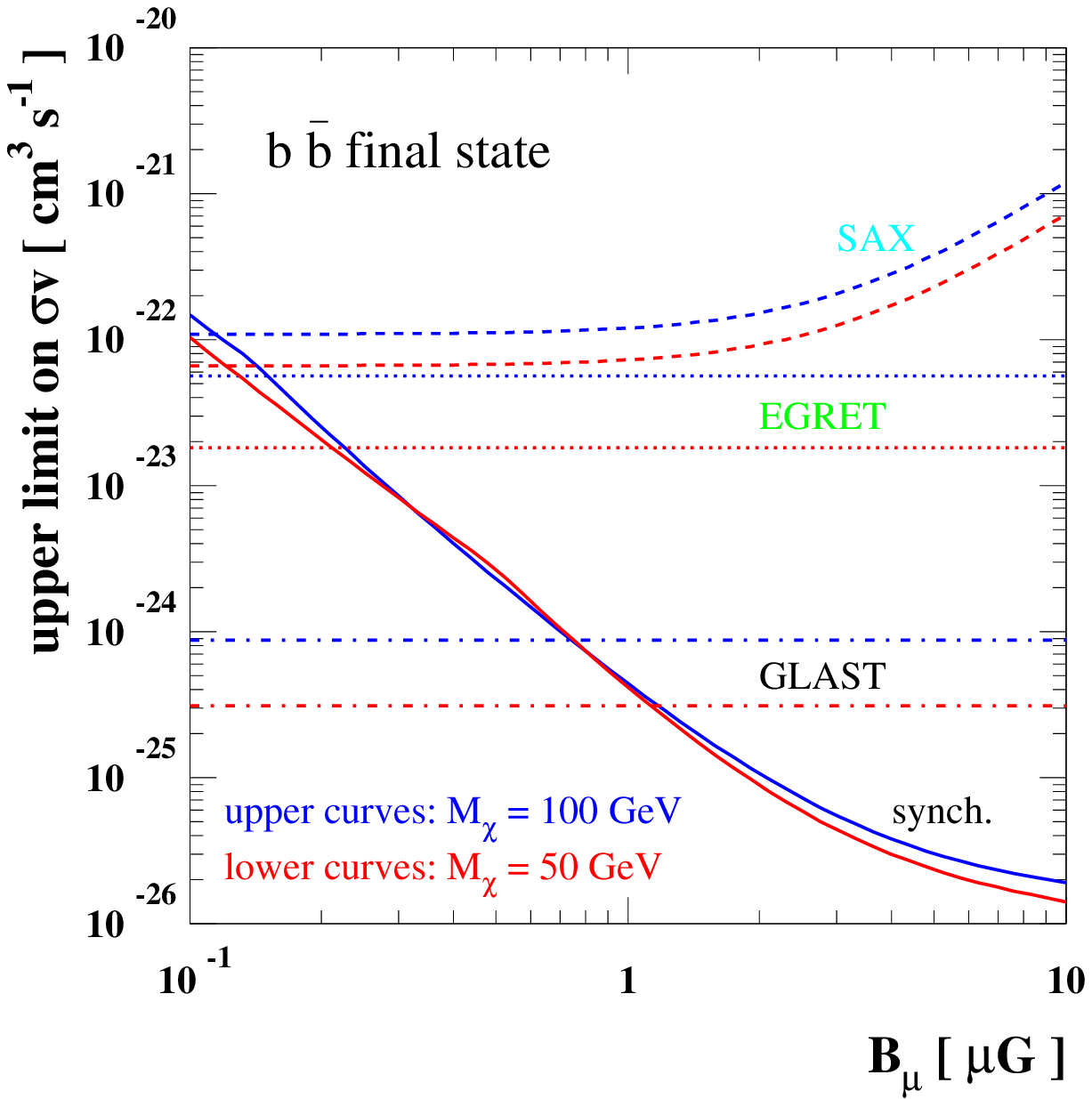}\\
\end{center}
\caption{Scaling of the multi-wavelength spectrum and of relative bounds on the particle
physics model with the assumed value for the mean magnetic field in Coma. Left panel: we
have chosen a few sample values for the magnetic field and varied freely pair
annihilation cross section and WIMP mass to minimize the $\chi^2$ for the fit of radio
data (a $b\,\bar{b}$ final state is assumed); the decrease in the magnetic field must be
compensated by going to larger $M_\chi$ and $\langle\sigma v\rangle_0$, with a net
increase in $\langle\sigma v\rangle_0/M_\chi^2$, as it can be seen from the increase in
the $\pi^0$ component. The increase in the IC component is, at large values for the
magnetic field, significantly more rapid, since for large values of the magnetic field
synchrotron losses are the main energy loss mechanism for electrons and positrons and
tend to decrease the number density for the equilibrium population. Right panel: upper
limit on $\langle\sigma v\rangle_0$ as a function of the assumed value for the mean
magnetic field in Coma; at each wavelength the limit is derived assuming that the
predicted flux should be lower than the upper limit from each individual data point
(slight overestimate of the limit from radio data, but we do not need to decide the cut
on the reduced $\chi^2$ marking the overshooting of the radio flux). Two sample values of
$M_\chi$ are assumed. The lines marked GLAST refer to the GLAST projected sensitivity
assuming no other $\gamma$-ray component is present. In both panels the halo profile is
the best fit N04 profile: $M_{vir} = 0.9 \, 10^{15} \msun h^{-1}$ and $c_{vir} = 10$,
with subhalo setup as in Fig.~\protect{\ref{fig:npairs}}.}
 \label{fig:multifluxb}
\end{figure*}

It is possible to recover the EUV and HXR data on Coma with a IC flux by secondary
electrons by increasing the annihilation cross-sections by a factor $\sim 10^2$ (i.e., up
to values $\langle \sigma v \rangle_0 \approx 7 \cdot 10^{-23} cm^3 s^{-1}$) in the
best-fit $b {\bar b}$ soft WIMP model (at fixed $M_{\chi}=40$ GeV). However, in such a
case both the radio-halo flux and the hard gamma-ray flux at $\sim 1$ GeV as produced by
$\pi^0$ decay should increase by the same factor leading to a problematic picture: in
fact, while the radio-halo data would imply lower values of the magnetic field $B \sim
0.1$ \mug which might still be allowed by the data, the $\pi^0 \to \gamma \gamma$
gamma-ray flux at $E > 100$ MeV should exceed the EGRET limit on Coma. This option is
therefore excluded by the available data.\\
Alternatively, it would be possible to fit the EUV and HXR spectra of Coma with the
adopted value of $\langle \sigma v \rangle_0 \approx 7 \cdot 10^{-23} cm^3 s^{-1}$ for
the $b {\bar b}$ model with $M_{\chi}=40$ GeV, in the case we sensibly lower the mean
magnetic field. Values of the average magnetic field $\simlt 0.2 \mu$G are required to
fit the HXR flux of Coma under the constraint to fit at the same time the radio-halo
spectrum (see Fig.\ref{fig:multifluxb}, left panel), consistently with the general
description of the ratio between the synchrotron and IC emission powers in Coma (see,
{\em e.g.}, \cite{CMP2005}, \cite{Reimer2003}). Magnetic fields as low as $\sim 0.15
\mu$G can fit both the HXR and the EUV fluxes of Coma.
However, also in this case the $\pi^0 \to \gamma \gamma$ gamma-ray flux predicted by the
same model at $E > 100$ MeV exceeds the EGRET limit on Coma, rendering untenable this
alternative.
Actually, the EGRET upper limit on Coma set a strong constraint on the combination of
values $B$ and $\langle \sigma v \rangle_0$ (see Fig.\ref{fig:multifluxb}, right panel)
so that magnetic field larger than $\simgt 0.3 $ \mug are required for the parameter
setup of the $b {\bar b}$ model with $M_{\chi}=40$ GeV. Fig.\ref{fig:multifluxb} shows
the upper limits on the value of $\langle \sigma v \rangle$ as a function of the assumed
value of the mean magnetic field of Coma.
According to these results, it is impossible to fit all the available data on Coma for a
consistent choice of the DM model and of the cluster magnetic field. The EUV and HXR data
in particular require extreme conditions, i.e. low values of the magnetic field and/or
high values of the annihilation cross section, which violate the EGRET gamma-ray limit.
Thus, realistic DM models that are consistent with the radio and gamma-ray constraints
predict IC emission which falls short of fitting the EUV and HXR data of Coma.

An appealing property of the WIMP model worked out here is that it can reproduce both the
spatial distribution of the radio-halo surface brightness of Coma and, in principle, also
the spatial profile of the EUV emission (see, {\em e.g.}, \cite{Bowyeretal2004}) which
seems more concentrated than the radio-halo surface brightness.
As for the radio-halo surface brightness profile, it seems necessary for this WIMP model
- due to the shape of the DM halo profile - to invoke a radial distribution of the
magnetic field with a mild decrease towards the Coma center to counterbalance the
centrally peaked DM profile, and with an exponential cutoff at large radii to
counterbalance the effect of the subhalo distribution.
We notice here that such a specific $B(r)$ spatial distribution is - interestingly
enough - able to reproduce the radial distribution of the RMs in Coma (see Sect.5.1).
While the (Synchrotron) radio surface brightness depends strongly on the magnetic field
radial profile, the (ICS) EUV radial profile only depends on the DM halo profile and on
the secondary electron properties and is, hence, more concentrated (see
Fig.~\ref{fig:bestfit} for an example). Thus, the radial distribution of the EUV emission
could be reasonably reproduced by the WIMP model which best fits the radio data but with
a very low value of the magnetic field of order of $\simlt 0.15$ $\mu$G.
We already noticed, however, that models with values of the average magnetic field in
Coma which are $\simlt 0.3 \mu$G produce a gamma-ray flux which exceeds the EGRET upper
limit of Coma (see Fig.~\ref{fig:multifluxb}), rendering these models untenable.

\begin{figure*}[!t]
\begin{center}
\includegraphics[scale=0.5]{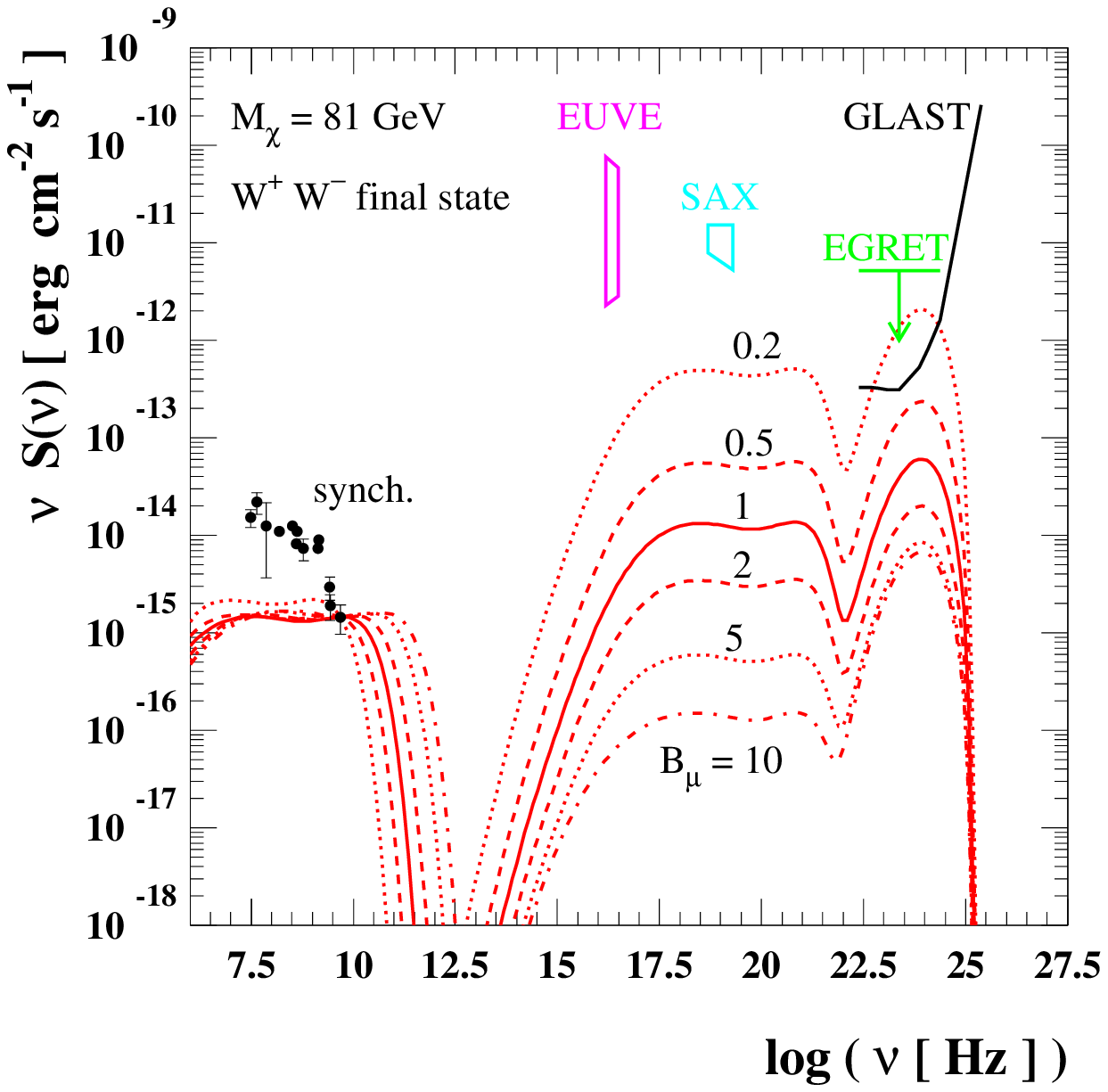}
\quad \hspace*{-1.cm}
\includegraphics[scale=0.5]{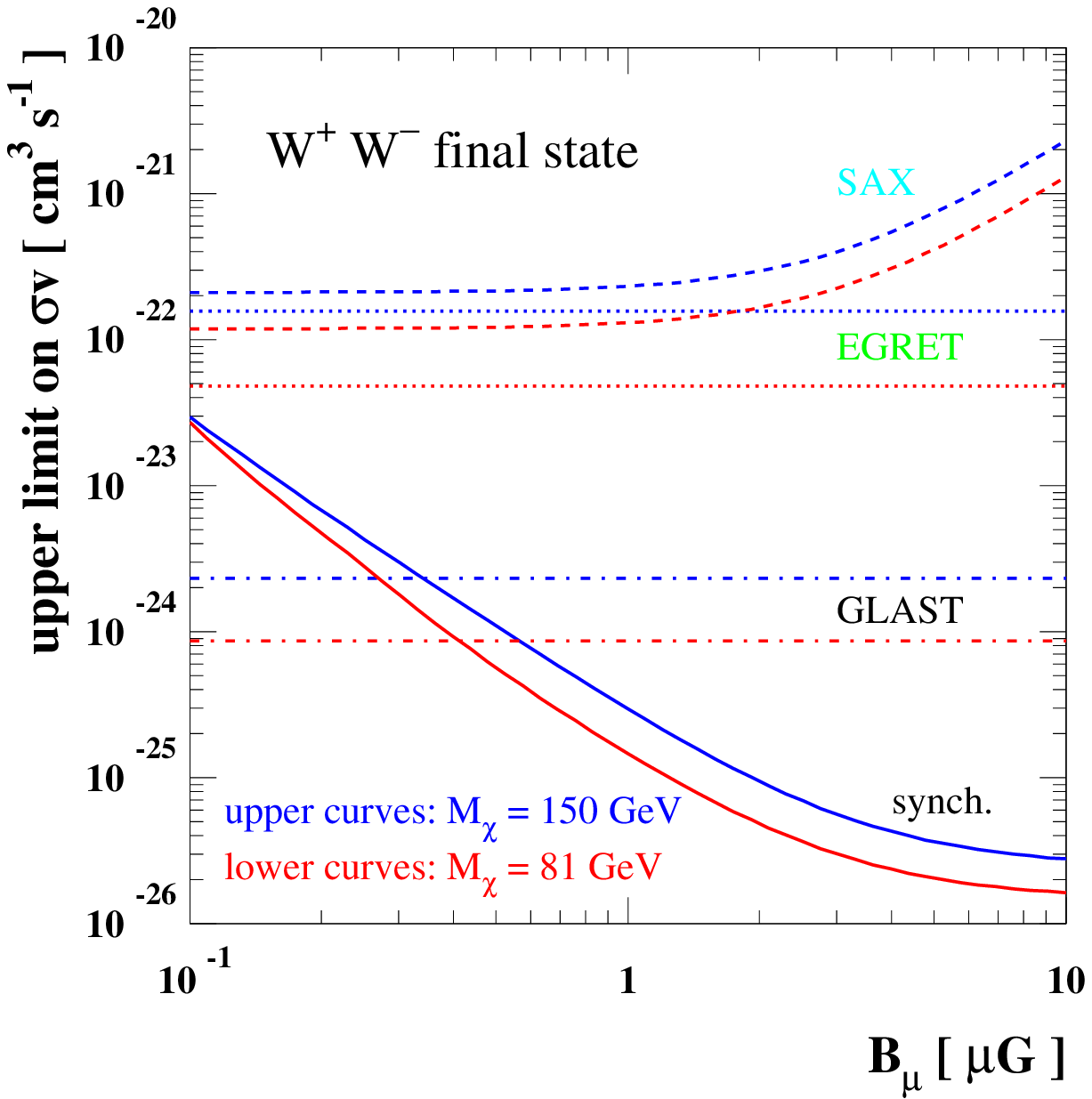}\\
\end{center}
\caption{The analogous of Fig.~\protect{\ref{fig:multifluxb}}, but now for the $W^+ W^-$ final state
and fixing the WIMP mass to 81~GeV. In the left panel, for each value of the magnetic field, the value
of $\langle\sigma v\rangle_0$ is obtained by normalizing the radio flux at the value of the flux at
the highest frequency point in the available dataset. Note again that on fair fit of the full radio
dataset can be derived for this hard channel.}
 \label{fig:sedww}
\end{figure*}

We summarize all the constraints on the neutralino models set by the magnetic field and
by the annihilation cross-section in Figs.~\ref{fig:multifluxb} and \ref{fig:sedww}.
The available data set constraints on the WIMP annihilation rate. Figs.22 and 23 show the
upper limits on $\langle\sigma v\rangle_0$ as a function of the assumed value for the
mean magnetic field in Coma. The EGRET limit proves to be the more constraining at the
moment with respect to the HXR and EUV data. These limits are able to test directly the
annihilation rate since they are independent of the magnetic field value. Nonetheless,
the combination of the gamma-ray and/or HXR constraints with the radio constraints will
be able to determine the full setup of the relevant quantities whose combination is able
to fit the overall Coma SED. In this context, it is clear that the possible GLAST
observations of Coma, combined with the radio data, will increase by far the constraints
in the $\langle\sigma v\rangle_0-B$ plane.

\begin{figure*}[!t]
\begin{center}
\includegraphics[scale=0.5]{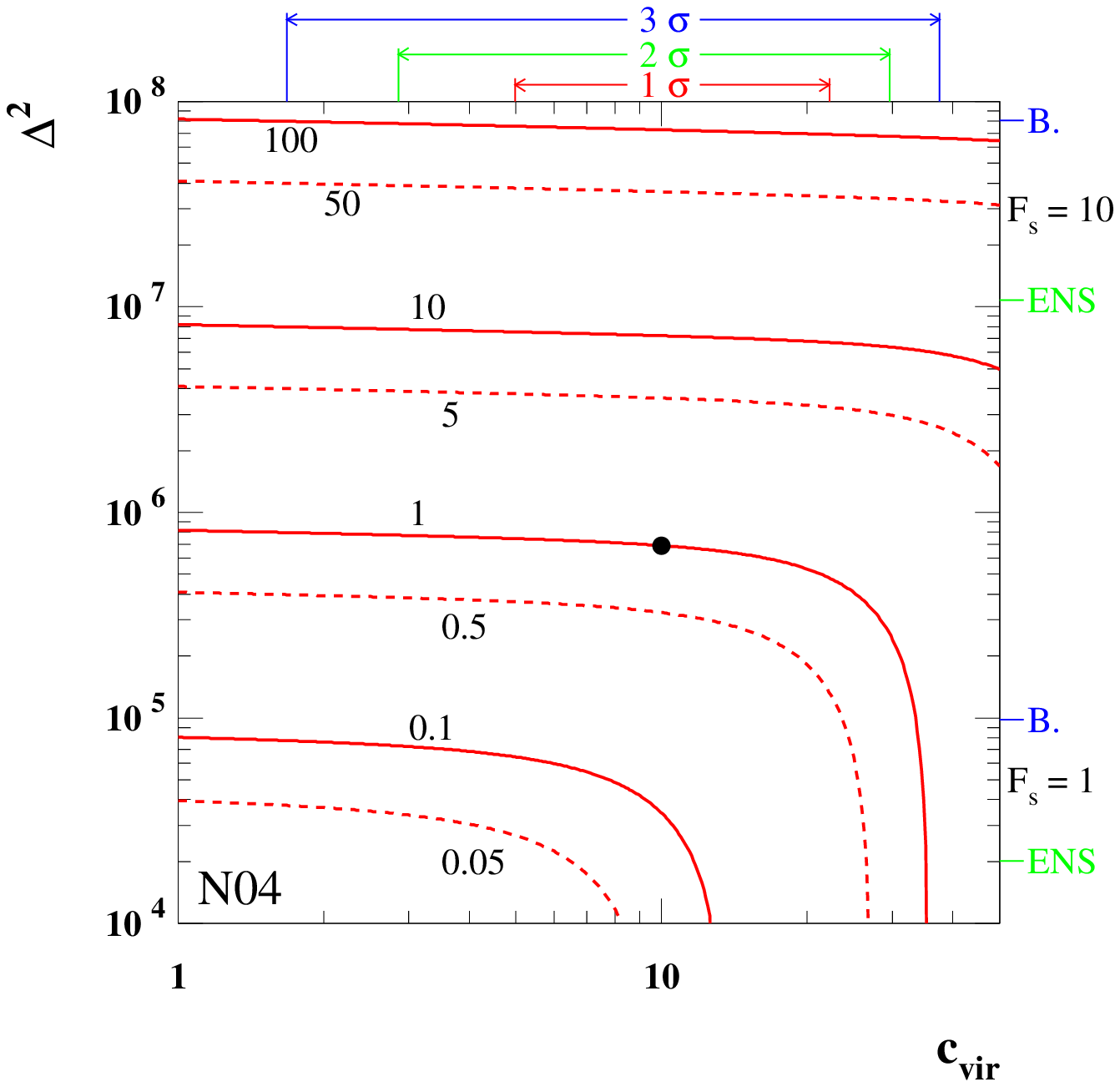}
\quad \hspace*{-1.cm}
\includegraphics[scale=0.5]{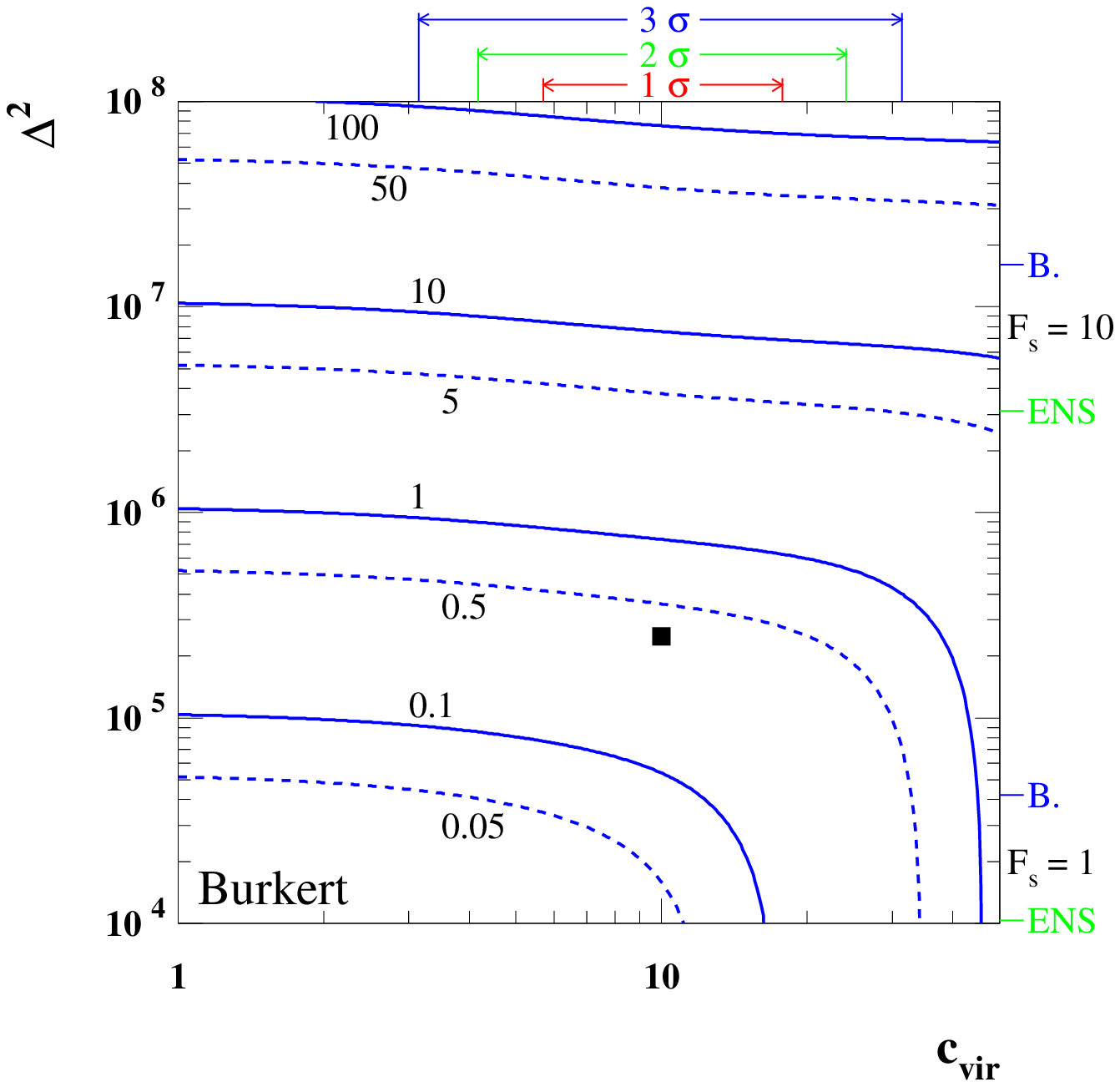}\\
\end{center}
\caption{Scaling of fluxes with the assumptions on the halo model for Coma. In the plane $c_{vir}$ --
$\Delta^2$ we plot isolevel curves for fluxes normalized to the corresponding values within the setup
as for the N04 profile in Fig.~\ref{fig:npairs}, marked with a dot in the left panel, {\em i.e.} the halo
model we have assumed so far as reference model; the model marked with square in the right panel
corresponds to the Burkert profile selected in Fig.~\ref{fig:npairs}. For all examples displayed we
have fixed $M_{vir} = 0.9 \, 10^{15} \msun h^{-1}$ and a 50\% mass in substructures. The left panel
refers to N04 profiles, the right panel to Burkert profiles; reference values for $c_{vir}$ and
$\Delta^2$, as obtained by fitting the corresponding halo profiles and from our discussion on the
substructure role, are marked on the axis. Switching to one of the halo models displayed here is
equivalent to shifting all values of $\langle\sigma v\rangle_0$ plotted in figures to $\langle\sigma v\rangle_0$ divided by the
scaling value shown here.}
 \label{fig:haloscale}
\end{figure*}

The results of our analysis also depend on the assumed DM halo density profile.
Fig.~\ref{fig:haloscale} shows the scaling of fluxes with the assumptions on the halo
model for Coma. We compare here, for the sake of illustration, the scalings of the N04
and of the Burkert model.
Switching to one of the halo models displayed here is equivalent to shifting all values
of $\langle\sigma v\rangle_0$ plotted in the figures to $\langle\sigma v\rangle_0$
divided by the scaling value shown here. This analysis allows us to compare correctly the
results of the multi-frequency analysis we have presented in this paper in terms of
substructure enhancement and halo density profile.

\begin{figure}[!t]
\begin{center}
\includegraphics[scale=0.5]{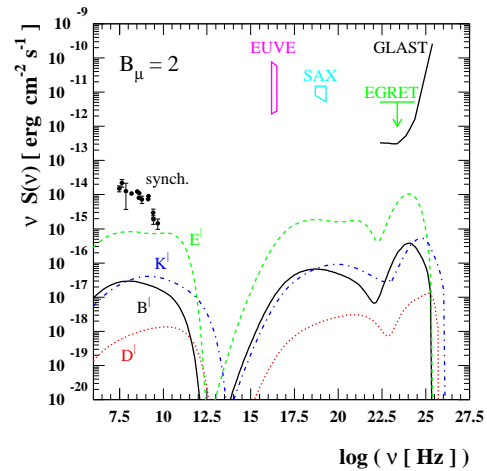}
\end{center}
\caption{Multi-wavelength spectra for the four benchmark models described in the text. The prediction
is shown for the best fit N04 profile, and our reference choice for subhalo parameters, and for a mean
magnetic field equal to $2 \mu G$.}
 \label{fig:benchSED}
\end{figure}

Table \ref{tab:detection} also shows the typical values for the annihilation
cross-section $\langle\sigma v\rangle_0$ and the relative signals expected at different
frequencies for the Coma best-fit model ($b {\bar b}$ neutralino model with $M_{\chi}=40$
GeV and $B= 1$ $\mu$G) that we explored in this paper.
\begin{table}[!b]
\begin{center}
\begin{tabular}{|c|c|c|}
 \hline
 & {\bf Cuspy profile} & {\bf Cored profile}\\
 \hline
 \hline
 {\bf Coma cluster} & $\langle\sigma v\rangle_0$ (${\rm cm^3}\ {\rm s^{-1}}$) &
 $\langle\sigma v\rangle_0$ (${\rm cm^3}\ {\rm s^{-1}}$)\\
 \hline
 Radio  & $5 \cdot 10^{-25}$   & $5 \cdot 10^{-24}$\\
 EUV    & $3 \cdot 10^{-23}$   & $3 \cdot 10^{-22}$\\
 HXR    & $2.5 \cdot 10^{-23}$ & $2.5 \cdot 10^{-22}$\\
 $\gamma$-ray (EGRET limit) & $1.3 \cdot 10^{-23}$ & $1.3 \cdot 10^{-22}$\\
 $\gamma$-ray (GLAST limit) & $7.5 \cdot 10^{-25}$ & $7.5 \cdot 10^{-24}$\\
 \hline
 \hline
 {\bf Milky Way} & $\langle\sigma v\rangle_0$ (${\rm cm^3}\ {\rm s^{-1}}$) &
 $\langle\sigma v\rangle_0$ (${\rm cm^3}\ {\rm s^{-1}}$)\\
 \hline
 Positrons     & $\sim 10^{-26}$ & $\sim 10^{-25}$\\
 Antiprotons   & $\sim 10^{-27}$ & $\sim 10^{-26}$\\
 Antideuterons & $\sim 10^{-27}$ & $\sim 10^{-26}$\\
 $\gamma$-ray (GLAST limit)    & $\sim 10^{-28}$ & $\sim 10^{-23}$\\
 \hline
\end{tabular}
\end{center}
\caption{Order-of-magnitude estimates of the values of the cross-section, $\langle\sigma
v\rangle_0$, in units of ${\rm cm^3}\ {\rm s^{-1}}$, needed to reproduce the detected
non-thermal emission features in Coma and to produce sizable signals for future indirect
dark matter search experiments in the Milky Way.}
 \label{tab:detection}
\end{table}
%
Large neutralino pair annihilation cross section will, in general, produce sizable
signals also for other indirect detection techniques, including antimatter searches and
gamma rays from the center of the Milky Way.
Antimatter and gamma-ray fluxes also largely depend on the Milky Way dark-matter halo and
on the specific neutralino model ({\em e.g.} through the antimatter yield per neutralino
annihilation). Existing analysis
(\cite{Baer:2005tw,Profumo:2004ty,Baer:2005ky,Profumo:2005xd}) make it possible to draw
some {\em qualitative} estimates of the cross-section, $\langle\sigma v\rangle_0$,
required to produce sensible signals at future DM search experiments. We provide in
Table~\ref{tab:detection} order-of-magnitude estimates for the value of $\langle\sigma
v\rangle_0$ expected to give observable signals in space-based antimatter (AMS-02,
Pamela, GAPS) and gamma-ray search experiments (GLAST), for two extreme choices of the
galactic dark-matter halo: a cuspy profile (such as the N04 profile) and a cored profile
(such as the Burkert profile).
Different search techniques, such as direct detection, or neutrino flux detection from
the core of the Sun induced by the annihilation of captured neutralinos, critically
depend upon the scattering cross section of neutralinos off nucleons, and the resulting
detection rates are therefore unrelated, in general, to the pair annihilation process
discussed here.

We show in Fig.~\ref{fig:benchSED} the overall SED of Coma as expected from the four
benchmark models described in the Sect.3.2. The predictions are shown for the best fit
N04 profile and for our reference choice for subhalo parameters, and for a mean magnetic
field of $2 \mu G$. None of these benchmark model may provide a reasonable fit to the
radio data. Notice, in addition, the quite dim multi-frequency SED predicted for Coma in
these benchmark models. The largest fluxes are, not surprisingly, obtained for the model
lying in the focus point region ($E^\prime$). In that case, neutralinos mostly annihilate
into gauge bosons, as can be inferred from the spectral shape, which closely resembles
that in Fig.~\ref{fig:sedww}. We therefore conclude that the expectation of astrophysical
signatures from neutralino DM annihilations in the Coma cluster (with natural assumptions
on the dark halo profile, substructures and magnetic field of Coma) is not promising in
the context of the commonly discussed minimal supergravity scenario.

It would be interesting to compare the predictions of the WIMP annihilation for Coma with
the implication of the presence of another population of cosmic rays of different origin
like that, often invoked, produced by acceleration processes in the atmosphere of Coma.
Acceleration scenarios usually produce power-law spectra for the electrons which are
primarily accelerated by shocks or turbulence and are, hence remarkably different from
the source spectra produced by neutralino acceleration. Specifically, acceleration models
do not exhibit a cut-off at the neutralino mass and do not produce the peculiar peaked
$\pi^0$ gamma-ray emission which remains a distinctive feature of neutralino DM models. A
continuum $\pi^0$ gamma-ray emission can be produced in secondary models where the
electrons are produced by proton-proton collision in the cluster atmosphere. But even in
this case the gamma-ray spectrum is likely to be resembled by a power-law shape which
keeps memory of the original acceleration events for the hadrons.
Thus, it will be possible to separate DM annihilation models from acceleration models
based on multi-frequency observations of the hadronic and leptonic components of the
cluster SED.

Finally, It should be noticed that the same problem with the consistent fitting of both
the synchrotron and IC components to the radio and EUV/HXR data of Coma still remains in
both DM and acceleration models, pointing to the fact that these spectral features, if
real, have probably different physical origin.

\section{Summary and conclusions}\label{sec:conclusions}

WIMP annihilations in galaxy cluster inevitably produce high-energy secondary particles
which are able, in turn, to produce a wide SED extended over more than 18 orders of
magnitude in frequency, from radio to gamma-rays.

A consistent analysis of the DM distribution and of its annihilation in the Coma cluster
shows that WIMP annihilation is able to reproduce both the spectral and the spatial
features of the Coma radio halo under reasonable assumptions for the structure of the
intracluster magnetic field. The mild decrease of the magnetic field towards the Coma
center, which reproduces the radial trend of the observed RM distribution in Coma, could
be better tested with a larger dataset of Faraday rotation measures of background radio
sources obtainable with the next generation sensitive radio telescopes (LOFAR, SKA), and
with the help of numerical MHD simulations. Radio data are the main constraint, so far,
to WIMP models.

The ICS emission produced by the same secondary electrons is able, in principle, to
reproduce both the spectrum and the spatial distribution of the EUV emission observed in
Coma, provided that a quite small average magnetic field $B \sim 0.15$ \mug is assumed.
Such low value of the B field is also able to make the radio data and the hard X-ray data
of Coma consistent within a Synchrotron/IC model for their origins.
However, such low magnetic field values in Coma produce an unacceptably large gamma-ray flux, which
exceeds the EGRET upper limit. The gamma-ray constraints are thus the most stringent ones for the
analysis of the astrophysical features of DM annihilations.

In conclusion, the viable models of WIMP annihilation which are consistent with the
available data for Coma yield a nice fit to the radio data but produce relatively low
intensity emission at EUV, X-ray and gamma-ray frequencies. The hadronic gamma-ray
emission could be, nonetheless, detected by the GLAST-LAT detector. These models also
produce negligible heating rates for the kind of non-singular halo profile we worked out
in this paper. It is interesting that the best-fit ($b \bar{b}$) WIMP model with
$M_{\chi}=40$ GeV predicts a detectable SZ effect (with a peculiar spectrum very
different from that of the thermal SZ effect) at the level of $\sim$ 40 to 10 $\mu$K in
the frequency range $\sim 10-200$ GHz, which could be observable with the next generation
high-sensitivity bolometric arrays, space and balloon-borne microwave experiments, like
PLANCK, OLIMPO, APEX, ALMA.

The observational "panorama" offered by the next coming radio, SZ, and gamma-ray
astronomical experiments might produce further constraints on the viable SUSY model for
Coma and for other large-scale cosmic structures.
Direct DM detection experiments have already explored large regions of the most
optimistic SUSY models, and the planned increase in sensitivity of the next-generation
experiments will probably be able to explore even the core of the SUSY models. In this
context, we have shown that indirect DM detection proves to be not only complementary,
but also hardly competitive, especially when a full multi-frequency approach is chosen.
When combined with future accelerator results, such multi-frequency astrophysical
searches might greatly help us to unveil the elusive nature of dark matter.

\acknowledgements{We thank the Referee for the useful comments and suggestions that
allowed us to improve the presentation of our results.
S.C. acknowledges support by PRIN-MIUR under contract No.2004027755$\_$003.}

\appendix

\section{A solution to the diffusion equation}
\label{sec:secondary}

To understand quantitatively the role of the various populations of secondary particles
emitting in the Coma cluster, we have to describe in details their transport, diffusion
and energy loss.
We consider the following diffusion equation ({\em i.e.} neglecting convection and
re-acceleration effects):
\begin{eqnarray}
\frac{\partial}{\partial t}\frac{dn_e}{dE} & = & \nabla \left[ D(E,\vec{x})
\nabla\frac{dn_e}{dE}\right] + \frac{\partial}{\partial E} \left[ b(E,\vec{x})
\frac{dn_e}{dE}\right] \nonumber \\
 & & + Q_e(E,\vec{x})\;.
 \label{diffeq}
\end{eqnarray}
We search for an analytic solution of the diffusion equation in the case of diffusion
coefficient and energy loss term that do not depend on the spatial coordinates, {\em
i.e.} we take:
\begin{eqnarray}
 D & = & D(E) \\
 b & = & b(E)
\end{eqnarray}
and we implement a slight variant of the method introduced in \cite{Baltz1998} and
\cite{Baltz2004}.\\
Let us define the variable $u$ as:
\begin{equation}
b(E) \frac{dn_e}{dE} = - \frac{dn_e}{du}
\end{equation}
which yields
\begin{equation}
u =\int_E^{E_{\rm max}} \frac{dE^{\prime}}{b(E^{\prime})}
\end{equation}
Then, it follows that $b(E) = E/\tau_{loss}$ in terms of the time scale $\tau_{loss}$ for
the energy loss of the relativistic particles, which, for $E_{\rm max} = \infty$, gives
$u=\tau$.\\
The diffusion equation can be rewritten as
\begin{equation}
\left[ - {\frac{\partial}{\partial t}} + D(E) \Delta - \frac{\partial}{\partial u}
\right] \frac{dn_e}{du} = b(E) Q_e(E,\vec{x})\;.
\end{equation}
We search for the Green function $G$ of the operator on the left-hand-side. Consider the
equation for its 4-dimensional Fourier transform ($t\rightarrow \omega$, $\vec{x}
\rightarrow \vec{k}$):
\begin{eqnarray}
\left[-i\omega + D(E) k^2 - {\frac{\partial}{\partial u}} \right] \tilde{G} & = &
\frac{1}{(2\pi)^2} \exp{ \left[-i(\omega t'+\vec{k}\cdot\vec{x}')\right]} \nonumber \\
 & & \cdot \delta(u-u')\; ,
\end{eqnarray}
which has the solution
\begin{eqnarray}
\tilde{G} & = & - \frac{1}{(2\pi)^2} \exp \bigg[-i(\omega t'+\vec{k}\cdot\vec{x}') - i
\omega (u-u') \nonumber \\
 & & - k^2\, \int_{u'}^{u} d\tilde{u} D(\tilde{u})\bigg] \; .
\end{eqnarray}
Transforming back from the Fourier space we find:
\begin{eqnarray}
G_{\rm free} & = & - \frac{1}{(4 \pi (v-v'))^{3/2}}\,
\exp{\left[-\frac{|\vec{x}-\vec{x}'|^2}{4 (v-v')}\right]} \nonumber \\
 & & \cdot \delta\left((t-t')-(u-u')\right)\;
\end{eqnarray}
where we defined $dv \equiv D(u) du$, {\em i.e.} $v=\int_{u_{\rm min}}^u d\tilde{u}
D(\tilde{u})$. The suffix 'free' refers to the fact that there are no boundary conditions
yet. These are implemented with the image charges method.
To apply this technique to galaxy clusters, we can consider the approximation of
spherical symmetry with Green function vanishing at the radius $r_h$. Introducing the set
of image charges $(r_n,\theta_n,\phi_n)=((-1)^n r + 2 n r_h,\theta,\phi)$, one can verify
that
\begin{equation}
G(\vec{r},\vec{Y}) = \sum_{n=-\infty}^{+\infty} (-1)^n G_{\rm free}(\vec{r}_n,\vec{Y})
\end{equation}
fulfills such boundary condition (here $\vec{Y}$ labels the other variables in the Green
function). Moreover, we choose the reference frame in such way that we look at the signal
along the $z$ polar axis ($\cos\theta=1$) so that $|\vec{x}'-\vec{x}_n|^2 =
(r')^2+r_n^2-2\cos\theta' r' r_n$. If the source function does not depend on $\theta'$
and $\phi'$, the integral on these two variables can be performed explicitly and we find
\begin{eqnarray}
\frac{dn_e}{dE}& = & \frac{1}{b(E)} \int_E^{M_\chi}  dE' \frac{1}{[4\pi(v-v')]^{1/2}}
\sum_{n=-\infty}^{+\infty} (-1)^n \int_0^{r_h} dr' \frac{r'}{r_n} \cdot  \nonumber\\
 & & \left[\exp{\left(-\frac{(r'-r_n)^2}{4\,(v-v')}\right)}-
\exp{\left(-\frac{(r'+r_n)^2}{4\,(v-v')}\right)}\right] Q_e(r',E',t')
 \label{eq:full}
\end{eqnarray}
with $t' = t - (u-u')$ (or no time dependence for stationary source). Note that $E'>E$
(energy is lost) and hence $u'<u$, $v'<v$ and $t'<t$.

\subsection{Stationary limit and role of spatial diffusion in Coma}

In the limit of time-independence of the source and electron number density that has
already reached equilibrium, Eq.~(\ref{eq:full}) takes the form:
\begin{equation}
\frac{dn_e}{dE}\left(r,E \right) =  \frac{1}{b(E)} \int_E^{M_\chi}  dE' \;
\widehat{G}\left(r,v-v' \right) Q_e(r,E') \label{eq:full2}
\end{equation}
with
\begin{eqnarray}
\widehat{G}\left(r, \Delta v\right) & = & \frac{1}{[4\pi(\Delta v)]^{1/2}}
 \sum_{n=-\infty}^{+\infty} (-1)^n
\int_0^{r_h} dr' \frac{r'}{r_n} \cdot \\ \nonumber
 & &
\left[\exp{\left(-\frac{(r'-r_n)^2}{4\,\Delta v}\right)}-
\exp{\left(-\frac{(r'+r_n)^2}{4\,\Delta v}\right)}\right]
\frac{n_\chi^2(r')}{n_\chi^2(r)}\;.
 \label{eq:rescaling}
\end{eqnarray}
In the limit in which electrons and positrons lose energy on a timescale much shorter
than the timescale for spatial diffusion, {\em i.e.} if the first term on the r.h.s. of
Eq.~(\ref{diffeq}) can be neglected, the expression for equilibrium number density
becomes:
\begin{equation}
\left({\frac{dn_e}{dE}}\right)_{nsd}\left(r,E \right) =  \frac{1}{b(E)} \int_E^{M_\chi}
dE'  \; Q_e(r,E')\;. \label{eq:nodiff}
\end{equation}
This is analogous to the form in Eq.~(\ref{eq:full2}), except for the factor
$\widehat{G}\left(r,v-v' \right)$ in the integrand: it follows that the latter is the
Green function term which  we need to study to understand whether spatial diffusion is
important or not.

Since we have encoded the dependence on the energy loss term and the diffusion
coefficient in the definition of the variable $v$, we preliminarily study what range of
$\Delta v$ is relevant in the discussion. To do that, we need to specify $D(E)$ and
$b(E)$. For the diffusion coefficient we assume the form:
\begin{equation}
D(E) =  D_0  \frac{d_{B}^{2/3}}{B_\mu^{1/3}} \left(\frac{E}{\rm{1\;GeV}}\right)^{1/3}\;,
\end{equation}
(\cite{ColafrancescoBlasi1998,BlasiColafrancesco1999}) where $d_{B}$ is the minimum scale
of uniformity of the magnetic field in kpc (throughout the paper we assume $d_{B} \simeq
20$ for Coma), $B_\mu$ is the average magnetic field in $\mu$G units, and $D_0$ some
constant that we estimate as $D_0 = 3.1 \times 10^{28}\; \rm{cm}^2 \rm{s}^{-1}$.

The energy loss term is the sum of effects due to Inverse Compton, synchrotron radiation,
Coulomb losses and Bremsstrahlung:
\begin{eqnarray}
 b(E) & = &  b_{IC}(E) + b_{syn}(E) + b_{Coul}(E) + b_{brem}(E) \nonumber \\
 & = & b_{IC}^0 \left(\frac{E}{\rm{1\;GeV}}\right)^{2} + b_{syn}^0 B_\mu^2 \left(\frac{E}{\rm{1\;GeV}}\right)^{2}
 \nonumber \\
 & & + b_{Coul}^0 n \left(1+\log(\gamma/n)/75 \right)  \nonumber \\
 & & + b_{brem}^0 n \left(\log(\gamma/n)+0.36 \right) \;.
 \end{eqnarray}
Here $n$ is the mean number density of thermal electrons in $\rm{cm}^{-3}$ (see
Eq.~(\ref{eq:gas}), the average over space gives about $n \simeq 1.3 \; 10^{-3}$),
$\gamma \equiv E/m_e$ and we find $b_{IC}^0 \simeq 0.25$, $b_{syn}^0 \simeq 0.0254$,
$b_{Coul}^0 \simeq 6.13$ and $b_{brem}^0 \simeq 1.51$, all in units of $10^{-16}\;
\rm{GeV}\, \rm{s}^{-1}$. For GeV electrons and positrons the Inverse Compton and
synchrotron terms dominate (see Fig.\ref{fig:losses}).

To get a feeling about what is the electron/positron energy range which will be of
interest when considering the radio emissivity, we can resort to the "monochromatic"
approximation, with relativistic particles of a given energy $E$ radiating at a single
frequency, namely the peak frequency:
\begin{equation}
\nu \simeq 0.29 \frac{3}{2} \frac{e B}{2\pi m_e c}  \simeq (4.7 \rm{MHz}) B_\mu
\left(\frac{E}{\rm{GeV}}\right)^{2} \,. \label{eq:mono}
\end{equation}
Since radio data on Coma extend down to about $30 \rm{MHz}$, for magnetic fields not much
larger than 10~$\mu$G, this translates into radiating particles with energies larger than
about 1~GeV.

\begin{figure*}[!t]
\begin{center}
\hspace*{-0.5cm}\includegraphics[scale=0.55]{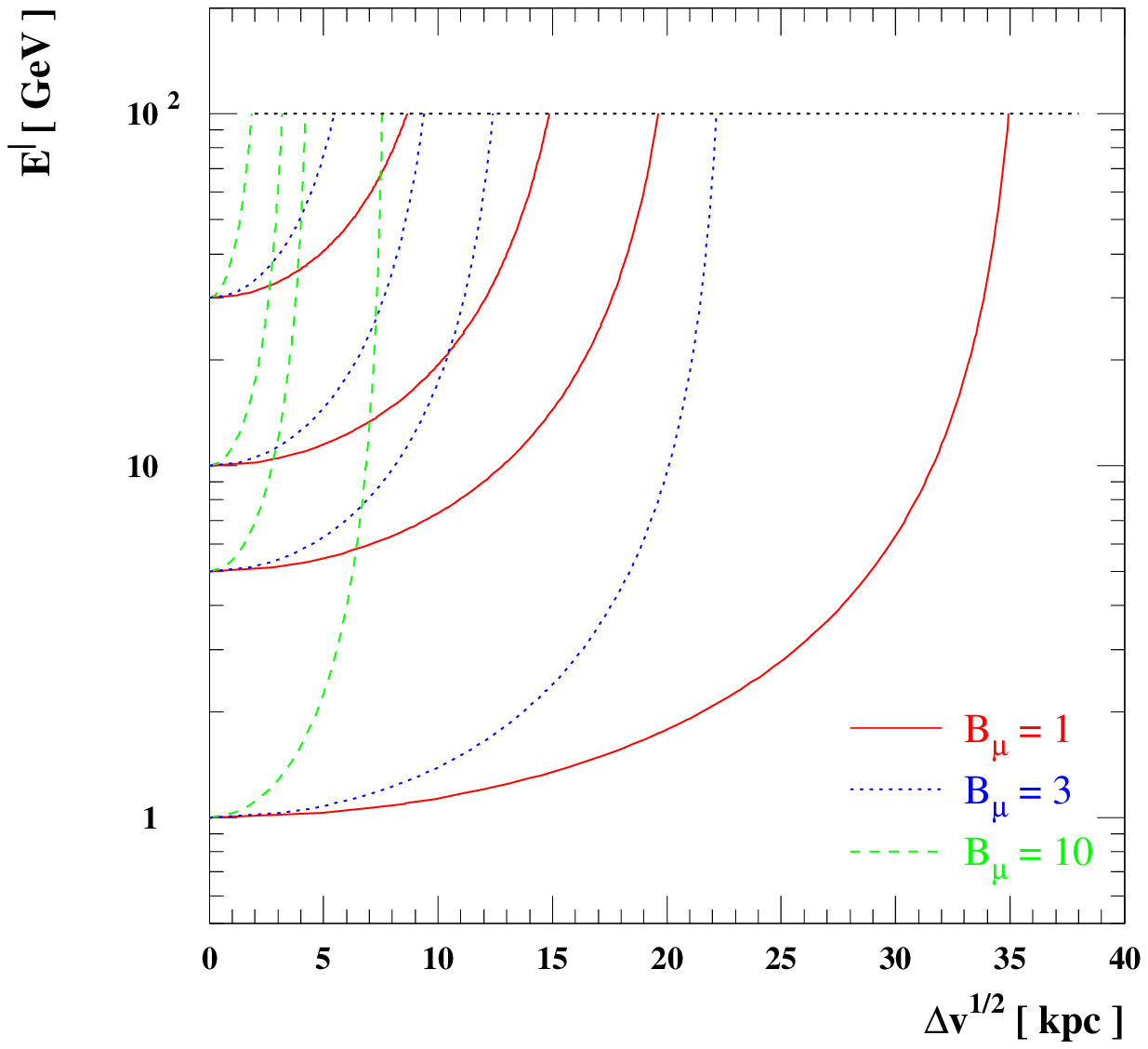}
\quad\includegraphics[scale=0.55]{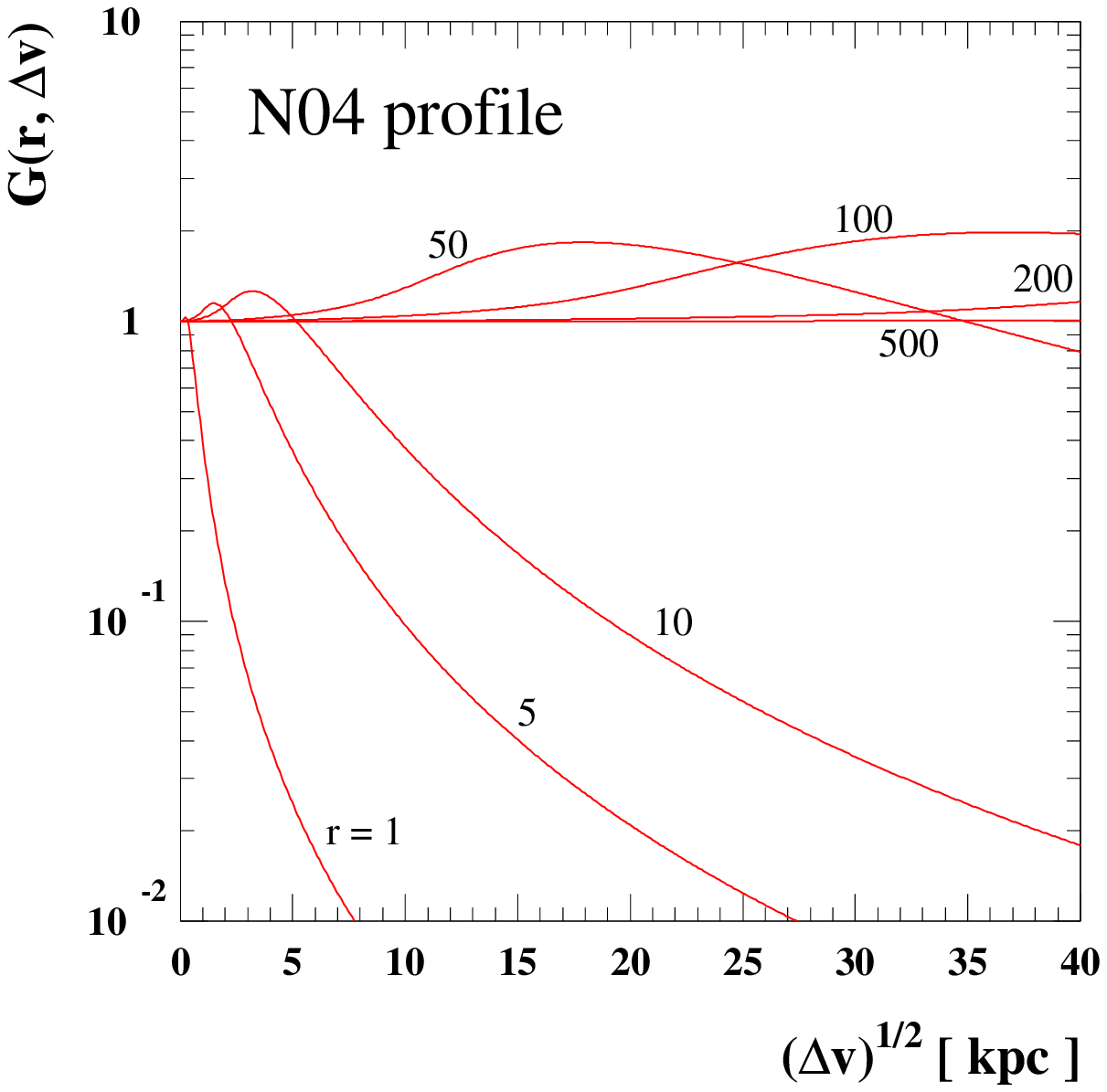}\\
\end{center}
\caption{Left: The figure shows the distance $(\Delta v)^{1/2}$ which, on average, a
electron covers while losing energy from its energy at emission $E^{\prime}$ and the
energy when it interacts $E$, for a few values of $E$: 30~GeV, 10~GeV, 5~GeV and 1~GeV,
and for a few values of the magnetic field (in $\mu$G); we are focusing on a WIMP of mass
100~GeV, hence cutting $E^{\prime} < 100$~GeV. Right: Green function $\widehat{G}$ as a
function of $(\Delta v)^{1/2}$, for a few values of the radial coordinate  $r$ (in kpc)
and in case the DM halo of Coma is described by a N04 profile with $M_{vir} = 0.9 \,
10^{15} \msun h^{-1}$ and $c_{vir} = 10$.} \label{fig:deltav}
\end{figure*}
%
\begin{figure*}[!t]
\begin{center}
\hspace*{-0.5cm}\includegraphics[scale=0.55]{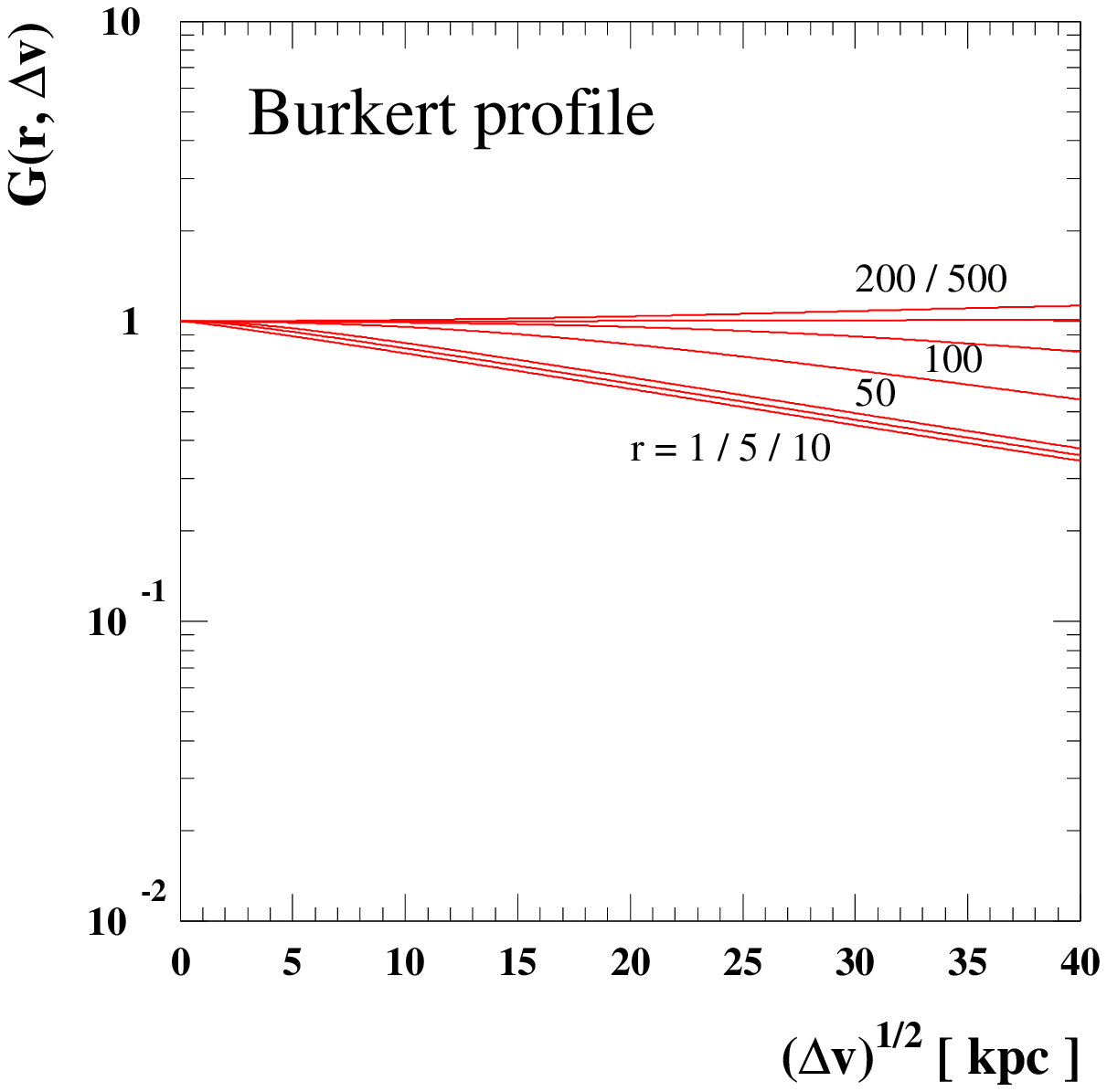}
\quad\includegraphics[scale=0.55]{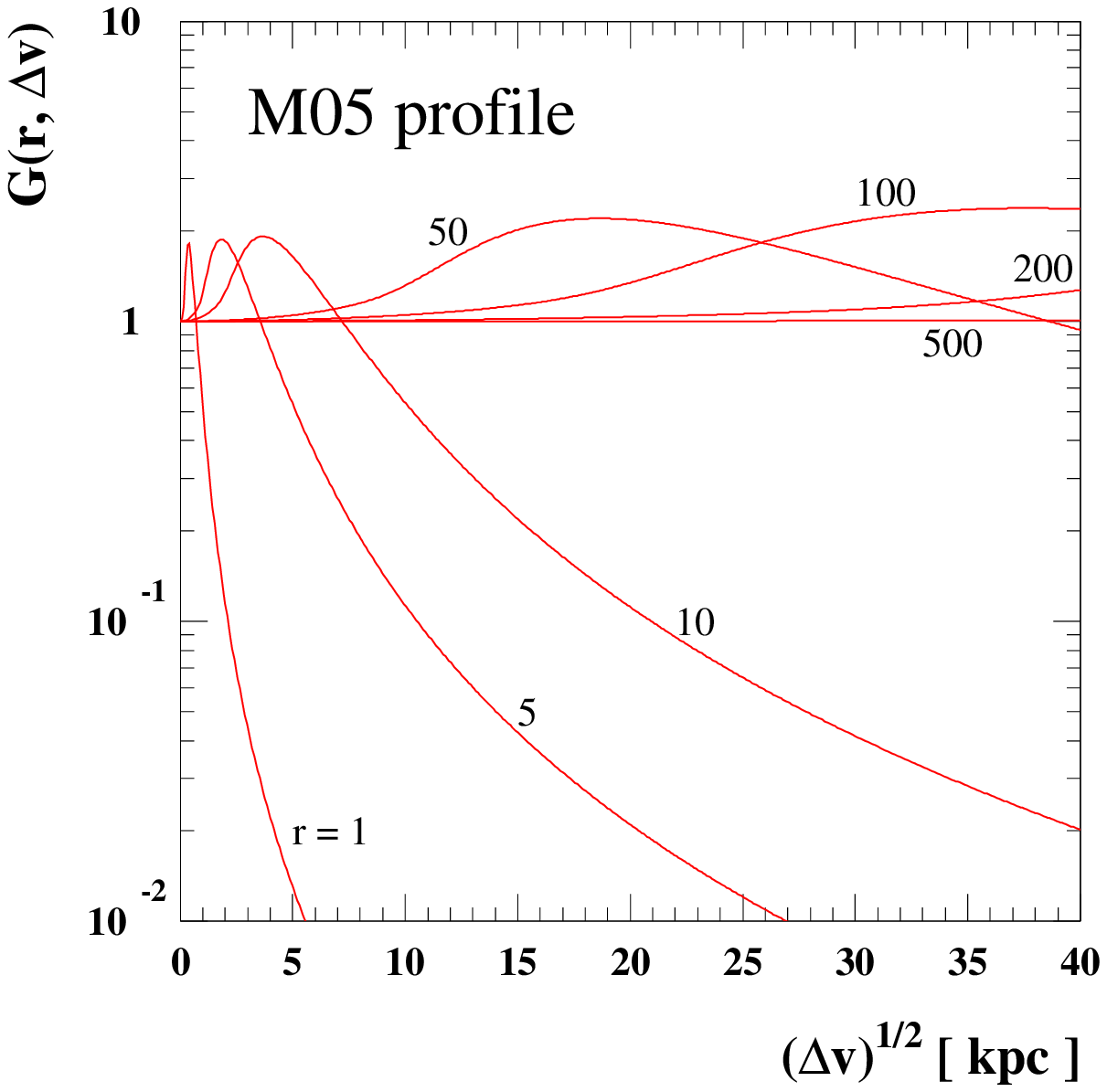}\\
\end{center}
\caption{Green function $\widehat{G}$ as a function of $(\Delta v)^{1/2}$, for a few
values of the radial coordinate  $r$ (in kpc) and in case the DM halo of Coma is
described by a cored Burkert profile with $M_{vir} = 0.9 \, 10^{15} \msun h^{-1}$ and
$c_{vir} = 10$ and for the Diemand et al. profile with the same parameters.}
\label{fig:deltav2}
\end{figure*}

As a sample case, in Fig.~\ref{fig:deltav} we consider a WIMP mass $M_\chi = 100$~GeV and
sketch the mapping between the energy $E' \in (E,M_\chi)$, with $E$ some reference energy
after diffusion, and the square root of $\Delta v = v - v' \equiv v(E) - v(E')$, for a
few values of $E$ and of the mean magnetic field $B_\mu$. We find as largest value
$(\Delta v)^{1/2} \sim 35$~kpc, corresponding to $E = 1$~GeV and $B_\mu = 1$~$\mu$G; the
maximum value of $\Delta v$ diminishes rapidly when increasing $E$ or $B_\mu$.

On the right-hand side of Fig.~\ref{fig:deltav}, we plot $\widehat{G}$ as a function of
$(\Delta v)^{1/2}$ for a few values of the radial coordinate  $r$ and in case the DM halo
of Coma is described by a N04 profile. In the very central part of the halo, there are
significant departures of the value of $\widehat{G}$ from unity, on scales $(\Delta
v)^{1/2}$ at which, for the given radius $r$, the mean squared value of the DM profile is
significantly different from the square of the value of the profile at $r$. Note,
however, that this effect is confined in the innermost region of the cluster,
corresponding to an angular size of $\approx 1-2$ arcmin. We then expect that taking into
account spatial diffusion will modify only slightly the predictions for the radio surface
brightness distribution from moderate to large radial distances in Coma.

In Fig.~\ref{fig:deltav2}, we plot $\widehat{G}$ as a function of $(\Delta v)^{1/2}$ for
the same values of the radial coordinate  $r$ as in Fig.~\ref{fig:deltav} but now in case
the DM halo of Coma is described by a Burkert profile. It is clear that departures from
unity are essentially negligible even in the inner portion of the halo and, hence,
spatial diffusion can be safely neglected for all practical purposes in this case.

To get a more physical insight on the reason why spatial diffusion can be neglected, it
is useful to consider the  following qualitative solution (see, {\em e.g.},
\cite{Colafrancesco2005a}) for the average electron density
 \be
 {d n_e(E,r) \over dE} \approx [Q_e(E,r) \tau_{loss}] \times {V_{s} \over V_s + V_o} \times {\tau_{D} \over \tau_{D}+ \tau_{loss}}
 \label{eq.solution.qualitative}
 \ee
which resumes the relevant aspects of the transport equation (Eq.~\ref{diffeq}). Here,
$V_s \propto R^3_h$ and $V_o \propto \lambda^3(E)$ are the volumes occupied by the DM
source and the one occupied by the diffusing electrons which travel a distance
$\lambda(E) \approx [D(E) \cdot \tau_{loss}(E)]^{1/2}$ before loosing much of their
initial energy. The relevant time scales in Eq.~(\ref{diffeq}) are the diffusion
time-scale, $\tau_D \approx R^2_h/ D(E)$, and the energy loss time-scale $\tau_{loss} =
E/b_e(E)$, where $D$ is again the diffusion coefficient for which we can assume the
generic scaling $D(E) = \tilde{D}_0 (E/E_0)^{\gamma} B^{- \gamma}$, and $b(E)$ the
electron energy loss per unit time at energy $E$.
%

\begin{figure}[!t]
\begin{center}
\hspace*{-0.5cm}\includegraphics[scale=0.35]{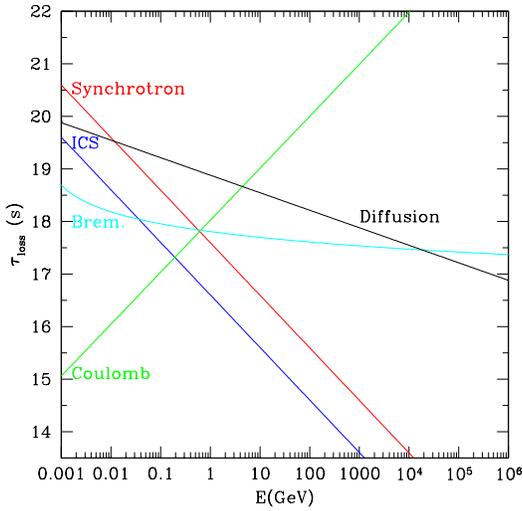}
\end{center}
\caption{A comparison among the time scales for the energy losses due to various
mechanisms (as labeled in the figure) and the time scale for diffusion (black solid
curve) in a cluster of size $r_h=1$ Mpc. A uniform magnetic field of value $B = 1 \mu$G
and a thermal gas density $n=1.3 \cdot 10^{-3}$ cm$^{-3}$ have been assumed in the
computations.}
 \label{fig:losses}
\end{figure}
%
For $E > E_* = (\tilde{D}_0 E_0/R^2_h b_0 B_{\mu}^{\gamma})^{1/(1-\gamma)}$ (for
simplicity we have kept leading terms only, implementing $b(E) \simeq b_0(B_{\mu}) (E
/GeV)^2+ b_{Coul}$), the condition $\tau_D > \tau_{loss}$ (and consistently $\lambda(E) <
R_h$) holds, the diffusion is not relevant and the solution of Eq.~(\ref{diffeq}) is
$dn_e / dE \sim Q_e(E,r) \tau_{loss}$ and shows an energy spectrum $\sim Q(E) \cdot
E^{-1}$. This situation ($\lambda(E) < R_h$, $\tau_D > \tau_{loss}$) applies to the
regime of galaxy clusters which we discuss here for the specific case of Coma, as one can
see from Fig.~\ref{fig:losses}.\\
For $E < E_*$, the condition $\tau_D < \tau_{loss}$ (and consistently $\lambda(E)
> R_h$) holds, the diffusion is relevant and the solution of Eq.~(\ref{diffeq}) is $dn_e / dE \sim [Q_e(E,r)
\tau_D] \times (V_{s} / V_o)$ and shows an energy spectrum $\sim Q(E) \cdot E^{(2-5
\gamma)/2}$ which is flatter or equal to the previous case for reasonable values $\gamma
= 1/3 - 1$. This last situation ($\lambda(E) > R_h$, $\tau_D < \tau_{loss}$) applies to
the regime of dwarf galaxies and we will discuss this case more specifically elsewhere
(\cite{inprep}).

\begin{figure}[!t]
\begin{center}
\hspace*{-0.5cm}\includegraphics[scale=0.35]{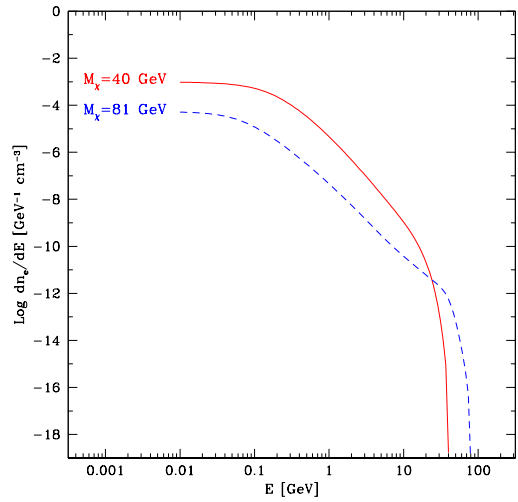}\\
\end{center}
\caption{The electron equilibrium spectra calculated at the center of Coma as obtained
for a soft spectrum due to a $b \bar{b}$ annihilation final state (solid line, model with
$M_{\chi} = 40$~GeV) and of a hard spectrum due to a $W^+ W^-$ channel (dashed line,
model with $M_{\chi} = 81$~GeV). }
 \label{fig:eqspectra}
\end{figure}

Fig.\ref{fig:eqspectra} shows the energy shape of the electron equilibrium spectra
derived in our approach for a ($b \bar{b}$) model with $M_{\chi} = 40$ GeV and for a
$W^+W^-$ model with $M_{\chi} = 81$ GeV. The astrophysical predictions of these two
models will be extensively discussed in the following.
We notice that the energy losses in the diffusion equation erase almost completely the
details of the electron source spectra (see Fig.~\ref{fig:esource}). The equilibrium
spectra are generally characterized by three different regions: i) a low-energy plateau
at $E \simlt 0.1$ GeV with a constant value of $dn_e / dE$ which remains almost constant
down to the electron rest-mass energy; ii) an almost power-law branch at $ 0.1 M_{\chi}
\simlt E \simlt 0.5 M_{\chi}$ which is steeper in the softer $b \bar{b}$ annihilation
final state with respect to the hard spectrum due to a $W^+ W^-$ channel; and iii) a
sharp cut-off at the energy corresponding to the neutralino mass which marks the natural
maximum energy of the secondary electron spectra.
We will show in the next Sect.\ref{sec:multiwave} how these three branches of the
electron equilibrium spectra will provide observable features in the multi-frequency
spectrum of Coma and can, consequently, be used to constrain the neutralino model.



\end{document}